\documentclass[12pt]{article}
\pdfoutput=1
\usepackage{latexsym,amsmath,amssymb, graphicx, subfigure}
\renewcommand{\thefootnote}{\fnsymbol{footnote}}

\usepackage{mathrsfs }
\usepackage{amsmath}
\usepackage{amsfonts}

\numberwithin{equation}{section}
	
\def\doubleset#1#2{\bgroup%
\def\doit#1#2{%
\setbox\dblsetbox=\hbox{$\cstyle #1$}%
\raise#2\ht\dblsetbox\copy\dblsetbox%
\hskip-\wd\dblsetbox%
\raise-#2\ht\dblsetbox\box\dblsetbox}%
\mathchoice%
{\def\cstyle{\displaystyle}\doit#1#2}%
{\def\cstyle{\textstyle}\doit#1#2}%
{\def\cstyle{\scriptstyle}\doit#1#2}%
{\def\cstyle{\scriptscriptstyle}\doit#1#2}\egroup}
\def\underarrow#1{\vbox{\ialign{##\crcr$\hfil\displaystyle
 {#1}\hfil$\crcr\noalign{\kern1pt\nointerlineskip}$\longrightarrow$\crcr}}}

\def\Tr{{\rm Tr}}
\def\vol{{\rm vol}}

\def\IL{\relax{\rm I\kern-.18em L}}
\def\IH{\relax{\rm I\kern-.18em H}}
\def\IR{{\mathbb R}}
\def\IC{\mathbb C}
\def\IB{\relax{\rm I\kern-.18em B}}
\def\ID{\relax{\rm I\kern-.18em D}}
\def\IE{\relax{\rm I\kern-.18em E}}
\def\IF{\relax{\rm I\kern-.18em F}}
\def\IZ{{\mathbb Z}}
\def\IG{\relax\hbox{$\inbar\kern-.3em{\rm G}$}}
\def\IGa{\relax\hbox{${\rm I}\kern-.18em\Gamma$}}
\def\IH{\relax{\rm I\kern-.18em H}}
\def\II{\relax{\rm I\kern-.18em I}}
\def\IK{\relax{\rm I\kern-.18em K}}
\def\IP{\relax{\rm I\kern-.18em P}}
\def\IQ{\relax\hbox{$\inbar\kern-.3em{\rm Q}$}}

\def\hat{\widehat}
\def\CM {{\cal M}}
\def\CN {{\cal N}}

\def\CD {{\cal D}}
\def\CF {{\cal F}}
\def\CJ {{\cal J}}

\def\CO {{\cal O}}

\def\CG {{\cal G}}
\def\CH {{\cal H}}

\def\CB {{\cal B}}

\def\CA{{\cal A}}

\def\CQ{{\cal Q}}

\def\mod{{\rm mod}}

\def\gof{ \Gamma^0(4)}

\def\dau{ {da \over  du} }

\def\inbar{\,\vrule height1.5ex width.4pt depth0pt}

\def\Xminus{{X \backslash D}}

\newbox\dblsetbox

\newcommand{\bR}{\mathbb{R}}

\newcommand{\bZ}{\mathbb{Z}}

\newcommand{\dirac}{D\mspace{-13mu}/\mspace{4mu}}

\newlength{\extraspace}
\newlength{\extraspaces}
\newcommand{\be}{\begin{equation}
\addtolength{\abovedisplayskip}{\extraspaces}
\addtolength{\belowdisplayskip}{\extraspaces}
\addtolength{\abovedisplayshortskip}{\extraspace}
\addtolength{\belowdisplayshortskip}{\extraspace}}
\newcommand{\ee}{\end{equation}}

\newcommand{\ba}{\begin{eqnarray}
\addtolength{\abovedisplayskip}{\extraspaces}
\addtolength{\belowdisplayskip}{\extraspaces}
\addtolength{\abovedisplayshortskip}{\extraspace}
\addtolength{\belowdisplayshortskip}{\extraspace}}
\newcommand{\ea}{\end{eqnarray}}

\newcommand{\bd}{\begin{displaymath}
\addtolength{\abovedisplayskip}{\extraspaces}
\addtolength{\belowdisplayskip}{\extraspaces}
\addtolength{\abovedisplayshortskip}{\extraspace}
\addtolength{\belowdisplayshortskip}{\extraspace}}
\newcommand{\ed}{\end{displaymath}}

\newcounter{saveeqn}

\newcommand{\newsection}[1]{
\vspace{12mm} \pagebreak[3] \addtocounter{section}{1}
\setcounter{equation}{0} \setcounter{subsection}{0}
\noindent{\bf \thesection. #1} \nopagebreak
\medskip
\nopagebreak
\addcontentsline{toc}{section}{\thesection. #1}}

\newcommand{\newsubsection}[1]{
\vspace{0.8cm} \pagebreak[3] \addtocounter{subsection}{1}
\setcounter{subsubsection}{0}
\noindent{ \it \thesubsection. #1} \nopagebreak \vspace{2mm}
\nopagebreak
\addcontentsline{toc}{subsection}{\thesubsection. #1}}

\begin{document}
\addtolength{\baselineskip}{1.5mm}

\thispagestyle{empty}

\vbox{} \vspace{1.5cm}

\begin{center}
\centerline{\LARGE{Integration Over The $u$-Plane In Donaldson}}
\bigskip
\centerline{\LARGE{ Theory With Surface Operators}}      
\bigskip

\vspace{1.0cm} 

{Meng-Chwan~Tan \footnote{email: mengchwan@theory.caltech.edu}}
\\[3mm]
{\it California Institute of Technology, \\
Pasadena, CA 91125, USA} \\ [2mm]
and \\
{\it Department of Physics\\
National University of Singapore \\
Singapore 119260}\\[8mm]
\end{center}


\vspace{1.0 cm}

\centerline{\bf Abstract}\smallskip \noindent

We generalize the analysis by Moore and Witten in~[hep-th/9709193], and consider integration over the $u$-plane in Donaldson theory  with surface operators on a smooth four-manifold $X$. Several novel aspects will be developed in the process; like a physical interpretation of the ``ramified'' Donaldson and Seiberg-Witten invariants, and the concept of curved surface operators which are necessarily topological at the outset.  Elegant physical proofs -- rooted in $R$-anomaly cancellations and modular invariance over the $u$-plane -- of various seminal results in four-dimensional geometric topology obtained by Kronheimer-Mrowka~\cite{structure, structure 1} -- such as a universal formula relating the ``ramified'' and ordinary Donaldson invariants, and a generalization of the celebrated Thom conjecture --  will be furnished.  Wall-crossing and blow-up formulas of these ``ramified'' invariants which have not been computed  in the mathematical literature before, as well as a generalization and  a Seiberg-Witten analog of the universal formula as implied by an electric-magnetic duality of trivially-embedded surface operators in $X$, will also be presented, among other things.

\newpage

\renewcommand{\thefootnote}{\arabic{footnote}}
\setcounter{footnote}{0}

\tableofcontents 

\newsection{Introduction And Summary}

It has been a while since the connection between Donaldson theory~\cite{donaldson} and twisted Yang-Mills theory with ${\cal N} =2$ supersymmetry in four-dimensions, was first elucidated in~\cite{Witten} via what is known today as Donaldson-Witten theory. Since then, there have been several important developments, especially on the physics side~\cite{SW}, which have culminated in our present understanding of the  Donaldson invariants of  four-manifolds in terms of the ``simpler'' Seiberg-Witten monopole invariants~\cite{monopoles, Moore-Witten}. 
 
The effort~\cite{monopoles, sym}  that led to this understanding, however, was partially motivated by a prescient structure theorem proved by Kronheimer and Mrowka in~\cite{structure, structure 1}. One of the central ingredients in their proof is this notion of  ``ramified'' Donaldson invariants; these invariants can be understood as extensions of the ordinary Donaldson invariants to four-manifold with embedded surfaces~\cite{KM1, KM2}.  Such embedded surfaces can be physically interpreted as two-dimensional analogs of the 't Hooft line or loop operator; they were first used in the physics literature some twenty years ago to probe the dynamics of gauge theory and black holes~\cite{Preskill}-\cite{Bucher}. Despite the apparent versatility of these embedded surfaces,  their physical and mathematical applications have been somewhat limited ever since, with the exception of a few examples~\cite{Braverman}.

Nevertheless, there has been a strong revival of interest in these objects~\cite{n=4} following the recent work of Gukov and Witten in~\cite{Gukov-Witten}; in their work, generalizations of such embedded surfaces -- coined thereafter as surface operators   -- were studied in the context of a  four-dimensional $\CN =4$ gauge-theoretic interpretation of the ``ramified''  geometric Langlands conjecture~\cite{langlands}. Of late, these surface operators have also been analyzed in string theory~\cite{2}-\cite{8},  as well as  in four-dimensional ${\cal N} =2$ gauge theories~\cite{mine, loop, davide}.  
 
 In light of these new developments, and the fact that the proof of the  structure theorem by Kronheimer and Mrowka lies squarely on the use of such embedded surfaces, it would be timely and interesting to generalize the analysis in~\cite{Moore-Witten} to include arbitrarily-embedded surface operators. This is the main motivation for our present work. 
 

 \medskip\noindent{\it Organization And Summary Of The Paper}

The organization of the paper and its main results can be summarized as follows. In $\S$2, we will elucidate the relation between Donaldson-Witten theory with surface operators and the ``ramified'' Donaldson invariants. In particular, we will provide a physical interpretation of the ``ramified'' Donaldson invariants -- as defined mathematically by Kronheimer and Mrowka in~\cite{structure, KM1, KM2} -- in terms of certain correlation functions in Donaldson-Witten theory with surface operators. In turn, a physical proof --  based on $R$-anomaly cancellations -- of a minimal genus formula for  embedded two-surfaces in a simply-connected four-manifold obtained by Kronheimer and Mrowka in~\cite{KM1, KM2}, can be derived.

In $\S$3,  we study the connection between the  microscopic $SU(2)$ theory and the  macroscopic $U(1)$ theory. A careful analysis reveals that the discrete global $R$-symmetries of the ordinary theory without surface operators carry over to the ``ramified'' theory with surface operators. Consequently, the exact forms of the ``ramified'' and ordinary 4d low-energy prepotential are found to be similar. As such, the Coulomb branch of the ``ramified'' theory can be described using the same $u$-plane technology employed by Moore and Witten in~\cite{Moore-Witten}. A physical interpretation of the ``ramfied'' Seiberg-Witten invariants will also be appropriately furnished at this point. We then proceed to analyze a string-theoretic realization --   introduced earlier by L.F.~Alday et al. in~\cite{loop} -- of our field-theoretic setup. In doing so, we will be able to corroborate some of our field-theoretic results regarding the low-energy vacuum structure of the ``ramified'' gauge theory,  in terms of M2 and M5-branes from a purely geometrical viewpoint. In the process, a detailed account of some of the claims made in~\cite{loop} will also be presented. Last but not least, we will, from a D-branes perspective, also introduce the concept of curved surface operators which are necessarily topological at the outset. 

In $\S$4, we will ascertain the effective theory on the $u$-plane and its pertinent attributes. In particular, we find that the correlation function -- which corresponds to the generating function of the ``ramified'' Donaldson invariants -- will receive contributions from a generic region in the $u$-plane if and only if $b^+_2(X) \leq 1$. The additional contact terms in the low-energy operator observables arising from the insertion of a surface operator will also be determined.     

In $\S$5, we will derive the explicit form of the ``ramified'' $u$-plane integral. We will also verify -- using the technology of modular forms and generalized theta-functions -- the modular invariance of the ``ramified'' $u$-plane integral that is expected on physical grounds. In doing so, we will be led to a special condition on the ``quantum'' parameter of a nontrivially-embedded surface operator that was also obtained in $\S$3; this condition must be satisfied in order for modular invariance to hold.

In $\S$6, we will compute some ``ramified'' wall-crossing formulas that will be essential to ascertaining the complete formulas of the generating function of the ``ramified'' Donaldson invariants in $\S$7. In particular, we will obtain wall-crossing formulas of the ``ramified'' Donaldson invariants -- which have not been computed in the mathematical literature before -- that nonetheless reduce,  in the appropriate limit,  to well-established  wall-crossing formulas of the ordinary Donaldson invariants.  

In $\S$7, we will determine the ``ramified'' Seiberg-Witten contributions at the monopole and dyonic points, and obtain the complete formulas of the generating function of the ``ramified'' Donaldson invariants. We will also present  ``ramified'' generalizations of Witten's ``magic formula'' that underlie the relation between the ``ramified'' Donaldson and Seiberg-Witten invariants. 

In $\S$8 and $\S$9, as a check and an application of our physical computations obtained hitherto,  we will first furnish elegant physical proofs of various seminal results in four-dimensional geometric topology obtained by Kronheimer and Mrowka in~\cite{structure, structure 1, KM1, KM2}. The crucial ingredients in these proofs are the special condition on the ``quantum'' parameter of a nontrivially-embedded surface operator found in $\S$3 and $\S$5, and the vanishing of the generic $u$-plane contribution to the generating function of the ``ramified'' Donaldson invariants when $b^+_2(X) > 1$. Then, by considering certain surface operators which are trivially-embedded in $X$, we find that electric-magnetic duality will imply that Kronheimer and Mrowka's universal formula in~\cite{structure} ought to generalize to include embedded two-surfaces with \emph{zero} self-intersection number; furthermore, it will also imply that a Seiberg-Witten analog of the universal formula ought to exist -- a result which is consistent with and generalizes an independent mathematical computation performed in~\cite{Appendix 1}.

And lastly, in $\S$10 and $\S$11, we will --  as means of computing the ``ramified'' Donaldson invariants exactly with the aid of the wall-crossing formulas obtained earlier -- establish the vanishing of the ``ramified'' $u$-plane integral in some chambers and derive an associated blowup formula.  

An appendix of some useful expressions concerning elliptic curves, congruence subgroups, and modular forms, will also be provided.

\newsection{A Physical Interpretation Of The ``Ramified'' Donaldson Invariants}
 
 In this section, we will furnish a physical interpretation of the ``ramified''  Donaldson invariants and their corresponding generating function, in terms of the relevant correlation functions of (non)-local observables in Donaldson-Witten theory with surface operators. We then proceed to use our results to derive a physical proof of a minimal genus formula for embedded two-manifolds  obtained by Kronheimer and Mrowka in~\cite{KM2}.

 \newsubsection{Embedded Surfaces And The ``Ramified'' Donaldson Invariants}

Let us first review the mathematical definition of embedded surfaces and the ``ramified'' Donaldson invariants by Kronheimer and Mrowka (denoted as KM henceforth) in~\cite{structure}.  To this end, let $X$ be a smooth, compact, simply-connected, oriented 
four-manifold with Riemannian metric $\bar g$, and let $E\rightarrow X$
be an $SO(3)$-bundle over $X$ (that is, a rank-three real vector
bundle with a metric).

\bigskip\noindent{\it Embedded Surfaces}

An embedded surface $D$ is characterized by a two-submanifold of $X$  that is a complex curve of genus $g$ and self-intersection number $Q(D)$. Consider  the case where the second Stiefel-Whitney class $w_2(E) =0$; the structure group of $E$ can then be lifted to its $SU(2)$ double-cover. In the neighborhood of $D$, one can choose a decomposition of  $E$ as
\be
E = {L} \oplus {L}^{-1},
\label{decompose}
\ee
where $L$ is a complex line bundle over $X$. In the presence of $D$, the connection matrix of $E$ restricted to $X \backslash  D$ (in the \emph{real} Lie algebra) will look  like 
\be
A = \alpha  d \theta + \cdots,
\label{connection}
\ee
where $\alpha$ is a real number valued in the generator 
\be
\begin{bmatrix} 
 1 & \ 0\\ 
0 & -1
\end{bmatrix}
\label{frak t}
\ee
of the Cartan subalgebra $\frak t$, $\theta$ is the angular variable of the coordinate $z = re^{i \theta}$ of the plane normal to $D$, and the ellipses refer to the ordinary terms that are regular near $D$. Notice that since $d\theta = i dz / z$, the connection is singular as $z \to 0$, that is,  as one approaches $D$. In any case, the singularity in the connection induces the following gauge-invariant holonomy 
\be
\textrm{exp} ( 2 \pi  \alpha) 
\label{holonomy}
\ee
around any small circle that links $D$. Hence, if the holonomy is trivial, we are back to considering ordinary connections on $E$. Therefore, $\alpha$ actually takes values in  $\mathbb T$, the maximal torus of the gauge group with Lie algebra $\frak t$. As we shall see shortly, this mathematical definition of embedded surfaces will coincide with our (more general) physical definition of surface operators. 

\bigskip\noindent{\it The ``Ramified'' Donaldson Invariants}

In analogy with the original formulation of Donaldson theory~\cite{donaldson},  KM introduced the notion of ``ramified'' Donaldson invariants -- that is, Donaldson invariants of $X$ with an embedded surface $D$. According to KM~\cite{structure, KM2}, the ``ramified'' Donaldson polynomials $\CD'_E$ can be defined as polynomials
on the homology of $\Xminus$ with real coefficients:\footnote{To be precise, KM actually defines the ``ramified'' Donaldson polynomials to be the map $\CD'_E : \textrm{Sym}[H_0(\Xminus,\IR) \oplus H_2(\Xminus,\IR)] \otimes \wedge*H_1(\Xminus, \IR)  \rightarrow \IR$. However, their definition can be truncated as shown, in accordance with Donaldson's original formulation in~\cite{donaldson}.}
\be
\CD'_E: H_0(\Xminus,\IR)
\oplus H_2(\Xminus,\IR) \rightarrow \IR .
\ee
Assigning degree  $4$ to
$p\in H_0(\Xminus,\IR)$ and $2$ to $S \in H_2(\Xminus,\IR)$,
the degree $s$  polynomial  may be expanded as
\be{
\CD'_E(p,S) = \sum_{2m + 4 t = s} S^m p^t d^{k'}_{m,t},
\label{Donaldson polynomial}
}\ee
where $s$ is the dimension of the moduli space ${\cal M}'$
of gauge-inequivalent classes of anti-self-dual  connections  on $E$ restricted to $X \backslash D$ with first Pontrjagin number $p_1(E)$ and instanton number $k' = - \int_{X \backslash D} p_1(E) /4$. The numbers $d^{k'}_{m,t}$ -- in other words, the ``ramified'' Donaldson invariants of X -- can be defined as in  Donaldson theory in terms of intersection theory on the moduli space ${\cal M}'$; for maps 
\begin{eqnarray}
\label{map of defn}
p \in  H_0(\Xminus,\IR) & \rightarrow & \Omega^0(p) \in H^4(\CM'), \nonumber \\
S \in H_2(\Xminus,\IR)  & \rightarrow & \Omega^2 (S) \in H^2 (\CM'), 
\end{eqnarray}
the ``ramified'' Donaldson invariants can be written as
\be
d^{k'}_{m,t} = \int_{\CM'} [\Omega^0(p)]^t \wedge \Omega^2 (S_{i_1}) \wedge \dots \wedge \Omega^2(S_{i_m}).
\label{ram DW}
\ee
Moreover, one can also
package the ``ramified'' Donaldson polynomials into
a generating function: by summing over all topological
types of bundle $E$ with fixed $\xi=w_2(E)$ (where $\xi$ may be non-vanishing in general) but varying $k'$, the generating function can be defined as 
\be
{\bf Z}'_{\xi, \bar g}(p,S) =
\sum^{\infty}_{k' =0}\sum_{m \geq 0, t\geq 0} {S^m \over  m!}{ p^t \over  t!} d^{k'}_{m,t}.
\label{Donaldson generating function}
\ee
Clearly, ${\bf Z}'_{\xi, \bar g}$ depends on the class $w_2(E)$
 but not on the instanton number $k'$  (as this
has been summed over).

\bigskip\noindent{\it About The Moduli Space Of ``Ramified'' Instantons}

Another relevant result by KM is the following. Assuming that there are no reducible connections on $E$ restricted to $X \backslash D$,  ${\cal M}'$ -- which we will hereon refer to as the moduli space of ``ramified'' instantons -- will be a smooth manifold  of finite dimension
\be
s = 8k - {3 \over 2}(\chi + \sigma) + 4 {l} - 2( g-1)
\ee
for $\it any$ nontrivial value of $\alpha$.  Here, $\chi$ and $\sigma$ are the Euler characteristic and signature of $X$, and for $\xi =0$, the integer $k$ is given by 
\be
k = -{1 \over 8 \pi^2} \int_X \textrm{Tr} \hspace{0.1cm} F \wedge F,
\ee  
where $F$ is the curvature of the bundle $E$ over $X$, and $\textrm{Tr}$ is the trace in the two-dimensional representation of $SU(2)$.  The integer $l$ -- called the monopole number by KM -- is given by
\be
{l} = - \int_D c_1(L).
\ee
Here, $c_1(L) = - F_L / 2\pi$, where $F_L$ is the curvature of $L$; thus, $l$ measures the degree of the reduction of $E$ near $D$. 

As we shall see below, $l$ will depend explicitly on $\alpha$ because the singular term in the connection $A$ will result in a singularity proportional to $\alpha$ along $D$ in the field strength (extended over $D$).  Likewise, $k$ will also depend explicitly on $\alpha$. Thus, the invariance of $s$ must mean that both $l$ and $k$ will vary with $\alpha$  in such a way as to keep it fixed for any nontrivial value of $\alpha$.  

\bigskip\noindent{\it Topological Invariance Of ${\bf Z}'_{\xi, \bar g}$}

Let $b^+_2$ denote the self-dual part of the second Betti number of $X$. According to KM (see $\S$7 of~\cite{KM2}), if $b^+_2  > 1$, ${\bf Z}'_{\xi, \bar g}$ is independent of the metric $\bar g$ and hence,  just like the generating function of the ordinary Donaldson invariants, defines invariants of the smooth structure of $X$. This is consistent with the fact that for $b^+_2 \geq 3$ (and $b_1 =0$), the ``ramified'' Donaldson invariants can be expressed solely in terms of the ordinary Donaldson invariants (see Theorem 5.10 of~\cite{structure}). We will give a physical proof of this remarkable mathematical result in $\S$8.  

However, if $b_2^+=1$, we run into the phenomenon of chambers; ${\bf Z}'_{\xi, \bar g}$ will jump as we move across a ``wall'' in the space of metrics on $X$. We will demonstrate this mathematical phenomenon via a purely physical approach in $\S$6.

\newsubsection{Surface Operators In Pure $SU(2)$ Theory With ${\CN} =2$ Supersymmetry}

\bigskip\noindent{\it Supersymmetric Surface Operators}

As in~\cite{mine}, we would like to define surface operators along $D$ which are compatible with ${\cal N} =2$ supersymmetry.  In other words,  they should be characterized by solutions to the supersymmetric field configurations of the underlying gauge theory on $X$ that are singular along $D$. 

In order to ascertain what these solutions are, first note that any supersymmetric field configuration of a theory must obey the conditions implied by setting the supersymmetric variations of the fermions to zero. In the original (untwisted) theory without surface operators, this implies that any supersymmetric field configuration must obey $F =0$ and $\nabla_\mu a = 0$, where $a$ is a scalar field in the ${\cal N} = 2$ vector multiplet~\cite{Marcos}. Let us assume for simplicity the trivial solution $a=0$ to the condition $\nabla_\mu a =0$ (so that the relevant moduli space is non-singular); this means that any supersymmetric field configuration must be consistent with $\it{irreducible}$ flat connections on $X$ that obey $F=0$. Consequently, any surface operator along $D$ that is supposed to be supersymmetric and compatible with the underlying ${\cal N} =2$ supersymmetry,  ought to correspond to a $\it{flat}$ irreducible connection on $E$ restricted to $X \backslash D$ which has the required singularity  along $D$.\footnote{This prescription of considering connections on the bundle $E$ restricted to $X \backslash D$ whenever one inserts a surface operator that introduces a field singularity along $D$,  is just a two-dimensional analog of the prescription one adopts when inserting an 't Hooft loop operator in the theory. See $\S$10.1 of~\cite{QFT2} for a detailed explanation of the latter.}  Let us for convenience choose  the singularity of the connection along $D$ to be of the form shown in (\ref{connection}). Then,  since $d (\alpha d \theta) = 2 \pi \alpha \delta_D$, where $\delta_D$ is a delta two-form  which is Poincar\'e dual to $D$,  our surface operator will equivalently correspond to a $\it{flat}$ irreducible connection on a bundle $E'$ over $X$ whose field strength is $F' = F - 2\pi \alpha \delta_D$, where $F$ is the field strength of the bundle $E$ over $X$.\footnote{To justify this statement, note that the instanton number $\tilde k$ of the bundle $E$ over $X \backslash D$ is (in the mathematical convention)  given by $\tilde k = k +  2 \alpha {l} - \alpha^2 D \cap D$, where $k$ is the instanton number of the bundle $E$ over $X$ with curvature $F$, and $l$ is the  monopole number  (cf.~eqn.~(1.7) of~\cite{KM1}). On the other hand,  the instanton number $k'$ of the bundle $E'$ over $X$ with curvature $F' = F - 2\pi \alpha \delta_D$ is (in the physical convention) given by $k' = - {1\over 8 \pi^2} \int_X \textrm {Tr} F' \wedge F' = k +  2 \alpha {l} - \alpha^2 D \cap D$. Hence, we find that the expressions for $\tilde k$ and $k'$ coincide, reinforcing the notion that the bundle $E$ over $X \backslash D$ can be equivalently interpreted as the bundle $E'$ over $X$. Of course, for $F'$ to qualify as a nontrivial field strength, $D$ must be a homology cycle of $X$, so that $\delta_D$ (like $F$) is in an appropriate cohomology class  of $X$.}  In other words, a supersymmetric surface operator will correspond to a gauge field solution over  $X$ that satisfies 
\be
F = 2 \pi \alpha \delta_D
\label{surface operator}
\ee
along $D$.  Indeed,  the singular term in $A$ of (\ref{connection}) that is associated with the inclusion of an embedded surface, is such a solution. Thus, our physical definition of supersymmetric surface operators coincides with the mathematical definition of embedded surfaces. 

Some comments on (\ref{surface operator}) are in order. Note that even though $\alpha$ is formally defined in (\ref{connection}) to be $\frak t$-valued, we saw that it actually takes values in the maximal torus $\mathbb T$.
Since $\mathbb T = {\frak t} / \Lambda_{\textrm{cochar}}$, where $\Lambda_{\textrm{cochar}}$ is the cocharacter lattice of the underlying gauge group,  (\ref{surface operator}) appears to be unnatural, since one is free to subject $F$ to a shift by an element of $\Lambda_{\textrm{cochar}}$. This can be remedied by lifting $\alpha$ in (\ref{surface operator}) from ${\frak t} / \Lambda_{\textrm{cochar}}$ to ${\frak t}$. Equivalently, this corresponds to a choice of an extension of the bundle $E$  over $D$ -- something that was implicit in our preceding discussion. 

\bigskip\noindent{The ``Quantum'' Parameter $\eta$}

With an extension of the bundle $E$ over $D$, the restriction of the field strength $F$ to $D$ will be $\frak t$-valued. Hence, we roughly have an abelian gauge theory in two dimensions along $D$. As such, one can generalize the physical definition of the surface operator, and introduce a two-dimensional theta-like angle $\eta$ as an additional ``quantum'' parameter which enters in the Euclidean path-integral via the phase
\be
\textrm{exp}\left( 2 \pi i \hspace{0.00cm} \textrm{Tr} \hspace{0.1cm}  \eta \frak m \right),
\label{eta term}
\ee
where $\frak m = \int_D F/ 2 \pi$. Since $F$ restricted to $D$ is $\frak t$-valued, and since the monopole number $l =  \int_D F_L / 2\pi$ is an integer, it will mean that $\frak m$ must take values in the subset of diagonal, traceless $2 \times 2$ matrices -- which generate the maximal torus $\mathbb T$ -- that have \emph{integer} entries only; that is, ${\frak m} \in \Lambda_{\textrm{cochar}}$. Also, values of $\eta$  that correspond to a nontrivial phase must be such that $\textrm{Tr} \hspace{0.1cm} \eta \frak m$ is \emph{non-integral}. Because $\textrm{Tr} \hspace{0.1cm}  {\frak m}' {\frak m}$ is an integer if ${\frak m}' \in \Lambda_{\textrm{cochar}}$, it will mean that  $\eta$  must takes values in ${\frak t} / \Lambda_{\textrm{cochar}} = \mathbb T$. Just like $\alpha$, one can shift $\eta$ by an element of $\Lambda_{\textrm{cochar}}$ whilst leaving the theory invariant.\footnote{This characteristic of $\eta$ is consistent with an $S$-duality in the corresponding, low-energy  effective  abelian theory which maps  $(\alpha_{\textrm{eff}}, \eta_{\textrm{eff}}) \to (\eta_{\textrm{eff}}, - \alpha_{\textrm{eff}})$~\cite{mine}.}  As we shall see in a later section, $\eta$ will play a crucial role in our physical proof of KM's relation between the ``ramified'' and ordinary Donaldson invariants.

\bigskip\noindent{\it A Point On Nontrivially-Embedded Surface Operators}

  More can also be said about the ``classical'' parameter $\alpha$ as follows. In the case when the surface operator is trivially-embedded in $X$ -- that is, $X= D' \times D$ and the normal bundle to $D$ is hence trivial -- the self-intersection number
\be
D \cap D = \int_{X} \delta_D \wedge \delta_D
\label{DcapD}
\ee
vanishes. On the other hand, for a nontrivially-embedded surface operator supported on $D \subset X$, the normal bundle is nontrivial, and the intersection number is non-zero. The surface operator is then defined by the gauge field with singularity in (\ref{connection}) in each normal plane.

When the surface operators are nontrivially-embedded, there is a condition on the allowed gauge transformations that one can invoke in the physical theory~\cite{Gukov-Witten}. Let us explain this for when the underlying gauge group is $U(1)$ with gauge bundle $L$, first. Since there is a singularity of $2 \pi \alpha \delta_D$ in the (abelian) field strength $F_L$ restricted to $D$, we find, using (\ref{DcapD}), that $\int_D F_L/2 \pi  = \alpha  D \cap D \ \textrm{mod}  \ \mathbb Z$. Since $\int_D F_L/2 \pi = l$ is always an integer, we must have
\be
\alpha  D \cap D \in \mathbb Z.
\label{intersection number}
\ee
This observation has a generalization to the non-abelian case of interest: if $\alpha \to f(\alpha)$ is any real-valued linear function on $\frak t$ that takes integer values on the cocharacter lattice $\Lambda_{\textrm{cochar}}$, then
\be
f(\alpha) D \cap D \in \mathbb Z.
\label{intersection number 2}
\ee

Now consider a gauge transformation -- in the normal plane -- by the following $\mathbb T$-valued function 
\be
(r, \theta) \to \textrm{exp} (\theta u),
\label{twisted gauge tx}
\ee
where $u \in \frak t$;  its effect is to shift $\alpha \to \alpha + u$ whilst leaving the holonomy of the gauge connection in (\ref{holonomy}) -- that underlies the \emph{effective} ``ramification'' of the theory -- unchanged. Clearly, the only gauge transformations of this kind which can be globally-defined along $D$, are those whereby the corresponding shifts in $\alpha$ are compatible with (\ref{intersection number 2}) (or (\ref{intersection number}), if the gauge group is $U(1)$). For nontrivial $\alpha$, since $D\cap D \in \mathbb Z$, the relevant gauge transformations are such that $u \notin \Lambda_{\textrm{cochar}}$ -- in other words, $\textrm{exp} (2 \pi u) \neq 1$, and the gauge transformations are not single-valued under $\theta \to \theta + 2\pi$. 

For a non-simply-connected gauge group such as $SO(3)$, we have $\textrm{exp} (2 \pi u) = y$, where $y$ is a nontrivial element of the center of $SU(2)$. In this case, the gauge transformation changes the topology of the $SO(3)$-bundle $E$ via the shift $w_2(E) \to w_2(E) + y [D]$, where $[D]$ is the cohomology class Poincar\'e dual to $D$. This means that $SO(3)$ theories with different values of $w_2(E)$ and suitably related values of $\alpha$, are physically equivalent. On the other hand, if the gauge group is the simply-connected $SU(2)$ double-cover of $SO(3)$ -- since $w_2(E) =0$, regardless -- we have $y[D] = 0$.

\bigskip\noindent{\it The Effective Field Strength In The Presence Of Surface Operators}

In any gauge theory, supersymmetric or not, the kinetic term of the gauge field has a positive-definite real part. As such, the Euclidean path-integral (which is what we will eventually be interested in) will be non-zero if and only if the contributions to the kinetic term  are strictly non-singular. Therefore, as a result of the singularity (\ref{surface operator}) when one includes a surface operator in the theory, the effective field strength in the Lagrangian that will contribute non-vanishingly to the path-integral must be a shifted version of the field strength $F$. In other words, whenever we have a surface operator along $D$, one ought to study the action with field strength $F' = F - 2\pi \alpha \delta_D$ instead of $F$. This means that the various fields of the theory are necessarily coupled to the gauge field $A'$  with field strength $F'$. This important fact was first pointed out in~\cite{Gukov-Witten}, and further exploited in~\cite{mine} to prove an $S$-duality in a general, abelian ${\CN} = 2$ theory without matter in the presence of surface operators.

Since the surface operator does not introduce any singularities in the other fields of the underlying theory, it suffices to modify only the field strength to obtain the effective Lagrangian. Nevertheless, in a different theory whereby supersymmetric configurations involve not just the field strength $F$ but also the other fields, a supersymmetric surface operator can give rise to a singularity in the other fields as well. For example, in the pure ${\cal N}=4$  theory considered in~\cite{Gukov-Witten}, supersymmetric configurations involve the Higgs field $\varsigma$ in addition to the field strength $F$. Consequently, the inclusion of a surface operator in the ${\cal N} =4$ theory that is supposedly supersymmetric, can also result in a singularity  in $\varsigma$ along $D$.

\newsubsection{The Topological Twist And The ``Descent'' Equations}

In order to make the paper as self-contained as possible, we shall now review some facts about Donaldson-Witten theory~\cite{Witten} and  the construction of its non-local observables via the ``descent'' equations. The familiar reader may wish to skip this subsection, although it should be emphasized that some less well-known but nonetheless relevant aspects of the theory will be discussed in what follows. 

\bigskip\noindent{\it The Topological Twist}

\def\1{{\bf 1}}
\def\2{{\bf 2}}
\def\R{{\bf R}}

 The  recipe employed in~\cite{Witten} towards constructing a $\it{pure}$ $\CN =2$ topological field theory on $X$ with gauge group $SU(2)$ or $SO(3)$, also known as Donaldson-Witten theory, can be explained as follows. Firstly, let us recall some standard facts about $\CN=2$ supersymmetric theories in four dimensions. Firstly, note that on flat ${\bf R}^4$, the double cover $Spin(4)$ of the rotation group is isomorphic to $SU(2)_-\times SU(2)_+$.  The two factors of $SU(2)$ act respectively on the $-$ and $+$ spin representations of $Spin(4)$; let us label these as $S_-$ and $S_+$.  In the case where there are no central charges, the relevant $\CN=2$ theories also possess an additional $U(2) \cong SU(2) \times U(1)$ group of $R$-symmetries; let us call it $SU(2)_R \times U(1)_R$.  Under
$SU(2)_-\times SU(2)_+\times SU(2)_R \times U(1)_R$, the eight supercharges transform
as $(\2,\1,\2)^{-1}\oplus (\1,\2,\2)^{1}$, where $\1$ and $\2$ represent
the trivial  and  two-dimensional representations of $SU(2)$. By introducing $SU(2)_-$ indices $A,B,C=1,2$, $SU(2)_+$ indices
$\dot A,\dot B,\dot C=1,2$, and $SU(2)_R$ indices $I,J,K=1,2$,
we can therefore denote the four supercharges that transform as $(\2,\1,\2)^{-1}$ by $Q_A^I$, and the other four that transform as $(\1,\2,\2)^{1}$ by $\overline Q_{\dot A J}$.  
They satisfy an important anticommutator relation
\be
\{ Q_{A I}, \overline Q_{\dot B J}  \} = 2 \epsilon_{IJ} \sigma^\mu_{A \dot B} P_\mu,
\label{impt relation}
\ee
where $\mu = 0,\dots, 3$ are the coordinate indices on ${\R}^4$, $\epsilon_{IJ}$ is the $SL(2, \IC)$ invariant tensor, $\sigma^\mu_{A \dot B} = (1, \sigma^1, \sigma^2, \sigma^3)$ (where the $\sigma^i$'s are the usual Pauli matrices), and $P_\mu = \partial / \partial{x^\mu}$ is the momentum generator. 

To obtain a topological field theory that can be defined on an arbitrary Riemannian  four-manifold $X$, one will need a prescription of ``twisting'' that enables at least one linear combination of the eight supercharges to transform as a pure singlet under the Poincar\'e group; the supersymmetries generated by these supercharges can then be globally-defined on $\it any$ $X$. One  such means of doing so is to  consider a diagonal
subgroup $SU(2)'$ of $SU(2)_+\times SU(2)_R$, and  to introduce a ``new'' definition
of the Poincar\'e group of $\R^4$ in which the rotation group is $Spin(4)'=  SU(2)_-\times SU(2)'$ instead of $Spin(4)=SU(2)_-\times SU(2)_+$.  This step is tantamount to identifying the $SU(2)_R$ indices $I, J, K$ with the $SU(2)_+$ indices $\dot A,\dot B,\dot C$ -- consequently, the supercharges can be written as $Q_A^{\dot B}$ and $\overline Q_{\dot A \dot B}$. Among them,  one can define a $Spin(4)'$-invariant combination $\overline \CQ=\epsilon^{\dot A \dot B} \overline Q_{\dot A \dot B} $,
and a $Spin(4)'$-vector $G_{\mu} = {i \over  4} (\overline \sigma_{\mu})^{\dot B A}Q_{A \dot B}$, where the $\overline \sigma_\mu$'s are (up to a sign) given by the usual Pauli matrices. They obey
\be
\overline \CQ^2 = 0
\label{Q^2=0}
\ee    
and 
\be
\{ G_\mu, G_\nu \} = 0.
\label{GG}
\ee
In addition, from (\ref{impt relation}), we find that 
\be
\{\overline \CQ, G_\mu \} = \partial_\mu. 
\label{QG=d}
\ee
Note that (\ref{QG=d}) can also be viewed as a specialization of the relation
 \be
T'_{\mu \nu} = \{ \overline \CQ, G_{\mu \nu} \}
\label{stress tensor}
\ee
-- where $T'_{\mu \nu}$ is the ``new'' stress tensor --  to
$
T'_{0 \mu} = \{ \overline \CQ, G_{ 0 \mu} \} = P_\mu,
$
where $G_{0 \mu} = G_\mu$. 

Note that the ``new'' stress tensor $T'$ differs from the original stress tensor $T$ by a derivative term that is irrelevant on flat ${\bf R}^4$;  in this case, the twist is just a different way of looking at the same theory, as implied by the above prescription. However, on curved $X$, the twisted theory is really different from the original theory. Nevertheless, the relations  (\ref{Q^2=0}), (\ref{QG=d}) and (\ref{stress tensor}) will continue to hold, regardless.

Since $\overline \CQ$ generates a global supersymmetry that is well-defined on all of $X$, one can certainly find  $\overline \CQ$-invariant observables in the twisted theory. Moreover, since $\overline \CQ^2 =0$, one can define a $\overline \CQ$-cohomology class of observables which are equivalent to one another modulo $\overline \CQ$-exact terms; in other words, quantities of the form $\{\overline \CQ, \dots \}$  are cohomologically equal to zero in any correlation function. Consequently, since the stress tensor $T'_{\mu \nu}$ -- which quantifies the variation of the action under a change in the metric on $X$ -- is of the form $ T'_{\mu\nu} = \{ \overline \CQ, G_{\mu \nu}\}$, any correlation function of $\overline \CQ$-invariant observables is invariant under deformations of the metric on $X$.  Moreover, this also means that the correlation function is independent of the point of insertion of the observables in $X$; the theory is thus topological from a physical point of view. 

\bigskip\noindent{\it The ``Descent'' Equations}

\def\cq{{\overline \CQ}}

An important consequence of (\ref{GG}) and (\ref{QG=d}) is the following. Assume that we start with a $\cq$-invariant operator $\CO^{(0)}(x)$.  Then, consider the operator
\be
\CO^{(n)}_{\mu_1 \mu_2 \dots \mu_n} (x)  = G_{\mu_1} G_{\mu_2} \dots G_{\mu_n} \CO^{(0)}(x), \quad n = 1, \dots, \textrm{dim} X,
\ee
where $G_\mu \CO$ is shorthand for $[G_\mu, \CO \} = G_\mu \CO- (-1)^{\CO} \CO G_\mu$. From (\ref{GG}), we learn that the $G_{\mu_i}$'s  anticommute. Hence, we can interpret
\be
\CO^{(n)} = {1 \over n!} \CO^{(n)}_{\mu_1 \mu_2 \dots \mu_n}  dx^{\mu_1} \wedge \dots \wedge dx^{\mu_n} 
\ee
as an $n$-form in spacetime. By using (\ref{QG=d}), and noting the $\cq$-invariance of $\CO^{(0)}(x)$, we find that the above forms satisfy the ``descent'' equations
\be
d \CO^{(n)} = [ \cq, \CO^{(n+1)} \}, \quad n \geq 0, 
\label{descent eqns}
\ee
where $d$ denotes the spacetime exterior derivative on $X$. The operators $\CO^{(n)}$ are called the topological ``descendants'' of $\CO^{(0)}$. 

Because of (\ref{descent eqns}), one can construct the following $\textrm {\it non-local}$ $\cq$-invariant observables 
\be
{I}_n ({\gamma_n}) = \int_{\gamma_n} \CO^{(n)}, 
\ee
where $\gamma_n \in H_n(X)$. Indeed, 
$
[ \cq, {I}_n ({\gamma_n}) \} = \int_{\gamma_n} [ \cq, \CO^{(n)} \} = \int_{\gamma_n} d \CO^{(n-1)} = \int_{\partial\gamma_n}\CO^{(n-1)} = 0,
$
since $\partial{\gamma_n} = 0$. Likewise, if $\gamma_n$ is trivial in homology (that is, if it is $\partial$-exact), then ${I}_n ({\gamma_n})$ is $\cq$-exact because of (\ref{descent eqns}). Thus, we can construct a family of  $\cq$-invariant observables 
\be
{I}_n (\gamma_{i_n}), \quad i_n = 1, \dots, b_n; \quad n = 1, \dots, \textrm{dim} X,  
\ee
that are in one-to-one correspondence with homology classes of $X$. 

An important point to emphasize about the above ``descent'' procedure is that it is independent of the Lagrangian description and the scale of the theory. Therefore, even though the Lagrangian of the corresponding macroscopic theory that will concern us later is not unique and subject to duality transformations, one can still employ the ``descent'' procedure to construct non-local observables in the theory, which, can then, be canonically matched to the microscopic non-local observables constructed above, once the low-energy counterpart of $\CO^{(0)}$ is identified.

Also,  all other consistent choices of diagonal subgroups lead to the same underlying topological theory -- that is, there is only $\it one$ way to ``twist'' the theory. Moreover, the ``twist'' does not modify the ($U(1)$) $R$-charge or ``ghost'' number of the supercharges. As such, since $Q$ and $\overline Q$ have $R$-charge $-1$ and $1$, respectively, $G_\mu$  and $\cq$ will have $R$-charge $-1$ and $1$, too. The $U(1)_R$ symmetry is a classical symmetry that may be anomalous in the quantum theory because of instanton effects; 
we shall elaborate on this in the next section. 

\newsubsection{$\cq$-Invariant Observables In Donaldson-Witten Theory With Surface Operators}

The fields in the corresponding Donaldson-Witten (DW) theory with surface operators are a gauge field $A'_{\mu}$ and  a fermi field $\psi_\mu$ that are spacetime one-forms,  a complex bose field $\phi$ and a fermi field $\zeta$ that are spacetime scalars, and a fermi field $\chi_{\mu \nu}$ and a bose auxiliary field $K_{\mu \nu}$ that are $\textrm{\it self-dual}$ spacetime two-forms. These fields are all part of the underlying $\CN =2$ vector multiplet, and are therefore valued in the adjoint representation of the Lie algebra of the $SU(2)$ or $SO(3)$ gauge group. Since the spinor fields are now all differential forms on $X$ after ``twisting'', there  no longer is an obstruction to their existence on non-Spin $X$; thus, we can certainly consider examples where $w_2(X) \neq 0$. The supersymmetry transformations of these fields that leave the topological Lagrangian invariant are (cf.~\cite{Marcos}) 
 \begin{equation}
 \label{susy tx - non-abelian}
\begin{matrix}
[\cq, A' ] =  \psi,
\quad & \quad
\{\cq, \psi \}   =  2 \sqrt{2} \nabla' \phi, \\ \\
[\cq, \phi]   = 0,
\quad & \quad
[\cq,\phi^ \dagger] =   2 \sqrt{2} i \zeta, \\ \\
\{\cq, \zeta\} = [\phi, \phi^\dagger],
\quad & \qquad
 \{\cq, \chi_+\}   = i( F'_+  - K_+),  \\\\
 \hspace{2.0cm} [\cq, K_+]   =   (\nabla' \psi)_+ + \sqrt 2 [\phi, \chi]_+,   \\
\end{matrix}
\end{equation}
where the $+$ subscript indicates the self-dual components of the indicated fields, and $\nabla'$ refers to the covariant derivative with respect to the $SU(2)$ or $SO(3)$ gauge field $A'$. Notice that the field strength in the above transformation relations is $F' = F - 2\pi \alpha \delta_D$ and not $F$; as explained earlier, $F$ needs to be replaced by $F'$ in the Lagrangian when one includes surface operators in the theory.  The topological Lagrangian is thus invariant under transformations involving $F'$. 

\bigskip\noindent{\it Non-Local Observables From The Topological ``Descendants''}

We now wish to employ the ``descent'' procedure to construct $\cq$-invariant non-local observables explicitly.  To do so, one would need the action of $G_\mu$ on the above fields; it is given by (cf.~\cite{Marcos}) 

\begin{equation}
 \label{G action}
\begin{matrix}
[G , \phi]   = {1 \over 2 \sqrt{2}} \psi,
\quad & \quad
[G, \phi^{\dagger}]   = 0,  \\ \\
\{G , \psi \} = - 2(F'_-  + K_+ ),
 \quad & \quad
[G,  A']   = i \zeta - 2i \chi,  \\ \\
\{G, \zeta\}   = -{i \sqrt{2} \over 4} \nabla' \phi^{\dagger},
\quad & \quad
\{G, \chi_+ \}   = -{3i \sqrt{2} \over 8} * \nabla' \phi^{\dagger},  \\ \\
[G, K_+]  = -{3i \over 4} * \nabla' \zeta
+ {3 i \over 2} \nabla' \chi_+, \cr 
\end{matrix}
\end{equation}
where $\ast$ denotes the Hodge-dual of the relevant fields.

An appropriate choice of $\CO^{(0)}$ is a gauge-invariant local operator that is $\cq$-closed but not $\cq$-exact. Since $[\cq, \phi] =0$ and $\phi \neq [\cq, \dots \}$, a suitable candidate would be an $SU(2)$-invariant polynomial in $\phi$ -- that is, $\CO^{(0)} = {1 \over 8 \pi^2}\textrm{Tr} (\phi^2)$.\footnote{For gauge group $SU(N)$, ``Tr'' means the trace in the $N$-dimensional representation. Equivalently, for  $SU(2)$ or $SO(3)$, ``Tr'' is $1/4$ of the trace in the adjoint representation.} 

Since the action of $G_\mu$ commutes with a gauge transformation, one can use (\ref{G action}) to construct the following gauge-invariant topological ``descendants'' of $\CO^{(0)}$:
\begin{eqnarray}
\CO^{(1)}  & = &  {1 \over  8 \sqrt 2  \pi^2}\textrm{Tr} \left(\phi \psi \right),  \nonumber \\
\CO^{(2)} & = & {1 \over 4 \pi^2} \textrm{Tr} \left (   {1 \over 8} \psi \wedge \psi   - {1 \over \sqrt 2} \phi [F'_-  + K_+ ] \right).
\label{descendants in SU(2)}
\end{eqnarray}
Of course, one can continue with the ``descent'' procedure to construct $\CO^{(3)}$ and $\CO^{(4)}$ on a  four-manifold $X$. However, since our main objective is to furnish a physical interpretation of the ``ramified'' Donaldson invariants as defined by KM, it suffices for us to consider just the  observables $I_0(P) = \CO^{(0)}(P)$ and $I_2(\gamma_2) = \int_{\gamma_2} \CO^{(2)}$ for $\it any$ $P \in H_0(X)$ and $\gamma_2 \in H_2(X)$. Moreover, in this paper, we shall be concerned with examples where $b_1 =0$ only; thus, there are no one-cycles $\gamma_1$ to consider.   

As we shall explain shortly, one can take the semiclassical limit when computing the desired correlation functions of the observables $I_0(P)$ and $I_2(\gamma_2)$. In this limit, it suffices to consider -- in the correlation functions -- only the quadratic fluctuations of the $\phi$ field and the classical configurations of the $A'$ and $\psi$ fields which minimize the action. These classical configurations obey the constraints obtained by setting the variations of the fermi fields in (\ref{susy tx - non-abelian}) to zero. Therefore, from $\{\cq, \chi_+\}  = 0$, we find that we can, for all our purposes, make the substitution $K_+ = F'_+$ in the expression for $\CO^{(2)}$ in (\ref{descendants in SU(2)}).  This means that one can actually consider $I_2(\gamma_2)$ to be given by
\be
I_2(\gamma_2) = {1 \over 4 \pi^2} \int_X \delta_{\gamma_2} \wedge  \textrm{Tr} \left ( {1 \over 8} \psi \wedge \psi  - {1 \over \sqrt 2} \phi [F - 2 \pi \alpha \delta_D] \right).
\label{I_2(gamma)}
\ee

In light of the map (\ref{map of defn}) which underlies the mathematical definition of the ``ramified'' Donaldson invariants, let us take $P = p$ and $\gamma_2 = S$, where $p \in H_0(X \backslash D)$ and $S \in H_2(X \backslash D)$.  Then, since $\int_X \delta_S \wedge \delta_D =0$, from (\ref{I_2(gamma)}), we see that the relevant $\cq$-invariant  observables ought to be given by 
\begin{eqnarray}
I_0(p) & = &{1 \over 8 \pi^2}\textrm{Tr} \left(\phi^2\right)(p),  \nonumber \\ 
I_2(S) & = & {1 \over 4 \pi^2} \int_S  \textrm{Tr} \left ( {1 \over 8} \psi \wedge \psi   - {1 \over \sqrt 2} \phi F \right).
\label{I's}
\end{eqnarray}
As desired, the definition of  $I_2(S)$ is $\it independent$ of the extension of the bundle $E$ over $D$. 

Last but not least,  note that any gauge field -- and in our case  $A'$ -- necessarily has $R$-charge 0.  Thus, since $\cq$ and $G_\mu$ have $R$-charge $1$ and $-1$,  one can, according to (\ref{susy tx - non-abelian}) and (\ref{G action}),  consistently define the complex scalar $\phi$ and the fermi field $\psi$ to have  $R$-charge   2 and 1. In turn, this implies that $I_0(p)$ and $I_2(S)$ have $R$-charge 4 and 2, respectively. This coincides with the degree of 4 and $2$ assigned to $p$ and $S$ in the ``ramified'' Donaldson polynomial $\CD'_E(p,S)$ of (\ref{Donaldson polynomial}). This indicates that a consistent physical interpretation of the ``ramified'' Donaldson invariants in terms of the observables $I_0(p)$ and $I_2(S)$ ought to exist.

\newsubsection{A Physical Interpretation Of The ``Ramified'' Donaldson Invariants} 

\def\mo{{\mathscr O}}


\medskip\noindent{\it Correlation Functions Of $\cq$-Invariant Observables}

\def\mI{{\mathscr I}}

Another important feature of the twisted theory is that the action can be written as (cf.~\cite{Marcos})
\be
S_E =  { \{ \cq , V \}  \over e^2}  +   {i\Theta \over 8 \pi^2} \int_X \textrm{Tr} \hspace{0.1cm} F' \wedge F'  -  i  \int_X  \textrm{Tr}  \hspace{0.1cm} \eta  \delta_D \wedge F'
\label{S}
\ee
for some fermionic operator $V$ of $R$-charge -1 and scaling dimension 0, and complexified gauge coupling $\tau = {4 \pi i \over e^2} + {\Theta \over 2 \pi}$. The action is thus $\cq$-exact up to purely topological terms that are therefore metric-independent and $\cq$-invariant. In fact, since the topological terms are $\cq$-invariant, it will mean -- via the interpretation of $\cq$ as a $d$-operator in (\ref{descent eqns}), and Poincar\'e's lemma -- that one can express them locally as $\{\cq, \dots \}$. In other words, we can write $S_E = \{\cq, V'\} /e^2$, where $V'$ has the same properties as $V$. (For an explicit demonstration of this in the ordinary case, see~\cite{Witten}). This allows for a simplification of the problem at hand, as follows. 

Consider the set of $\cq$-invariant observables ${\mathscr O}_i$ and their correlation function
\be
\langle \mo_1 \dots \mo_n \rangle = \int \CD \Phi \   \mo_1 \dots \mo_n  e^{- {S_E}},
\label{CF}
\ee
where $\CD \Phi$ denotes the total path-integral measure in all fields. Since the action $S_E$ is $\cq$-exact, a differentiation of the correlation function with respect to the gauge coupling $e$ yields
$
{\partial \over \partial e} \langle \mo_1 \dots \mo_n \rangle  = {2 \over e^3} \langle \mo_1 \dots \mo_n  \{ \cq , V' \}  \rangle  = {2 \over e^3} \langle \{ \cq, \mo_1 \dots \mo_n V' \}  \rangle = 0,
$
where we have made use of the fact that $\langle \{ \cq, \dots \} \rangle = 0$ since $\cq$ generates a (super)symmetry of the theory. In other words, the correlation function of $\cq$-invariant observables is independent of the gauge coupling $e$; as such, the semiclassical approximation to its computation will be exact. In this approximation, one can freely send $e$ to a very small value in the correlation function. Consequently, from (\ref{S}) and (\ref{CF}), we see that the non-zero contributions to the correlation function will be centered around classical field configurations -- or the zero-modes of the fields -- which minimize $\{ \cq , V' \}$ and therefore the action. Thus, it suffices to consider quadratic fluctuations around these zero-modes, as mentioned earlier.

Let us first consider the fluctuations. Assuming that $I_0(p)$ and $I_2(S)$ can be expressed purely in terms of the relevant zero-modes, the path-integral over the fluctuations of the fields in the kinetic terms of the action give rise to determinants of the corresponding kinetic operators. Due to supersymmetry, the determinants resulting from the bose and fermi fields cancel up to an irrelevant sign~\cite{Witten}.  

Let us now consider the zero-modes. The bosonic zero-modes obey the constraints obtained by setting the supersymmetric variation of the fermi fields to zero. From (\ref{susy tx - non-abelian}), we find that these constraints are $F'_+ = K_+$,  $\nabla' \phi =0$ and $[\phi, \phi^{\dagger}] =0$. If we assume the trivial solution $\phi =0$ to the constraints $\nabla' \phi =0$ and $[\phi, \phi^{\dagger}] =0$, it will mean that the zero-modes of $A'$  do $\textrm{\it not}$ correspond to reducible connections, and that there are $\it no$ zero-modes of $\phi$. Also, $K_+ =0$ on-shell. Hence, the first constraint can actually be written as $F'_+ = 0$. Altogether, this means that the only bosonic zero-modes come from the gauge field $A'$, and that they correspond to irreducible, anti-self-dual connections which are characterized by the relation
\be
(F - 2 \pi \alpha \delta_D)_+ = 0.
\label{ram instantons}
\ee
Again, recall that the bundle $E'$  over $X$ with curvature $F' = F- 2\pi \alpha \delta_D$ can be equivalently viewed as the bundle $E$ with curvature $F$ restricted to $X \backslash D$. Hence, the constraint (\ref{ram instantons}) just defines  anti-self-dual connections on the bundle $E$ restricted to $X \backslash D$, whose holonomies around small circles linking $D$ are as given in (\ref{holonomy}). In other words, modulo gauge transformations that leave (\ref{ram instantons}) invariant, the expansion coefficients of the zero-modes of $A'$ that appear in the path-integral measure will correspond to the collective coordinates on $\CM'$ - the moduli space of ``ramified'' instantons.  

Since we have restricted ourselves to connections $A'$ that are irreducible, it will mean that the relation $\nabla' \zeta =0$ only has the trivial solution $\zeta =0$. Therefore, there are no zero-modes for $\zeta$. Moreover, if we assume $A'$ to be regular as well, then the cokernel of the self-dual projection of the covariant derivative $\nabla'$, is zero~\cite{Marcos}. This in turn implies that $\chi$ has no zero-modes either.  Thus, the only fermionic zero-modes whose expansion coefficients contribute to the path-integral measure come from $\psi$. 

The number of bosonic zero-modes is, according to our analysis above, given by the dimension $s$ of $\CM'$. What about the number of zero-modes of $\psi$? Well, (since there are no zero-modes for $\zeta$ and $\chi$),    
it is given by the index of its kinetic operator that appears in the (``ramified'') Lagrangian. Generalizing the relevant analysis in~\cite{Witten} to our ``ramified'' case, we find that  this index also counts the number of infinitesimal connections $\delta A'$ where gauge-inequivalent classes of $A' + \delta A'$ satisfy $F'_+ = 0$, that is,  (\ref{ram instantons}). In other words, the number of zero-modes of $\psi$ will also be given by the dimension $s$ of $\CM'$. Altogether, this means that after integrating out the non-zero modes,  we can write the remaining part of the measure in the expansion coefficients $a'_i$ and $\psi_i$ of the zero-modes of $A'$ and $\psi$ as
\be
\Pi^s_{i=1} da'_i d\psi_i.
\label{measure} 
\ee
Notice that the $s$ distinct $d\psi_i$'s anti-commute. Hence, (\ref{measure}) can be interpreted as a natural measure for the integration of differential forms on $\CM'$.  This is an important observation, as we shall see below.

\bigskip\noindent{\it Correlation Functions Of $I'_0(p)$ And $I'_2(S)$ And The ``Ramified'' Donaldson Invariants} 

Now consider the correlation functions of the $\cq$-invariant observables $I_0(p)$ and $I_2(S)$. Since our above analysis assumes that $I_0(p)$ and $I_2(S)$ are expressed in terms of  zero-modes only, and since we know that $\phi$ -- which appears in $I_0(p)$ and $I_2(S)$ -- consists purely of non-zero modes, we must find a way to express $\phi$ in terms of just zero-modes. 

One means of doing so is to ``integrate out'' the fluctuations in $\phi$~\cite{Witten}. This entails replacing $\phi$ in $I_0(p)$ and $I_2(S)$ with its expectation value 
\be
\langle \phi \rangle = \int \CD \Phi \ \phi  \ \textrm{exp} \left(- S (\phi, \phi^\dagger) / e^2 \right)
\ee   
before considering them in any correlation function, where $S(\phi, \phi^\dagger)$ denotes the relevant terms in the action given by\footnote{Relevant in the sense that one can have at most single contractions between the $\phi$ and $\phi^\dagger$ fields in the contributions to the expectation value.  This is because we are considering only quadratic fluctuations in the semiclassical approximation.}
\be
S(\phi, \phi^\dagger) = \int_X d^4 x \ \textrm{Tr} \left(  \nabla' \phi  \nabla' \phi^\dagger  - {i \over \sqrt 2} \phi^\dagger [\psi, \psi] \right).
\ee 
Indeed, by expanding the exponential of the second term in $-S(\phi, \phi^\dagger) / e^2$ as $1  + {i \over e^2 \sqrt 2} \int_X d^4x \, \phi^\dagger [\psi, \psi] + \dots$, and the fact that we have the free-field operator product expansion $\phi (x) \cdot \phi^{\dagger} (y)  = -  e^2 G (x-y)$, where $G(x-y)$ is the unique solution to the relation $\nabla'^2 G(x-y) = \delta^4(x-y)$, we find that we have
\be
\langle \phi (x) \rangle = - {i \over \sqrt 2} \int_X d^4y \,  G(x-y)[\psi(x), \psi(y)]_0 
\label{<phi>}
\ee
up to the lowest order in $e$ in the semiclassical approximation, where the subscript ``0'' denotes the restriction to zero-modes of the indicated expression. Hence, in replacing $\phi$ with $\langle \phi \rangle$ in (\ref{I's}),  we have 
\begin{eqnarray}
I'_0(p) & = &{1 \over 8 \pi^2}\textrm{Tr}{ \langle \phi (p) \rangle}^2,  \nonumber \\ 
I'_2(S) & = & {1 \over 4 \pi^2} \int_S  \textrm{Tr} \left ( {1 \over 8} \psi \wedge \psi   - {1 \over \sqrt 2} \langle \phi \rangle F \right),
\label{I's zero-modes}
\end{eqnarray}
for any $p \in H_0(X \backslash D)$ and $S \in H_2(X \backslash D)$, where $I'_0(p)$ and $I'_2(S)$ are express purely in terms of zero-modes, as desired. Note that the expression (\ref{<phi>}) is physically consistent; it is independent of $e$ and its has $R$-charge 2, as in the case with $\phi$. Thus,  $I'_0(p)$ and $I'_2(S)$ continue to be $e$-independent with $R$-charge $4$ and $2$. Moreover, based on our above discussion about  (\ref{measure}) being a natural measure for the integration of differential forms on $\CM'$, we see that $I'_0(p)$ and $I'_2(S)$ (which contain 4 and 2 zero-modes of $\psi$, respectively) can be interpreted as 4-forms and 2-forms in $\CM'$. 

Finally, let us compute an arbitrary correlation function in the $\cq$-invariant observables $I'_0(p)$ and $I'_2(S)$. For the correlation function to be non-vanishing, the $d\psi_i$'s in the remaining measure (\ref{measure})  have to be  ``soaked up'' by the  zero-modes of $\psi$ -- that appear in the combined operator whose correlation function we wish to consider --  $\it exactly$. This translates to the fact that the combined operator ought to correspond to a top-form (of degree $s$) in $\CM'$. Therefore, if such a combined operator is given by $[{I'_0}(p)]^t I'_2(S_{i_1}) \dots I'_2(S_{i_m})$,\footnote{Notice that we are considering a combined operator in which there are $t$ operators $I'_0(p_{i_1}), \dots, I'_0(p_{i_t})$ that coincide at one particular point $p$ in $X$. Such a combined operator can be consistently defined in any physical correlation function; this is because the  $I'_0(p_{i_k})$'s consist only of non-interacting zero-modes and moreover, any correlation function of the topological theory is itself independent of the insertion points of the operators.} its $\it topological$ correlation function -- for ``$\it ramified$'' instanton number $k' = -  \int_X p_1(E') / 4$ -- can be written as
\be
\langle  [{I'_0}(p)]^t I'_2(S_{i_1}) \dots I'_2(S_{i_m})   \rangle_{k'} = \int_{\CM'}  [{I'_0}(p)]^t \wedge I'_2(S_{i_1}) \wedge \dots \wedge 
I'_2(S_{i_m}),
\label{correlation function 1}
\ee
where $2m + 4t = s$. This coincides with the definition of the ``ramified'' Donaldson invariants $d^{k'}_{m,t}$ in (\ref{ram DW}). Thus, we have found, in  (\ref{correlation function 1}),  a physical interpretation of the ``ramified'' Donaldson invariants in terms of the correlation functions of $\cq$-invariants observables $I'_0(p)$ and $I'_2(S)$.

As a result, the generating function ${\bf Z}'_{\xi, \bar g}(p,S)$ in (\ref{Donaldson generating function}) can also be interpreted in terms of $I'_0$ and $I'_2$ as 
\be
{\bf Z}'_{\xi, \bar g}(p,S) =  \sum^{\infty}_{k'=0} \  \langle e^{p I'_0 +  I'_2(S)} \rangle_{k'}.
\label{Donaldson generating function - physical}
\ee

\newsubsection{A Minimal Genus Formula For Embedded Two-Manifolds - A Physical Proof}

We shall now  give a physical derivation of a minimal genus formula for embedded two-surfaces that was first proved by KM as Theorem 1.1 of~\cite{KM2}. To this end, it suffices for us to consider the case where $\xi =0$, that is, $G=SU(2)$. 

 The theorem can be stated as follows. For a simply-connected $X$ with non-vanishing Donaldson invariants and odd $b^+_2 \geq 3$, an orientable two-manifold $D$ that is smoothly embedded in $X$ with genus $g \geq 1$, will obey
\be
2 g -2 \geq D \cap D.
\label{inequality theorem}
\ee
Let us make some comments about the above theorem before we proceed further. Firstly, in our analysis so far, we have assumed $b_1$ to be arbitrary. Let us now consider $X$ with $b_1 =0$. This in turn implies that for manifolds with odd $b^+_2 \geq 3$, we have even $b^+_2 - b_1 +1 = {(\chi + \sigma) \over 2}  \geq 4$. Secondly, it is shown by KM in~\cite{structure, KM2} that the ``ramified'' Donaldson invariants is proportional to the ordinary Donaldson invariants (a remarkable result whose physical proof will be given in $\S$8). In other words, the above theorem applies to $X$ with non-vanishing $d^{k'}_{m,t}$'s also. Indeed, the condition that $(\chi + \sigma) /2$ must be an even integer implies that the dimension $s$ of the moduli space $\CM'$ must also be an even integer -- a requirement for $d^{k'}_{m,t}$  to be consistently defined. 

With the above points in mind, let us proceed to note that based on our earlier arguments, the expression for the index of the kinetic operator of $\psi$  that counts the number $N_{\psi}$ of $\psi$ zero-modes, is the expression for the index in the ordinary case but with gauge bundle $E'$. In other words, 
\begin{equation}
N_{\psi} = 8k' - {3 \over 2} (\chi + \sigma), 
\label{Npsi o}
\end{equation}
where the ``ramified'' instanton number $k' = - {1\over 8 \pi^2} \int_X \textrm{Tr} \hspace{0.1cm} F' \wedge F'$.  

Since $\alpha$ takes values in $\Lambda_{\textrm{cochar}} \otimes_\mathbb Z  \IR$ while  ${\frak m}$ takes values in $\Lambda_{\textrm{cochar}}$, the inner product $\textrm{Tr} \hspace{0.1cm} \alpha \frak m$ is well-defined. Thus, one can compute $k'$ and hence $N_{\psi}$ explicitly, as
\be
N_{\psi}  =  8k + 8 \textrm{Tr} \hspace{0.1cm} \alpha {\frak m} - 4 \textrm{Tr} \alpha^2 D \cap D - {3 \over 2} (\chi + \sigma), 
\ee
where the integer $k = - {1\over 8 \pi^2} \int_X \textrm{Tr} \hspace{0.1cm} F \wedge F$. 

By noting that $\frak t$ is generated by (\ref{frak t}), one can further simplify $N_{\psi}$ to
\be
N_{\psi}  =  8k +  16 \alpha {l} - 8 \alpha^2 D \cap D - {3 \over 2} (\chi + \sigma).
\label{Npsi}
\ee

In order for the correlation function (\ref{correlation function 1}) and therefore the $d^{k'}_{m,t}$'s to be non-vanishing, we must have $N_{\psi} = 2m + 4t = s$. In particular, this means that $d^{k'}_{m,t} = 0$ if $N_{\psi} < s$, that is, if
\be
16 \alpha {l} - 8 \alpha^2 D \cap D  <   4 {l} - 2( g-1).
\ee
Since the above must be true for all values of $\alpha$ in the interval $[0, {1\over 2}]$,\footnote{When $\alpha =0$, the holonomy (\ref{holonomy}) is 1. When $\alpha = {1\over 2}$, the holonomy is $-1$. However,  the $SU(2)$-bundle is just the $SO(3)$-bundle ${E}$ with $w_2({E}) =0$. Since $SU(2)$ is a double-cover of $SO(3)$, the holonomy of $-1$ is the same as 1 from the viewpoint of ${E}$. In other words, one can simply consider $\alpha$ to lie between $0$ and $1\over 2$, as did KM.}  near the limits $\alpha = 0$ and $\alpha = {1\over 2}$, we find that $d^{k'}_{m,t} =0$  if $4 {l} >  (2  g -2)$ and $4 {l}  < 2 D \cap D - (2  g -2)$, respectively. Thus, for $d^{k'}_{m,t} \neq 0$, $4 l$ must lie in the range 
\be
 2 D \cap D - (2  g -2) \leq 4 {l} \leq (2 g -2).
 \label{inequality proof}
\ee
Hence, we can infer from (\ref{inequality proof}) that for an orientable two-manifold $D$ with genus $g \geq 1$ that is smoothly-embedded in $X$ with non-vanishing (``ramified'') Donaldson invariants and odd $b^+_2 \geq 3$, we have
\be
\label{min genus}
2  g -2 \geq D \cap D.
\ee
This is just KM's theorem stated in (\ref{inequality theorem}). Our physical proof is thus complete.

\newsection{The Low-Energy Description Of The Non-Abelian Gauge Theory With Surface Operators}   

We shall now analyze the macroscopic description of the non-abelian theory in the presence of surface operators. The reason for doing so is that we would eventually like to relate the topological correlation functions (\ref{correlation function 1}) and their generating function (\ref{Donaldson generating function - physical}) to quantities associated with a ``simpler''  (topological) $\it abelian$ theory, as was done in~\cite{Moore-Witten}.  An interesting string-theoretic realization of our gauge-theoretic setup in flat space will also be considered in this section.

\newsubsection{The Low-Energy Vacuum Structure With Surface Operators} 

Let us start by elucidating the low-energy vacuum structure of the untwisted pure $SU(2)$ theory with surface operators in flat Minkowski space.  To this end, note that the classical action of the microscopic theory contains the scalar potential
\be
V = {1 \over 2 e^2} \textrm{Tr} [ \phi, \phi^{\dagger}]^2, 
\ee  
whereby the classical vacua is determined by $V=0$, that is, $[ \phi, \phi^{\dagger}] = 0$. Thus, $\phi$ must be valued in the Cartan subalgebra, such that one can write its vacuum expectation value as
\be
\langle \phi \rangle = {a \over 2}  \begin{bmatrix} 
 1 & \ 0\\ 
0 & -1
\end{bmatrix},
\label{phi cartan}
\ee   
where $a$ is a complex number. When $\langle \phi \rangle$ is of the form (\ref{phi cartan}) with non-zero $a$, the gauge group is spontaneously-broken to $U(1)$ at a scale below $a$. This is because a non-zero  $\langle \phi \rangle$ generates masses for  the $W^{\pm}$ bosons through the kinetic term $|\nabla' \phi|^2$ in an $\CN =2$ Higgs mechanism. Consequently, the only light degrees of freedom at low-energy come from a $U(1)$ $\CN =2$ vector multiplet.  

Next, notice that different vacuum expectation values of (\ref{phi cartan}) lead to physically inequivalent theories. Hence, since Weyl reflections map $a \to -a$, one can construct a gauge and Weyl-invariant operator $\textrm{Tr} \hspace{0.1cm} \phi^2$ whose expectation value 
\be
u = \langle \textrm{Tr} \hspace{0.1cm} \phi^2 \rangle
\label{u}
\ee  
parameterizes the moduli space of the theory. Clearly, we have $u = a^2 /2$, whereby the $SU(2)$ gauge symmetry is restored at the origin $u=0$. However, just as in the ordinary case without surface operators, there will be quantum corrections to the metric on moduli space, such that the $SU(2)$ gauge symmetry is nowhere restored over the $\it quantum$ moduli space $\CM_q$. As such,  $u = a^2 /2$ is only a classical relation that will be shifted in the quantum theory. To unravel the nature of $\CM_q$, we need to turn to the low-energy effective action of the $SU(2)$ theory. However, it will be useful to analyze the chiral anomaly of the theory first. 

\bigskip\noindent{\it The Chiral Anomaly}

\def\bz{{\bf Z}}

The $U(1)_R$ symmetry of the $SU(2)$ theory is actually anomalous; to one-loop order, under a $U(1)_R$ transformation, its physical \emph{Minkowskian} action $I$ effectively changes by 
\be
\delta I = -  {\gamma 4 N_c \over 8 \pi^2} \int_X  \textrm{Tr} \hspace{0.1cm} F' \wedge F',
\label{change I}
\ee
where $\gamma$ is the phase of the $U(1)_R$ transformation, and $N_c =2$. 

Notice that we can write $\delta I = (\gamma 4 N_c) k'$, where $k' = -{1 \over 8 \pi^2} \int_X \textrm{Tr} \hspace{0.1cm} F' \wedge F' = - {1 \over 8 \pi^2} \int_X \textrm{Tr} \hspace{0.1cm} F \wedge F + \textrm{Tr} \hspace{0.1cm} \alpha \frak m - {1\over 2} \textrm{Tr} \alpha^2 D \cap D$. The first term $-{1 \over 8 \pi^2} \int_X \textrm{Tr} \hspace{0.1cm} F \wedge F$ is always an integer, while the second and third terms $\textrm{Tr} \hspace{0.1cm} \alpha \frak m$ and  $-{1\over 2} \textrm{Tr} \alpha^2 D \cap D$ are \emph{a priori} not. However, recall that one is free to invoke a ``ramification''-preserving twisted gauge transformation of the kind in (\ref{twisted gauge tx}) such that (\ref{intersection number 2}) holds; in other words, $\alpha$ is gauge-equivalent to an element of $\Lambda_{\rm cochar}$, and since ${\frak m} \in \Lambda_{\rm cochar}$  and  $D \cap D \in \mathbb Z$, one can -- after noting that $\textrm{Tr} \hspace{0.1cm} {\frak n}' {\frak n}$ is an \emph{even} integer for any ${\frak n}', {\frak n} \in  \Lambda_{\rm cochar}$ -- regard $\textrm{Tr} \hspace{0.1cm} \alpha \frak m$ and  $-{1\over 2} \textrm{Tr} \alpha^2 D \cap D$ as integers. Thus, $k'$ in $\delta I$ is effectively an integer; in other words, we have $\delta I =  8 \gamma \mathbb Z$. This means that for the path-integral to be invariant under the $U(1)_R$ transformation, we must have $\gamma = 2 \pi /8$. Consequently, one can apply $\it eight$ consecutive discrete chiral transformations on any field with $R$-charge 1 before returning to the identity transformation; in other words, the continuous $U(1)_R$ symmetry is actually broken down to a $\bz_8$ symmetry. Thus, the global symmetry group is $SU(2)_R \times \bz_8$. However, the $\bz_2$ center of $SU(2)_R$ -- which acts as $-1$ on the fermi fields with $R$-charge 1 -- is contained in $\bz_8$. This means that the true global $R$-symmetry of the theory is
$
(SU(2)_R \times \bz_8) / \bz_2.
$

Now recall that the field $\phi$ has charge 2 under the classical $U(1)_R$ symmetry; this means that it gets transformed to $e^{2 \pi i \over 4} \phi$ under the nonperturbative $\bz_8$ symmetry. In other words,  the $\bz_8$ symmetry will be spontaneously-broken down to a $\bz_4$ symmetry at a point in moduli space where  $\langle \phi \rangle \neq 0$ -- that is, the final global $R$-symmetry of the $SU(2)$ gauge theory is
\be
(SU(2)_R \times \bz_4) / \bz_2.
\ee
Nevertheless, over the entire moduli space, one still has the full $\bz_8$ symmetry which acts nontrivially on its holomorphic coordinate $u$: since $u = \langle \textrm{Tr} \hspace{0.1cm} \phi^2 \rangle$, it  acts as a $\bz_2$ symmetry which maps $u \to -u$.

\bigskip\noindent{\it The Low-Energy Effective Action}

\def\ae{{\alpha_{\textrm {eff}}}}
\def\ne{{\eta_{\textrm {eff}}}}

Let us now proceed to analyze the low-energy physics of the $SU(2)$ theory. The low-energy theory does not contain any non-abelian gauge interactions; therefore, it exhibits smooth behavior in the infrared. Hence, there is no problem in constructing its effective action.

 Let $\CA$ and $W$  be an $\CN =1$ chiral superfield and chiral spinorial (abelian) superfield strength, respectively, whose collective components make up a $U(1)$ ${\cal N} =2$ vector multiplet; the lowest  component of this multiplet is a complex scalar $\varphi$ (which comes from $\CA$), and it has vacuum expectation value $\langle \varphi \rangle = a$.  Then, due to $\CN =2$ supersymmetry, the low-energy effective action of the $SU(2)$ theory can be written in $\CN =1$ superspace as~\cite{mine} 
\begin{eqnarray}
S_{\textrm{eff}} & = &  -{1\over 2\pi} \textrm{Im} \left[ \int d^4 x  d^2 \theta d^2 {\bar\theta} \ {\partial {\cal F}(\CA) \over \partial \CA} \bar \CA \right] - {1\over 4\pi}\textrm{Im} \left[\int d^4 x  d^2 \theta \ {\partial^2 {\cal F}(\CA)\over \partial \CA^2}  {W'}^{\alpha}{W'}_{\alpha} \right] \nonumber \\
&& \quad + {1 \over 2 \pi} \textrm{Im} \left [\int d^4 x d^2 \theta \ [i 2 \pi \eta_{\textrm{eff}} (\delta_D)_{\mu \nu} (\sigma^{\mu\nu}\theta)]^\alpha W'_\alpha \right],
\label{Seff}
\end{eqnarray}
where $W'_\alpha = W_\alpha - 2 \pi i \alpha_\textrm{eff} (\delta_D)_{\mu\nu} (\sigma^{\mu\nu}\theta)_\alpha$ is a chiral superfield with Weyl-spinor indices $\alpha = 1\dots 2$ and flat \emph{Minkowski} space indices $\mu, \nu = 0,\dots, 3$, ${\sigma^{\mu \nu}} = {1\over 4} ({\sigma^\mu} {\bar\sigma}^{\nu} - {\sigma}^\nu {\bar\sigma}^{\mu})$, and $\ae$ and $\ne$ -- which take values in $\IR / \mathbb Z$ -- are the effective surface operator parameters of the low-energy pure abelian theory. Also, $\cal F$ is a holomorphic function in $\CA$ commonly known as the prepotential.

By doing the $\theta$-integration in (\ref{Seff}), one is led to the expression for $S_\textrm{eff}$ in flat Minkowski space,  from which the $\it local$ metric on the one complex-dimensional moduli space is found to be given by
\be
(ds)^2 = g_{a\bar a} da d \bar a =  \textrm{Im} \tau(a) da d \bar a,
\label{metric on M}
\ee
where the holomorphic function $\tau(a)$ -- which is the complexified gauge coupling of the $U(1)$ theory -- can be written as
\be
\tau(a) = {4 \pi i \over e^2 (a)} + {\Theta (a) \over 2 \pi} =  {\partial^2 {\cal F}(a) \over {\partial a^2}}.
\label{tau(a)}
\ee

Classically, we have $\CF_{\textrm{cl}}(a) =   {a^2 \over 2} \tau_{\textrm cl}$, where $\tau_{\textrm cl} = {4 \pi i \over e^2} + {\Theta \over 2 \pi}$ is the bare complexified gauge coupling, and $u = a^2 /2$. Consequently, the metric on the $\it classical$ moduli space $\CM_c$ can be written as
$
ds^2 \sim {du d\bar u / |u|^2}. 
$ 
Hence, we see that $\CM_c$ is singular at the point $u=0$. This just corresponds to the fact that in the classical picture, the $SU(2)$ gauge symmetry is restored at $u=0$: since the $W^{\pm}$ bosons become massless at this point, we are effectively integrating out massless degrees of freedom in considering an effective $U(1)$ theory, and this results in the observed singularity in $\CM_c$ at $u=0$.   

Nevertheless, by taking into account all quantum corrections to $\CF_{\textrm{cl}}(a)$ and thus $\tau_{\textrm{cl}}$, one can  ascertain the exact form of $\CF(a)$ and hence, the exact form of the metric  $\textrm{Im} \tau(a)$ on the quantum moduli space $\CM_q$, whence one will find that the singularity at $u=0$ will cease to exist, among other things.

\bigskip\noindent{\it The Perturbative and Nonperturbative Corrections To  $\CF_{\textrm{cl}}(\CA)$}

Let us now determine the perturbative and non-perturbative corrections to $\CF_{\textrm{cl}}(\CA) = { \CA^2\over 2}\tau_{\textrm{cl}}$, following~\cite{Seiberg}. Firstly, note that due to non-renormalization theorems in $\CN =2$ supersymmetry, the holomorphic prepotential does not receive perturbative corrections beyond one-loop order; hence, it suffices to consider one-loop effects only. Secondly, note that the one-loop chiral anomaly of the underlying $SU(2)$ theory in (\ref{change I}) implies that under a $U(1)_R$ transformation, $S_{\textrm{eff}}$ will effectively change by     
\be
\delta S_{\textrm{eff}} =  - {\gamma 4 N_c \over 8 \pi^2}  \int_X F'_L \wedge F'_L.
\label{change Seff}
\ee
Here, $F'_L = F_L - 2 \pi \ae \delta_D$ is the $\it effective$ $U(1)$ field strength of the  low-energy theory defined over all of $X$; it can also be interpreted as the curvature of a $U(1)$-bundle defined over $X \backslash D$ with singularity $2 \pi \ae \delta_D$ along $D$.  As the notation implies, $F_L$ can indeed be identified as the curvature of  the complex line bundle $L$ in (\ref{decompose}). This is because an $SU(2)$-bundle $E$ that is associated with an $SU(2)$ gauge symmetry that is spontaneously-broken, is, in mathematical terms, reducible; consequently, $E$ can be decomposed as shown in $(\ref{decompose})$, with $L$ being the line bundle that corresponds to the unbroken $U(1)$ gauge symmetry.

The one-loop correction to $\CF_{\textrm{cl}}(\CA)$ can then be determined from the requirement that  under a $U(1)_R$ action on $\CA$ in (\ref{Seff}), $S_{\textrm{eff}}$ must change  to compensate for the anomaly in (\ref{change Seff}) $\it exactly$ -- that is, the effective action is required to be $U(1)_R$-invariant, at least perturbatively.   Notice that the second term in (\ref{Seff}) is the only term that will contribute to a change of the form (\ref{change Seff}) under a $U(1)_R$ action on $\CA$.  Therefore, since $\CA$ is an $\CN =1$ chiral superfield that necessarily has $R$-charge 2,  we must have 
\begin{eqnarray}
\label{F change}
- {1\over 4\pi}\textrm{Im} \left[\int d^4 x  d^2 \theta \ {\cal F}''(e^{2i \gamma}\CA) {W'}^{\alpha}{W'}_{\alpha} \right] & = &  - {1\over 4\pi}\textrm{Im} \left[\int d^4 x  d^2 \theta \ {\cal F}''(\CA)  {W'}^{\alpha}{W'}_{\alpha} \right] \nonumber \\
&& + {1\over 4\pi}\textrm{Im} \left[\int d^4 x  d^2 \theta \  ({2\gamma N_c \over \pi})    {W'}^{\alpha}{W'}_{\alpha} \right], 
\end{eqnarray}
where $\CF''(\CA) = \partial^2\CF / \partial \CA^2$. Note that we have used the identity $\textrm{Im} \left[\int d^4 x  d^2 \theta \  {W'}^{\alpha}{W'}_{\alpha} \right] = \int_X F'_L \wedge F'_L$ in deriving the above relation. 

From (\ref{F change}), we find that the one-loop corrected form of $\CF(\CA)$ is restricted by the condition
\be
{\cal F}''(e^{2i \gamma}\CA)  = {\cal F}''(\CA) - {2\gamma N_c \over \pi}.
\label{F''}
\ee
For an infinitesimal $\gamma$, whereby one can make the replacement $e^{2i \gamma} \CA = \CA + 2i \gamma\CA$, using Talyor's expansion, we find that  (\ref{F''}) can also be written as
\be
{\partial^3{\CF} \over \partial \CA^3}   =    { i N_c  \over \pi} {1  \over \CA}.
\label{F'''}
\ee
Noting that $N_c =2$, one can triply integrate (\ref{F'''}) and obtain the one-loop corrected form of $\CF(\CA)$ as 
\be
\CF_{\textrm{1-loop}} (\CA) = {1 \over 2} \tau_{\textrm{cl}} \CA^2 +  {i \over 2 \pi} \CA^2 \textrm{ln} \left({\CA^2 \over \Lambda^2}\right), 
\label{F 1-loop}
\ee
where $\Lambda$ represents a fixed dynamically-generated scale of the theory.  

From (\ref{F 1-loop}) and (\ref{tau(a)}), we obtain the one-loop corrected complexified gauge coupling as
\be
\tau_{\textrm{1-loop}} (a) = \tau_{\textrm{cl}} + {2i \over \pi} \ \textrm{ln} \left(  {a \over \Lambda'}  \right),
\label{tau 1-loop}
\ee
where $\Lambda' = e^{-3/2} \Lambda$. Note that $\tau_{\textrm{1-loop}} (a)$ actually coincides with the one-loop $\it running$ coupling constant of the underlying $SU(2)$ theory at scale $\mu =a$. This just corresponds to the fact that at scales below $a$, the massive $W^{\pm}$ bosons decouple from the theory, which thus halts the running of the coupling constant at scale $\mu =a$, since the effective $U(1)$ theory has \emph{vanishing} $\beta$-function.  

\def\Seff{{S_\textrm{eff}}}

Let us proceed to determine the nonperturbative corrections to $\CF_{\textrm{1-loop}} (\CA)$. Firstly, being nonperturbative, these corrections can only come from (``ramified'') $SU(2)$-instantons.\footnote{I would like to thank N.~Seiberg for clarifications related to this point.} Moreover, since the prepotential is $\it holomorphic$, anti-instantons $\it cannot$ contribute to these corrections. Secondly, as emphasized in~\cite{Seiberg}, these corrections to the prepotential ought to reflect the fact that nonperturbatively, $\Seff$  is no longer $U(1)_R$-invariant but rather, it is only invariant under the discrete symmetry that is left unbroken  in the underlying $SU(2)$ theory; in our case, we saw that this is a $\bz_8$ symmetry. Thirdly,  notice that the chiral anomaly (\ref{change Seff}) can be regarded as causing a shift in the $\Theta$-angle of the theory. Consequently, one can,  for simplicity,  use an appropriate chiral rotation of the fermions to set the $\Theta$-angle to zero. Finally, since the corrections coming from a configuration of (``ramified'') instanton number $k'$ must be proportional to the $k'$-instanton factor, which -- after setting the $\Theta$-angle to zero and using (\ref{tau 1-loop}) -- is given by $\textrm{exp}(- 8 \pi^2 k' / e^2) \sim \left(\Lambda / a \right)^{4k'}$, we find that the $k'$-instanton correction to $\CF_{\textrm{1-loop}} (\CA)$ ought to be proportional to 
\be
\CF_{{k'}}(\CA) =  \left(  {\Lambda \over \CA}  \right)^{4k'} \CA^2.
\label{inst factor}
\ee 
Indeed, since $k'$ -- which characterizes the unbroken $\bz_8$ symmetry via (\ref{change I}) --  can be effectively regarded as an integer  (as explained earlier), and since $\CA$ has $R$-charge 2, the $R$-charge of  the ($k'$-instanton contribution to the) first term in $\Seff$, namely, $\CF'_{k'}(\CA) \hspace{0.1cm} \overline \CA$, will be an integer multiple of 8. In other words, $\Seff$ is invariant under a $\bz_8$ chiral symmetry at most, in accordance with the second point mentioned above. Nevertheless, notice that one can restore the full $U(1)_R$ symmetry of the theory by defining $\Lambda$ to have $R$-charge 2 (as is customarily done in the ordinary case without surface operators),  such that the instanton-corrected prepotential will persist to have $R$-charge 2.

Hence, we can conclude from (\ref{F 1-loop}) and (\ref{inst factor}) that the full expression for the holomorphic prepotential must take the form
\be
\CF (\CA) = {1 \over 2} \tau_{\textrm{cl}} \CA^2 +  {i \over 2 \pi} \CA^2 \textrm{ln} \left({\CA^2 \over \Lambda}\right) + \sum^{\infty}_{l=1} f_{l} \left(  {\Lambda \over \CA}  \right)^{4l} \CA^2,
\label{full prepotential}
\ee
where -- due to the fact that in supersymmetric theories, instantons contribute to the path-integral through zero-modes only~\cite{divecchia} -- the coefficients $ f_{l}$ must be non-field-dependent constants.

\def\CMq{{\CM_q}}

\bigskip\noindent{\it The Metric On Quantum Moduli Space $\CMq$}

From (\ref{tau(a)}) and (\ref{full prepotential}), we find that the full expression for the complexified gauge coupling -- including perturbative and nonperturbative corrections -- will be given by  
\be
\tau(a) = \tau_{\textrm{cl}} + {2i \over \pi} \ \textrm{ln} \left(  {a \over \Lambda'}  \right) + \sum_{l=1}^{\infty} t_l \left( {\Lambda \over a} \right)^{4l},
\label{tau(a) full}
\ee
where the coefficients $t_l$ -- being proportional to $f_l$ in (\ref{full prepotential}) -- are also non-field-dependent constants. 

From (\ref{tau(a) full}), we see that the $\it form$ of the metric $\textrm{Im}\tau(a)$ on $\CMq$ is $\it identical$ to that in the ordinary case without surface operators; in other words, our careful analysis thus far suggests that the presence of surface operators should not modify the vacuum structure of the original theory.  Note that one can also corroborate this conclusion in a later subsection using  string theory, by interpreting surface operators in our gauge theory as D2-branes embedded in D4-branes which end on NS5-branes. Moreover, a surface operator can be viewed as a two-dimensional analog of a combination of a  Wilson and `t Hooft loop operator, and the inclusion of these latter operators are certainly not known to modify the vacuum structure of the underlying four-dimensional gauge theory, although they can be used to detect different phases of the theory or the presence of spontaneous symmetry breaking.


\def\imt{{\textrm{Im}\tau(a)}}
\def\imtd{{\textrm{Im}\tau_d(a_d)}}

At any rate, note that in the region $u \to \infty$, that is, where $|a|$ is large, we see from (\ref{tau(a) full}) that the nonperturbative corrections to $\tau(a)$ are suppressed, and that (for zero $\Theta$-angle) the effective coupling $e(a)$ is small. Hence, we can regard the region $u \to \infty$ as a semiclassical region  whereby perturbative computations in the $U(1)$ theory are reliable, and $u \approx a^2/2$. Also, notice that in this region,  $\tau(a) \approx i (\textrm{ln}(a^2 / {\varepsilon\Lambda}^2)  + 3) / \pi$ for some real positive constant $\varepsilon$.  As we circle around  $u$, there is a shift $\tau(a) \to \tau(a) - 2$. Since this shift is $\it real$-valued, we find that although $\tau(a)$ is multi-valued, the metric $\textrm{Im}\tau(a)$  is a single-valued positive function. Therefore, since $\textrm{Im} \tau(a) = \textrm{Im} ({\partial^2 \CF \over \partial a^2})$,  $\imt$ must be a harmonic function.  This means that  the condition $\imt > 0$ does not hold everywhere on $\CMq$. On the other hand, since the kinetic energy of the $\it dynamical$ complex scalar field $\varphi$ must always be greater than zero, it will mean that  in a physically sensible unitary theory, one must have $\imt > 0$ $\it everywhere$ on $\CMq$. The resolution to this conundrum (according to a proposal by Seiberg and Witten (SW) in~\cite{SW}), is that $\imt$ can only be defined $\it locally$, and that in regions of $\CMq$ whereby $\imt \to 0$,\footnote{Notice that $\textrm{Im}\tau(a)$ cannot be negative, as it is given by $4 \pi / e^2$.} the underlying effective $U(1)$ theory ought to be described in its  ``magnetic''  frame in the dual variables $a_d$ etc., for which the corresponding metric $\imtd > 0$. Nevertheless, the metric on $\CMq$ should be invariant under a change of frame, since the  geometry of $\CMq$ is fixed.

\bigskip\noindent{\it The ``Magnetic" Frame Of The Effective $U(1)$ Theory}

\def\Seffd{{S^d_\textrm{eff}}}

Generalizing the approach of SW in~\cite{SW} to our case with surface operators,  we find that the underlying $U(1)$ theory in the ``magnetic'' frame should be described by a ``magnetic'' action $\Seffd$ that is an $S$-dual transform  of the ``electric'' action $S_\textrm{eff}$. Such a ``magnetic'' action has been determined in~\cite{mine}, and it is given by
\begin{eqnarray}
S_\textrm{eff}^d & = &  -{1\over 2\pi} \textrm{Im} \left[ \int d^4 x  d^2 \theta d^2 {\bar\theta} \ {\partial \CF_d (\CA_d)  \over \partial \CA_d} {\bar \CA}_d \right] - {1\over 4\pi}\textrm{Im} \left[\int d^4 x  d^2 \theta \  {\partial^2 \CF_d (\CA_d)  \over \partial \CA^2_d}  (W'_d)^{\alpha}(W'_d)_{\alpha} \right] \nonumber \\
&& \quad + {1 \over 2 \pi} \textrm{Im} \left [\int d^4 x d^2 \theta \ [i 2 \pi \eta^d_{\textrm{eff}} (\delta_D)_{\mu \nu} (\sigma^{\mu\nu}\theta)]^\alpha (W_d)_\alpha \right],
\label{Seff_d}
\end{eqnarray}
where $W'_d = W_d - 2 \pi i \alpha^d_\textrm{eff} (\delta_D)_{\mu\nu} (\sigma^{\mu\nu}\theta)$ is a  ``magnetic''  chiral superfield, and $\alpha^d_\textrm{eff}$ and $\eta^d_\textrm{eff}$ are the  effective ``magnetic'' surface operator parameters that take values in $\IR / \mathbb Z$; they are related to the effective ``electric'' surface operator parameters via $(\alpha^d_\textrm{eff}, \eta^d_\textrm{eff}) = (\eta_\textrm{eff}, -\alpha_\textrm{eff})$.  In addition, we also have the relations
\be
\CA_d = {\partial \CF(\CA) \over \partial \CA}, \quad \CA =  -  {\partial \CF_d(\CA_d) \over \partial \CA_d}, 
\label{ad=a}
\ee
and the  complexified gauge coupling (holomorphic in $a_d$) of the ``magnetic'' $U(1)$ theory 
\be
\tau_d(a_d) = {4 \pi i \over e_d^2 (a_d)} + {\Theta_d (a_d) \over 2 \pi} =  {\partial^2 {\CF}_d(a_d) \over {\partial a_d^2}} = - {1 \over \tau(a)}.
\label{taud(ad)}
\ee
Notice that the complexified gauge coupling is inverted in the ``magnetic'' frame, as is expected of an $S$-duality transformation. 
One can therefore  interpret $\Seff$ and $\Seffd$ as  actions of theories of abelian gauge fields which can be consistently coupled to electrically and magnetically charged particles, respectively. In particular, one can couple the ``magnetic'' $U(1)$ photon in $\Seffd$ -- with field strength ${F^d_L}'$ --  to monopoles (in a ``magnetic'' supermultiplet), while in a`` dyonic'' frame, where the effective variable is a linear combination of $a$ and $a_d$,  one can couple the corresponding ``dyonic'' $U(1)$ photon -- with field strength that is a linear combination of $F'_L$ and ${F^d_L}'$ -- to dyons of simultaneously non-vanishing electric and magnetic charges. Indeed, as we shall discuss shortly,  in order for the physics to be consistent, these couplings are required at certain points over $\CMq$, as is well-known from SW theory~\cite{SW}. Such couplings are possible because in spontaneously-broken $SU(N)$ gauge theories with scalar fields in the adjoint representation, one can find monopoles and dyons in their semiclassical spectrum~\cite{coleman}.     

We now come to a very important point that was also emphasized by SW in the ordinary case without surface operators~\cite{SW}. The ``electric'' and ``magnetic'' frames are just different ways to describe the $\it same$ underlying $U(1)$ theory. As such, $\Seff$ and $\Seffd$ should be completely symmetric under the exchange of $\CF$ and $\CF_d$, $\CA$ and $\CA_d$, $(\ae, \ne)$ and $(\alpha^d_\textrm{eff}, \eta^d_\textrm{eff})$, and $W'$ and $W'_d$. However, notice that there is a slight asymmetry in the forms of $\Seff$ and $\Seffd$ in (\ref{Seff}) and (\ref{Seff_d}); the last term involving the ``quantum'' surface operator parameter is different in $\Seff$ and $\Seffd$ -- in the ``electric'' frame with action $\Seff$, the corresponding phase in the path-integral is given by $\textrm{exp} (i \ne \int_D F'_L)$, while in the ``magnetic'' frame with action $\Seffd$, it is given by $\textrm{exp} (i \eta^d_{\textrm{eff}} \int_D F^d_{L})$. In other words, the surface operator in the ``electric'' frame is of a $\it different$ type from that  in the ``magnetic'' frame. This difference is irrelevant for a trivially-embedded surface operator: one can write the ``electric'' phase as $\textrm{exp} (i \ne \int_D F_L)\cdot \textrm{exp} (-2 \pi i \ne \ae \hspace{0.1cm} D \cap D)$, which for $D \cap D = 0$, is just $\textrm{exp} (i \ne \int_D F_L)$. On the other hand, for a \emph{nontrivially-embedded}  surface operator where $D \cap D \neq 0$, one cannot ignore the term $\textrm{exp} (-2 \pi i \ne \ae \hspace{0.1cm} D \cap D)$.  Nevertheless, as highlighted in~\cite{mine}, since one has the condition (\ref{intersection number}), by restricting $\ne$ to be an $\it integer$,  the term  $\textrm{exp} (-2 \pi i \ne \ae \hspace{0.1cm} D \cap D)$ is essentially equal to 1; that is, the ``electric'' phase will be $\textrm{exp} (i \ne \int_D F_L)$, as required. 
This is an important observation which will eventually allow us to furnish,  in $\S$8,  a physical proof of KM's universal formula in~\cite{structure}, that expresses the ``ramified'' Donaldson invariants purely in terms of the ordinary Donaldson invariants. 

\def\ade{{ \alpha^d_\textrm{eff}}} 
\def\nde {{\eta^d_\textrm{eff}}}
 
\bigskip\noindent{\it An $SL(2,\mathbb Z)$'s Worth Of Equivalent Descriptions}

From (\ref{Seff}), (\ref{tau(a)}), (\ref{Seff_d}), (\ref{ad=a}) and (\ref{taud(ad)}), we see  that the transformations $W \to W_d$, $\CA \to \CA_d$, $ \CA_d \to -\CA$, $\tau(\CA) \to  - 1 / \tau(\CA)$ and $(\ae, \ne) \to (\ne, - \ae)$,  map the underlying $U(1)$ theory in its ``electric'' frame to $\it itself$ in its ``magnetic'' frame, and vice-versa. 

Now, let us for simplicity, consider $X$ to be Spin.  (Our following discussion can be straightforwardly generalized to non-Spin $X$). Then, it can be shown~\cite{mine} that the shift $\tau(a) \to \tau(a) + 1$ is a symmetry of the theory if it is accompanied by the transformation  $(\ae, \ne) \to (\ae, \ne - \ae)$.\footnote{For a nontrivially-embedded surface operator,  there is actually a residual phase of $\textrm{exp}( i \pi \IR)$  in the path-integral. Nevertheless, since this $c$-number term is independent of any fluctuating quantum fields, one can set it to 1 via an appropriate chiral rotation of the fermions.} 

Consequently, one can show that for $M = \left(\begin{array}{ccc} a & & b \\ c & & d \end{array} \right) \in SL(2,\bZ)$, the maps
\be
\tau \to {{(a \tau + b)} \over {(c \tau + d)}},
\label{Mtx tau}
\ee
\be
(\CA_d, \CA) \to (\CA_d, \CA) \hspace{0.1cm} {M}^{-1},
\label{Mtx A}
\ee
and
\be  
(\ae, \ne) \to (\ae, \ne) \hspace{0.1cm} {M}^{-1},
\label{Mtx parameters}
\ee
as well as
\be
\CF \to  \widetilde \CF = {1\over 2} bd \CA^2 + {1\over 2} ac \CA^2_d + bc \CA \CA_d + \CF, 
\label{Mtx F}
\ee
together define an $\it equivalence$ transformation of the underlying, effective $U(1)$ theory~\cite{mine}.  

Observe that $\ae$ and $\ne$ transform the way magnetic and electric charges do, respectively, under an $SL(2,\mathbb Z)$ duality. This is not a mere coincidence; rather, it is a reflection of the fact that as pointed out in~\cite{Gukov-Witten}, the surface operator can be interpreted as the worldsheet of a Dirac string with \emph{improperly-quantized} magnetic and electric charges $\ae$ and $\ne$. 
In particular, this means that for an ${\bf S}^2 \subset X$ that surrounds the Dirac monopole at the origin of $D$,  one will have $\int_{{\bf S}^2} F_L / 2\pi = \ae$ mod $ \IZ$, where $\ae \in \IR / \IZ$. Nevertheless, $\int_{{\bf S}^2} F'_L / 2\pi$ remains integral, as required of $F'_L$'s interpretation as a \emph{nonsingular} curvature field strength of a legitimate $U(1)$-bundle over all of $X$: for example, consider $X = {\bf S}^2 \times D$, where ${\bf S}^2$ surrounds the origin of $D$, and $D$ is a Riemann surface of any genus;  from $F'_L = F_L - 2\pi \ae \delta_D$, and the fact that ${\bf S}^2 \cap D = 1$, we find that $\int_{{\bf S}^2} F'_L / 2\pi \in \IZ$. 

On the other hand, notice that $\int_{{\bf S}^2} F_L  / 2\pi \notin \IZ$ contradicts the condition $c_1(L) \in H^2(X, \IZ)$ which underlies the integrality of the monopole number $l$. A resolution to this contradiction is that $\int_{{\bf S}^2} F_L  / 2\pi$ is actually gauge-equivalent to an integer because of a ``ramification''-preserving twisted gauge transformation of the kind in (\ref{twisted gauge tx}) that one is allowed to invoke in the low-energy $U(1)$ theory. In other words, pick \emph{any} integral homology 2-cycle $U \subset X$ (assuming, for simplicity, that $H_2(X, \mathbb Z)$ is torsion-free); that $-c_1(L) [U] = \int_U F_L / 2\pi = \ae (U \cap D) \ {\rm mod} \ \IZ$ is always an integer implies that $\ae$ ought to be equivalent to $\alpha'_{\rm eff}$ whereby $\alpha'_{\rm eff} (U \cap D) \in \mathbb Z$; indeed, for some $u_{\textrm{eff}} \in \mathbb R / \mathbb Z$, there is a gauge transformation by the $U(1)$-function $\textrm{exp} (\theta u_{\textrm{eff}})$ -- which shifts $\ae \to \alpha'_{\rm eff} = \ae + u_{\textrm{eff}}$ -- that respects the condition $\alpha'_{\rm eff} (U \cap D) \in \mathbb Z$, and for which $\ae$ and $\alpha'_{\rm eff}$ are physically-equivalent.


Coming back to our main discussion, notice that (\ref{Mtx A}) also implies that the map
\be
(a_d (u), a (u)) \to (a_d (u), a (u)) \hspace{0.1cm} {M}^{-1}
\ee
defines an equivalence transformation; in particular, one is free to switch to a ``dyonic'' frame of the theory, whereby the effective variable is a linear combination of $a$ and $a_d$.  Thus, we see that at each point $u \in \CM_q$, we have an $SL(2,\mathbb Z)$'s worth of equivalent descriptions of the same underlying theory.  The pair $(a_d (u), a (u))$ can therefore be interpreted as a section of a flat ${\bf C}^2$-bundle over $\CM_q$ with structure group $SL(2,\bZ)$, such that the corresponding symplectic form on ${\bf C}^2$ can be written as $\omega = \textrm{Im} \hspace{0.1cm} da_d \wedge d \bar a$. Hence, if one circles around a $\it singular$ point in $\CM_q$,  a nontrivial monodromy in the section  $(a_d (u), a (u))$ will be induced, and vice-versa. Also,  since (\ref{tau(a)}) and (\ref{ad=a}) imply that $\tau(a) =  {\partial_a a_d}$, the metric  on $\CMq$,
\be
ds^2 =  \textrm{Im}\tau(a) da d \bar a  = \textrm{Im} {da_d \over du} {d \bar a \over d \bar u} du d \bar u = -{i \over 2} ({da_d \over du} {d \bar a \over d \bar u} -  {d \bar a_d \over d \bar u}{da \over du}) du d \bar u,
\ee
is actually the K\"ahler metric given by the pull-back of $\omega$. Consequently, it is manifestly $SL(2,\mathbb Z)$-invariant, as required. 

\bigskip\noindent{\it Monodromies On $\CM_q$ And An Elliptic Curve}

At this point, it is clear that by ascertaining the characteristics of the section $(a_d(u), a(u))$, one can unravel the structure of $\CMq$. From (\ref{ad=a}), we find that $a_d = \partial \CF (a) / \partial a$ and $a = - \partial \CF_d (a_d) / \partial a_d$. Hence, with the expression for $\CF(a)$ and thus $\CF_d(a_d)$ at hand, we can proceed to study the behavior of  $(a_d(u), a(u))$ over $\CMq$. In any case, recall that we made an important observation earlier that the metric $\textrm{Im} \tau(a)$ takes the same form in the ordinary and ``ramified'' cases. This is a direct consequence of the fact that $\CF(a)$  takes the same form with or without surface operators; likewise for $\CF_d(a_d)$. Consequently, the analysis of $(a_d(u), a(u))$ follows that in~\cite{SW} for the ordinary case without surface operators. As such, for brevity, we shall not repeat the calculations here. Instead, we will aim to provide a pedagogical summary of the relevant results.    

Firstly, in circling around the point $u \approx \infty$,  $(a_d (u), a (u))^T$  undergoes a monodromy transformation to $M_\infty \cdot (a_d (u), a (u))^T$, where $M_\infty \in SL(2,\mathbb Z)$. This monodromy transformation implies the existence of other monodromy transformations elsewhere on $\CMq$. However, these other monodromy transformations cannot commute with $M_\infty$, as this would imply that $a(u)$ can be globally-defined, which in turn means that the positive kinetic energy condition of the scalar field $\varphi$ is violated. In other words, the monodromy group must be non-abelian. The minimal number of elements of a non-abelian group is three. Therefore, there are at least two other monodromy transformations away from $u \approx \infty$. Since a discrete chiral transformation of a gauge theory is incapable of removing singularities in its moduli space, one can suppose that the ${\bf Z}_2$ symmetry which maps $u \to -u$, must also map a singular point in $\CMq$ to another -- that is, the other pairs of singularities should occur at $u$ and $-u$.   Indeed, there are precisely two other nontrivial monodromy transformations of $(a_d (u), a (u))^T$ at $u=1$ and $-1$ which are effected by the $2 \times 2$ matrices $M_{1}$ and $M_{-1}$, and they form a closed subgroup of $SL(2, \bZ)$ with $M_\infty$ via the relation $M_1 M_{-1} = M_{\infty}$. This subgroup is $\Gamma(2)$: the group of unimodular matrices congruent to the identity modulo 2. 

\def\imtu{{\textrm{Im} \tau(u)}}

Secondly, a physically consistent theory should be free of singularities in its moduli space of vacua. Let us assume that the singular points at $1$, $-1$ and $\infty$ in $\CMq$ are effectively ``resolved'' or  equivalently, removed,  via some underlying physical mechanism (which we shall account for in a while) -- that is, $\CMq$ is effectively given by the $u$-plane punctured at $1$, $-1$ and $\infty$. Hence, since $\imtu$ is supposed to be positive-definite over the $u$-plane, we can make the identification $\CMq \cong H^+ / \Gamma(2)$, where $H^+$ is the upper complex plane. In other words, $\CMq$ is the modular curve of $\Gamma(2)$. 

Thirdly, notice that $\CMq$ can be identified with the moduli space of a genus one Riemann surface characterized by the equation
\be
y^2 = (x-1)(x+1)(x-u),
\label{curve}
\ee
where $y, x \in \IC$. The idea here is that there is an elliptic curve $E_u$ over every point in $\CMq$. This curve coincides with the SW curve of the ordinary theory without surface operators. This observation can again be corroborated in string theory: by embedding an M2-brane -- which represents a surface operator in the gauge theory -- in an M5-brane in an equivalent M-theory setup, one will find that the SW curve (which lives in some two complex-dimensional $s$-$v$ plane) that the M5-brane must wrap, will not get deformed; we will elaborate on this shortly in $\S$3.4. 

Last but not least, by virtue of a mathematical theorem called ``Riemann's second relation'', the ratio $w_d/ w$ of period integrals of $E_u$ is always positive. Since this corresponds to the condition $\imtu > 0$, one can make the identification
\be
\tau(u) = {w_d(u)  \over w (u)},
\label{period integrals}
\ee
where
\be
w_d(u) = \oint_{c_1} {dx \over y}, \quad  w(u) = \oint_{c_2} {dx \over y}.
\label{w integral}
\ee
Here, $c_1$ and $c_2$ are a canonical basis of homology cycles of the elliptic curve $E_u$. Moreover, from the relation $\tau = \partial_a a_d$, we infer that
\be
w_d(u) = {\partial a_d (u) \over \partial u}, \quad w(u) = {\partial a (u) \over \partial u},
\label{w's}
\ee
which implies that 
\be
a_d (u) = {\sqrt 2 \over  \pi} \int^u_{1}  {dx \sqrt {x-u}\over \sqrt{x^2 -1}}, \quad a(u) = {\sqrt 2 \over \pi} \int^1_{-1}  {dx \sqrt{x-u} \over \sqrt{x^2 -1}}. 
\label{ad,a}
\ee 
Since $(w_d(u), w(u))$ form a system of solutions to the Picard-Fuchs equations associated with the curve (\ref{curve}), one can express  $a_d(u)$ and $a(u)$ explicitly in terms of linear combinations of hypergeometric functions via (\ref{w's}). In turn,  this allows us to compute the prepotential 
\be
\CF(a) = \int_a a_d (a)
\label{F(a)}
\ee 
exactly, such that the values of $f_l$'s in (\ref{full prepotential}) can be obtained precisely. (See~\cite{Lerche} for an excellent review of such an approach).

\bigskip\noindent{\it ``Resolving'' The Singularities With Massless Monopoles And Dyons}

 Let us now explain the underlying physical mechanism which allows the singularities at $1$, $-1$ and $\infty$ to be ``resolved''.  Firstly, as seen earlier in the case with $\CM_c$, when we meet with singular points in the moduli space of vacua, it generally indicates that one has inadvertently integrated out fields that have become massless at those points.  Since the singularities are now at $\pm 1$ and $\infty$, one might be inclined to think that the classical singular point $u=0$ has shifted elsewhere due to quantum corrections, such that at $\pm 1$ or $\infty$, the $W^{\pm}$ bosons become massless. This, as argued in~\cite{SW}, cannot be the case: the asymptotic superconformal invariance expected in the IR if there are massless $W^\pm$ bosons contradicts the fact that there is an instanton anomaly in the $U(1)_R$ symmetry of the supposed full $\CN =2$ superconformal algebra. Hence, the only states that can become massless at the singular points must have spins $\leq 1/2$.

Secondly, recall that as noted earlier, one can find monopoles and dyons in the semiclassical spectrum of the gauge theory. These are generically massive states with spins $\leq 1/2$, and they are BPS in our context. Hence, they survive beyond the semiclassical limit in the exact quantum theory, and are furnished by $\CN =2$ hypermultiplets. From $\CN =2$ supersymmetry and  the central charges of  its algebra, the $SL(2,\mathbb Z)$ equivalence of descriptions, and the fact that the masses $m$ of these states are physically observable and therefore, should be invariant under the monodromies of the moduli space, we find that 
\be
m = |an_e + a_d n_m|,
\label{BPS mass}
\ee    
where $(n_m, n_e)$ are the (magnetic, electric) charges of the BPS states under consideration. 

Thirdly, note that the equation (\ref{curve}) describes a double cover of the $x$-plane branched over $1$, $-1$, $\infty$ and $u$. Consequently, certain linear combinations of the cycles $c_i$ of  $E_u$ shrink to zero when (and only when) two branch points coincide. This happens precisely for $u=1$, $-1$ and $\infty$, whereby the corresponding collapsing cycles are $c_1$, $(c_1 - c_2)$ and $c_2$ (with $c_1$ going to infinite radius). That the torus $E_\infty$ has cycles $c_2$ and $c_1$ of zero and infinite radii, respectively, corresponds to the fact that  at $u = \infty$, one is in the semiclassical regime where $\textrm{Im} \tau \to \infty$ (cf.~(\ref{tau(a) full}) for large a and (\ref{period integrals})). That the cycle $c_1$ shrinks over $u=1$ means from (\ref{w integral}) and (\ref{w's}) that $a_d \to 0$;  hence, from (\ref{BPS mass}), we find that   a monopole of charge $(1,0)$ will become massless at $u=1$. Likewise, the fact that the cycle $(c_1 - c_2)$ shrinks over $u= -1$ means that a dyon of charge $(1,-1)$ will become massless at $u= -1$.      
    
Hence, the singularities at $u = 1$ and $-1$ can be ``resolved'' by coupling the effective $U(1)$ theory to massless hypermultiplet monopoles and dyons of charge $(1,0)$ and $(1,-1)$ at those points. The singularity at $u = \infty$ however, does not result from integrating out massless states;  it cannot be ``resolved'' by coupling the theory to massless solitons. Nevertheless, notice that at $u = \infty$, we have $\textrm{Im} \tau_d (a_d) \to 0$; the singularity therefore implies that there is a preferred frame --  namely, the ``electric'' frame in the variable $a$ --  to describe the effective $U(1)$ theory in this semiclassical region of $\CMq$. 

In fact, there is a preferred frame at $u=1$ and $-1$ also: at $u=1$ and $-1$,  the effective $U(1)$ theory needs to be expressed in its ``magnetic'' and ``dyonic'' frames in the variables $a_d$ and $(a-a_d)$ before it can be consistently coupled to (massless) magnetic monopoles and dyons. Indeed, one can check from the monodromy transformations of the vector $(a_d(u), a(u))^T$ that the only variables which are\emph{ single-valued} (up to an irrelevant Weyl reflection since we are working on the $u$-plane) upon circling the points $u= \infty$, $1$ and $-1$, are $a$, $a_d$ and $(a-a_d)$; that is, these variables are the only good variables at $u=\infty$, $1$ and $-1$.  Nonetheless, one is free to express the theory locally in $\it any$ frame within an $SL(2,\mathbb Z)$ class throughout the rest of $\CMq$. 

\bigskip\noindent{\it Coupling To Massless $\CN =2$ Hypermultiplets}

\def\mM{{\mathscr M}}

To couple the effective $U(1)$ theory in its ``magnetic'' frame to massless monopoles from an $\CN =2$ hypermultiplet at $u=1$ in a fully supersymmetric fashion, one needs to add to the ``magnetic'' action $\Seffd$ in (\ref{Seff_d}) the term (cf.~\cite{Marcos})
\be
S^d_{\textrm{hyper}} = \int d^4 x d^2 \theta d^2 \bar \theta \ \left(\mM^{\dagger} e^{2V'_d} \mM + {\widetilde \mM}^\dagger e^{-2V'_d} \widetilde \mM \right)  + \sqrt 2  \int d^4 x d^2\theta \ \left(\widetilde \mM \CA_d \mM + \textrm{h.c.}\right).  
\label{S-hyper}  
\ee 
Here, $\mM$ and $\widetilde \mM$ are ordinary (local) $\CN =1$ chiral superfields in conjugate representations of the $U(1)$ gauge group, and $V'_d$ is a ``magnetic'' $\CN =1$ vector superfield with  ``ramified'' abelian superfield strength $(W'_d)_\alpha = -{1\over 4} {\overline D}^2 D_\alpha V'_d$, where $D$ and $\overline D$ are the superspace derivative and its hermitian conjugate, respectively.  $\mM$ and $\widetilde \mM$ supply an $SU(2)_R$ doublet of complex scalar fields $q_I$, where $I = 1 \dots 2$; two singlet Weyl-spinors $\chi_{q_1}$ and $\psi_{q_2}$ as their superpartners; and an $SU(2)_R$ doublet of auxiliary fields $F_J$, where $J = 1 \dots 2$. The monopole fields correspond to the complex scalars $q_v$.  

One can do the same for the effective $U(1)$ theory in its ``dyonic'' frame at $u=-1$. In this case, the effective uncoupled action would be expressed in terms of the variable $(a-a_d)$ instead of $a_d$ etc., and the hypermultiplet action that one must add in order to couple the theory to massless dyons of charge $(1,-1)$ would be (\ref{S-hyper}), albeit with $\CA_d$ replaced by $(\CA - \CA_d)$. 

\newsubsection{The ``Ramified'' Seiberg-Witten Theory}

 What we will be interested in eventually, is a $\it topological$ version of the above coupled theory with action  $\Seffd + S^d_{\textrm{hyper}}$. This theory can be obtained by first wick-rotating the action $\Seffd + S^d_{\textrm{hyper}}$ to Euclidean space, and then ``twisting'' it via the recipe laid out in $\S$2.3; this is possible because the untwisted theory, although abelian, is still $\CN =2$ supersymmetric with an $SU(2)_R$ global symmetry.  We shall henceforth denote the resulting topological abelian action as $S_{u=1}$, and name the theory it describes as the ``ramified'' Seiberg-Witten theory.  

Note that as required of a topological theory, $S_{u=1}$ can be written (on a Spin $X$) as (cf.~\cite{Marcos}) 
\be
S_{u=1} =   { \{ \cq_{\textrm{sw}} , W \} \over e_d^2}  +   {i\Theta_d \over 8 \pi^2} \int_X   {F^d_L}' \wedge {F^{d}_L}'  -  i  \eta_d \int_X     \delta_D \wedge {F^d_L}'
\label{Su=1}
\ee
(modulo topological couplings to gravity), where $ \cq_{\textrm{sw}}$ is a nilpotent supercharge that generates the supersymmetric transformations of all the fields -- both dynamical and auxiliary -- in the (twisted) $\CN =2$ vector multiplet and hypermultiplet; $W$ is some scaling dimension zero fermionic operator whose $R$-charge is negative of that of $ \cq_{\textrm{sw}}$; $(\alpha_d, \eta_d) = (\alpha^d_{\textrm{eff}}, \eta^d_{\textrm{eff}})_{u=1}$;\footnote{As we will explain in a later subsection, the effective parameters $(\ae, \ne)$ may depend on $u$. Hence our notation.}  and ${F^{d}_L}'  = F^d_L - 2 \pi \alpha_d \delta_D$ is the $\it effective$ field strength of the ``magnetic'' $U(1)$ photon defined over all of $X$ -- it can also be interpreted as the curvature of a $U(1)$-bundle over $X \backslash D$ with singularity $2 \pi \alpha_d \delta_D$ along $D$.  

Recall from $\S$2.3 that the ``twisting'' procedure is tantamount to identifying the $SU(2)_R$ indices $I,J,K$ of the fields with the $SU(2)_+$ indices $\dot A,\dot B,\dot C$. In particular, the twist will convert the complex scalars $q_I$ to spinors $M_{\dot A}$ -- in other words, the monopoles will correspond to the spinor fields $M_{\dot A}$ in the twisted theory. Consequently, the monopoles are apparently not well-defined on non-spin manifolds. However, because they are coupled to the effective ``magnetic'' $U(1)$ photon in the (twisted) theory, the monopoles \emph{do} in fact remain well-defined on non-spin manifolds. We refer the reader to \cite{Appendix 1} for an explanation of this point.

\bigskip\noindent{\it The ``Ramified'' Seiberg-Witten Equations And Invariants}

\def\mF{{\mathscr F}}

At any rate, since at $u=1$, the ``magnetic'' frame is the preferred frame, the underlying effective $U(1)$ theory is \emph{uniquely} described by $S_{u=1}$. Moreover, notice from (\ref{tau(a) full}) that at $u=1$, the ``electric'' gauge coupling $\tau(a)$ is strong; conversely, the ``magnetic'' gauge coupling $\tau_d(a_d)$ is weak. Hence, one has $e_d \to 0$ in $S_{u=1}$. Consequently, as in the case with the topological $SU(2)$ theory discussed in $\S$2, the non-zero contributions to the correlation functions of the theory at $u=1$ are centered around classical contributions which minimize the action. Repeating the analysis in $\S$2.4 and $\S$2.5 here, we find that the correlation functions will correspond to integrals of top-forms in the moduli space of supersymmetric configurations defined by setting the  $\cq_{\textrm{sw}}$-variations of the fermi fields to zero. 

From the relevant $\cq_{\textrm{sw}}$-variations of the fermi fields, we find that supersymmetric configurations correspond to solutions of the equations  (cf.~\cite{Marcos})
\be
(F^d_L - 2 \pi \alpha_d \delta_D)_+ = (\overline M M)_+
\label{SW1}
\ee  
and 
\be
\dirac M =0,
\label{SW2}
\ee
where $\dirac$ is the Dirac operator $\it coupled$ to the effective ``magnetic'' $U(1)$ photon with field strength ${F^d_L}'$. (\ref{SW1}) and (\ref{SW2}) together define the ``ramified'' Seiberg-Witten equations, whence the nontrivial $\it topological$ correlation functions can be written as
\be
 \langle  [{J^d_0}(p)]^{q} J^d_1(\delta_{i_1}) \dots J^d_1(\delta_{i_r})   \rangle_{x'} = \int_{\CM^{x'}_{\textrm{sw}}} [{J^d_0}(p)]^{q} \wedge J^d_1(\delta_{i_1}) \wedge \dots \wedge 
J^d_1(\delta_{i_r}) = SW_{x'} (\beta_{i_1} \wedge \dots \wedge \beta_{i_r}).
\label{correlation function SW}
\ee
Here, $q = {1\over 2} (d^{x'}_{\textrm {sw}} - r)$;  $d^{x'}_{\textrm{sw}}$ is the dimension of the moduli space $\CM^{x'}_{\textrm{sw}}$ of the ``ramified'' SW equations for some $x' = - 2c_1({L'_d})$, where $L'_d$ is a $U(1)$-bundle on $X$ with curvature field strength ${F^d_L}'$;  and
\be
d^{x'}_{\textrm{sw}} =  \left ( \alpha^2_d D\cap D -2 \alpha_d  l_d \right)  - 2\left ({2 \chi + 3 \sigma \over 8}  + k_d \right),
\ee
where $l_d = \int_D {F^d_L / 2\pi}$ and $k_d = -{1 \over 8 \pi^2} \int_X F^d_L \wedge F^d_L$ are integers.  Also, by utilizing a $U(1)$ version of (\ref{G action}), we have
\be
\label{a_d}
{J^d_0}(p) = \langle \varphi_d(p) \rangle = a_d \quad \textrm{and} \quad  J^d_1(\delta) = {1 \over 4 \sqrt 2} \int_{\delta} \psi_d,
\ee 
where $p \in H_0(X)$ and $\delta \in H_1(X)$ (with dual $\beta \in H^1(X)$).\footnote{Note that there is a two-form descendant operator $J^d_2(\Sigma) \sim \int_{\Sigma} {F^d_L}'$, where $\Sigma \in H_2(X)$. Hence, $J^d_2(\Sigma)$ is just a number, as it measures the first Chern class of the ``magnetic'' $U(1)$-bundle. Hence, it is trivial as an operator, and we do not consider it. Note also that in contrast to the physical definition of the correlation functions which correspond to the ``ramified'' Donaldson invariants, here, we need not restricted the zero and one-cycles $p$ and $\delta$ to $X \backslash D$. This is because the operators $a_d$ and $\psi_d$ -- unlike $F$, $F_L$ or $F^d_L$ -- do not contain singularities along $D$.} As required, ${J^d_0}(p)$ is expressed purely in terms of non-fluctuating zero-modes, and $\psi_d$ -- which is the fermi field in the same ``magnetic'' $\CN=2$ vector multiplet as the  complex scalar $\varphi_d$ -- is a spacetime one-form in the twisted theory. However,  ${J^d_0}(p)$ and $ J^d_1(\delta)$ have $R$-charge 2 and 1 (associated with an accidental $U(1)_R$ symmetry at low-energy); consequently, they are two-forms and  one-forms in $\CM^{x'}_{\textrm{sw}}$, respectively. We shall call $SW_{x'} (\beta_{i_1} \wedge \dots \wedge \beta_{i_r})$ the ``ramified'' Seiberg-Witten invariant (for the class $x'$); the rationale for this nomenclature, and the formulas of $d^{x'}_{\rm sw}$ and $x'$ etc., are explained in detail in~\cite{Appendix 1}.

Finally, note that the topological ``dyonic'' theory at $u=-1$ can also be obtained using the same methods elucidated above. However,  in favor of brevity, we shall skip the discussion. Moreover, as we shall see later, the results that we will need at $u=-1$ can be easily obtained from the results that we will get at $u=1$ via a switch of variables $a \to (a-a_d)$, and via a discrete global symmetry which relates the vacua at $u =1$ to that at $u= -1$.

\newsubsection{The Mapping Of High-Energy Observables To Low-Energy Observables.} 

\def\ip{{\widetilde I_0}}
\def\is{{\widetilde I_2}}

We shall now discuss the mapping of the relevant high-energy observables in the microscopic $SU(2)$ or SO(3) gauge theory  -- given by \emph{untwisted} correlation functions of the form in (\ref{Donaldson generating function - physical})  --  to their corresponding low-energy observables in the effective $U(1)$ theory over the $u$-plane. The reason for this discussion is the following: let $\Sigma_i$ be some homology two-cycle in spacetime;  then, at low energies, the nonlocal operators $\widetilde I_2(\Sigma_i)$ -- which are the low-energy counterparts of $I_2(\Sigma_i)$ obtained via the scale-invariant ``descent'' procedure -- can intersect one another at certain points in spacetime; consequently, (since $\widetilde I_2(\Sigma_i)$ can have non-zero modes in the untwisted $U(1)$ theory), this results in additional ``contact terms'' due to nontrivial propagators in the neighborhood of these points --  in other words, the pair $I_2(\Sigma_1) I_2(\Sigma_2)$ does not map  exactly to $\is(\Sigma_1)\is(\Sigma_2)$. Let us study this more closely.   

\bigskip\noindent{\it The Correlation Functions Of The Low-Energy Operators}

As is clear from the form of the correlation functions in (\ref{Donaldson generating function - physical}), it suffices for us to consider,  in the low-energy theory, the correlators of the  operators $\widetilde I_0(p_j)$ and $\widetilde I_2(\Sigma_i)$  amongst themselves, as well as between themselves. Let us first consider the correlator
\be
\langle \ip (p_1) \dots \ip(p_r) \rangle. 
\ee 
To go to the low-energy limit, one can just scale up the metric -- that is, make the replacement $\bar g \to t\bar g$, and take $t \to \infty$. In this limit, since the $\widetilde I_0(p_j)$'s  are local operators of codimension four in $X$, one can assume that they do not intersect one other and are separated on a distance scale that is larger than any Compton wavelength in the theory. Consequently, one can employ  cluster decomposition and write the correlator as
\be
\langle \ip(p_1) \dots \ip(p_r) \rangle ={\langle \ip \rangle_{\Omega}}^r \cdot \langle 1\rangle,  
\label{cluster decomposition}
\ee
where $\langle \ip \rangle_\Omega$ is the normalized vacuum expectation value of $\ip$ in the infinite-volume limit of spacetime, or rather, flat ${\bf R}^4$; and $\langle 1 \rangle$ is just the partition function. Since $\ip$ is an operator zero-form in spacetime, its non-vanishing vacuum expectation value does not violate Lorentz invariance. Furthermore, from (\ref{I's}), (\ref{I's zero-modes}) and (\ref{u}), one can make the identification $2 u = {\langle I_0 \rangle}_\Omega$.  
(Why a factor of two is introduced will be clarified momentarily in $\S$4.1.)  

Next, let us consider the correlator 
\be
\langle \is(\Sigma_1) \dots \is(\Sigma_r)\rangle. 
\label{correlator S}
\ee
Notice that since the $\Sigma_i$'s are  two-cycles of codimension two in spacetime, they may still intersect one another even if we scale the metric up. Nevertheless,  by perturbing the $\Sigma_i$'s slightly within their homology and even homotopy classes, one can arrange for them to intersect transversely over a finite number of points in a pairwise fashion. Moreover, even for pairs of operators that do not intersect at all, one cannot apply cluster decomposition to simplify (\ref{correlator S}) to the form shown in (\ref{cluster decomposition}) -- this is because 
\be
{\langle \is(\Sigma_i) \rangle}_{\Omega} = \int_{\Sigma_i} {\langle Z_{mn}  \rangle}_\Omega d \sigma^{mn} = 0,
\ee 
since $Z_{mn}$ is an operator two-form in spacetime and  ${\langle Z_{mn} \rangle}_\Omega = 0$ by Lorentz invariance. Here, $d \sigma^{mn}$ is the integration measure over $\Sigma_i$. At any rate, it is clear that it suffices for us to consider the two-point correlator 
\be
\langle \is(\Sigma_1) \is(\Sigma_2) \rangle = \int_{\Sigma_1 \times \Sigma_2} \langle Z_{mn} (x) Z_{pq} (y) \rangle \hspace{0.1cm} d\sigma^{mn}(x)  d\sigma^{pq}(y), 
\label{2-pt}
\ee
where $x$ and $y$ is \emph{any} point in $\Sigma_1$ and $\Sigma_2$, respectively.

As pointed out earlier,  the $\beta$-function of the low-energy effective $U(1)$ theory actually vanishes, as reflected by the halting of the running coupling constant of the non-abelian theory at scale $\mu = a$. Thus, we effectively have a scale-invariant theory over the $u$-plane at low energies. Consequently, the correlator $\langle Z_{mn} (x) Z_{pq} (y) \rangle$ can be expressed -- via a scale-independent operator product expansion -- as
\be
\langle Z_{mn} (x) Z_{pq} (y) \rangle \sim \delta^4 (x-y) \langle T(y) \rangle + \cdots,
\label{OPE}
\ee 
where $T(y)$ and $\delta^4(x-y)$ are an operator zero-form and a delta four-form in spacetime, respectively, while the ellipses refer to less singular terms whose explicit forms will not concern us for now. 

At points where $x \neq y$, the \emph{non-vanishing} two-point correlator will be as given in (\ref{2-pt}), with (\ref{OPE}) in mind. (This is in contrast to the case with a mass gap discussed in~\cite{sym}, where in the $\bar g \to \infty$  limit, the two-point correlator actually vanishes for any $x \neq y$.) However, at points where $x= y$, that is, where $\Sigma_1$ and $\Sigma_2$ intersect, the singularity of the delta four-form in (\ref{OPE}) implies that the contributions to the two-point correlator will take the form
\be
\sum_{P \in \Sigma_1 \cap \Sigma_2} \epsilon_P \langle T(P) \rangle,  
\label{add}
\ee
where $\epsilon_P$ is a constant associated with the point of intersection $P$. Since reversing the orientation of $\Sigma_i$ in $\is(\Sigma_i)$ changes its sign, it would mean that if $\Sigma_1$ and $\Sigma_2$ intersect with opposite relative orientations at some $P$, the two-point correlator would pick up a minus sign, and vice-versa. With an appropriate normalization, we can take $\epsilon_P$ to be $\pm 1$, depending on the relative orientation of the intersection of $\Sigma_1$ and $\Sigma_2$ at $P$. 

\bigskip\noindent{\it The Map Between High And Low-Energy Correlation Functions Of Operators}

Note that because of asymptotic freedom in the microscopic $SU(2)$ or $SO(3)$ theory, the coupling gets very weak at very small scales. If we restrict the definition of our untwisted two-point correlation functions in the microscopic operators $I_2(\Sigma_i)$  to such scales where the semiclassical approximation is exact,\footnote{Such a restriction is inconsequential, since we will eventually be interested in the correlators of the scale-invariant topological theory.} we will not have additional contributions of the form (\ref{add}), since all interaction terms of the kind (\ref{OPE}) are negligible. In this sense, the additional contributions given in (\ref{add})  are a manifestation of the low-energy theory at \emph{non-vanishing} coupling only.   Thus, this means that within correlation functions, we have the map 
\be
I_2(\Sigma_1) I_2(\Sigma_2) \to \is(\Sigma_1) \is(\Sigma_2) + \sum_{P \in \Sigma_1 \cap \Sigma_2} \epsilon_P \hspace{0.0cm} T(P),
\label{map}
\ee 
from high to low energy observables. 

Since $I_2 (\Sigma_i)$ and $\is(\Sigma_i)$ are constructed out of the canonical ``descent'' procedure, they are manifestly $\cq$-invariant; hence, so will $I_2(\Sigma_1) I_2(\Sigma_2)$ and $\is(\Sigma_1) \is(\Sigma_2)$ (in our off-shell formulation).  Thus, the map (\ref{map}) implies that $T(P)$ must also be $\cq$-invariant. Since $T(P)$ was determined to be an operator zero-form in spacetime, $\cq$-invariance will mean that $T(P)$ must be holomorphic in $u$ (cf.~(\ref{susy tx - non-abelian})).   

Last but not least, consider any correlator between the  operators $\ip(p_j)$ and $\is(\Sigma_i)$. Since $\ip(p_j)$ and $\is (\Sigma_i)$ are local and nonlocal operators of codimension four and two in spacetime, upon scaling the metric up, one can always deform the $\Sigma_i$'s within their homology and homotopy classes such that the operators do not intersect. Therefore, contributions  to the correlator of the kind (\ref{add}) are absent.  

With all the above understood, from (\ref{cluster decomposition}), the identification $2u = {\langle I_0 \rangle}_\Omega$, and  (\ref{map}), we find that \emph{within correlation functions}, we have the map 
\be
{\rm exp} (p I_0 + I_2(\Sigma) ) \to {\rm exp} (2pu + \is(\Sigma) + \Sigma^2 T_{\Sigma}(u)),
\ee
from high to low energy observables.  Here, $\Sigma^2 = \Sigma \cap \Sigma$, where $\Sigma = \sum_i a_i \Sigma_i$ with real constants $a_i$; also, $T_\Sigma$ is written in a $P$-independent way because the $\cq$-invariance of $T_\Sigma(u(P))$ will imply (in the topologically twisted theory of eventual interest) that the correlation functions are independent of the insertion point $P$.

\newsubsection{ A String-Theoretic Realization Of The Gauge Theory With Surface Operators} 

\def\br{{{\bf R}^{1,1}}}
\def\bR{{{\bf R}^{3,1}}}
\def\ns{{\rm NS5'}}

We shall now furnish a string-theoretic realization of our  pure $\CN =2$,  $SU(2)$ gauge theory with surface operators, following~\cite{loop}. The objectives are to corroborate some of the observations made earlier about the low-energy vacuum structure of the gauge theory, as well as to introduce the concept of curved surface operators which are necessarily topological at the outset.  An in-depth account of the relevant claims made in~\cite{loop} will be furnished in the course of our analysis, and the reader who is familiar with~\cite{loop} is nevertheless encouraged to read this subsection.    

\bigskip\noindent{\it A String-Theoretic Realization}

Consider a trivially-embedded and \emph{flat} surface operator along $D = {\bf R}^{1,1}$ in the $\CN =2$, $SU(2)$ gauge theory in ${\bf R}^{3,1}$. As argued in~\cite{loop}, 
the surface operator -- by virtue of its embedding in $\bR$ and hence, the subgroup of an associated superconformal group it preserves -- is found to be half-BPS; that is, the surface operator  will only preserve half of the eight supercharges of the $\CN=2$ gauge theory, along $D$. 
 
\begin{figure}
  \centering
  \subfigure[]{\includegraphics[width=.35\textwidth]{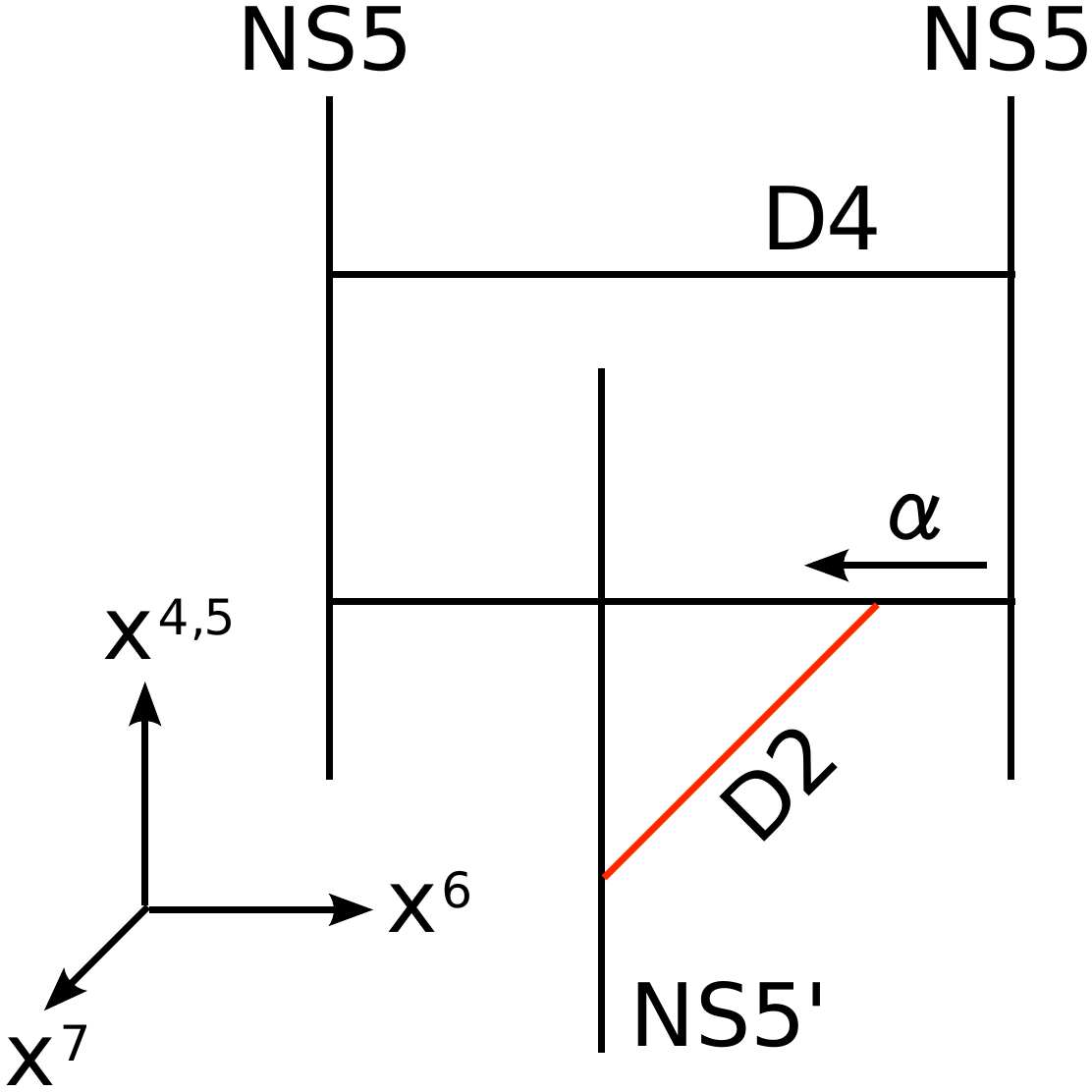}}
  \qquad \qquad
  \subfigure[]{\includegraphics[width=.35\textwidth]{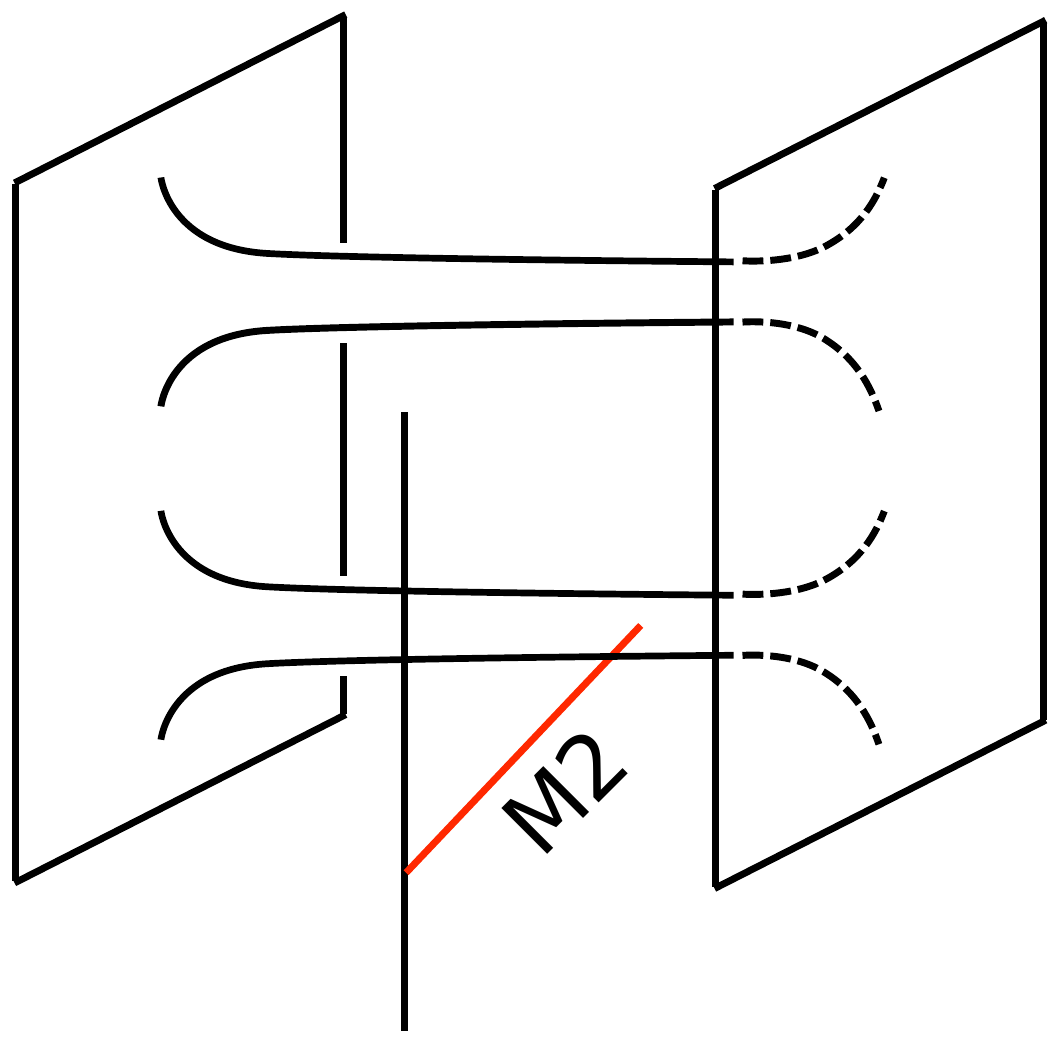}}
  \caption{The type IIA setup and its lift to M-Theory}
  \label{Fig1}
\end{figure}

With this in mind, now consider type IIA string theory and a stack of two D4-branes that are suspended between two NS5-branes with a D2-brane trivially-embedded in the worldvolume of the D4-branes along $\br$; let the other spatial direction of the D2-brane worldvolume end on an independent $\ns$-brane at  a finite distance; that is, consider the configuration:
\begin{eqnarray}
\label{IIA config}
{\rm  NS5s}: & \quad & 012345 \nonumber \\
{\rm D4}:  & \quad & 0123 \ \ \  \ 6 \nonumber \\
{\rm NS5'}: & \quad & 01 \ \ \  45 \ \ \ \ \ 89 \nonumber \\
{\rm D2}:  & \quad & 01 \ \ \ \ \ \ \ \ \ 7 
\end{eqnarray}
where the two NS5-branes, two D4-branes, $\ns$-brane and D2-brane are located at some value of $(x^6, x^7, x^8, x^9)$, $(x^4, x^5, x^7, x^8, x^9)$, $(x^2, x^3, x^6, x^7)$ and $(x^2, x^3, x^4, x^5, x^6, x^8, x^9)$, respectively. This is illustrated in Fig.~\ref{Fig1}(a) below.

Notice that the D4-branes worldvolume is finite in extent along $x^6$ with a length  $\Delta x^6$,  where  $\Delta x^6$ characterizes the separation of the two NS5-branes. At scales larger than  $\Delta x^6$ (or any finite spatial distances mentioned hereafter), the D4-branes worldvolume theory is effectively an  $\CN =2$ gauge theory in $\bR$. The D2-brane, being half-BPS, preserves only half of the eight supercharges of the  $\CN =2$ gauge theory along $\br$. Thus, it seems that we do have a string-theoretic realization of the above $\CN= 2$, $SU(2)$ gauge theory in $\bR$ as the effective worldvolume theory of the D4-branes,\footnote{Note that we have factored out the center-of-mass degree of freedom of the D4-branes, which thus leads to an $SU(2)$ gauge symmetry in the 4d theory.}  whereby the surface operator along $D$ corresponds to the trivially-embedded D2-brane in the $(x^0, x^1)$ directions. But what about the surface operator parameters? What is their interpretation in this string-theoretic setup? 

To answer this question, first note that along $D$ in the D2-brane  lives a $U(1)$ gauge theory with $\CN = (2,2)$ supersymmetry (plus some matter content). The finite distance $\Delta x^7|_\ns$  (along the second spatial direction of the D2-brane) between $D$ and the $\ns$-brane, means that the gauge coupling $e$ of this $\CN =(2,2)$, $U(1)$ theory -- given by $1 / e^2 \sim \Delta x^7|_\ns$ -- is non-vanishing; in other words, we have an \emph{interacting} theory along $D$.  Second, note that according to the arguments in~\cite{hori},  this interacting theory is characterized by a 2d $U(1)$  
twisted chiral multiplet; in the absence of twisted masses, its space of vacua is given by $\IC // U(1) \cong \IC {\bf P}^1$, and at the relevant scales, the theory is effectively an  $\CN =(2,2)$ non-linear sigma model (NLSM) with target space $\IC {\bf P}^1$.  Third, note that the Lagrangian of the NLSM has a term 
\be
{it \over 2 \pi} \int_{D} \Phi^* \left[{\omega^{\rm FS}}\right],
\label{t}
\ee
where $\Phi$ is a holomorphic map $\Phi: D \to \IC {\bf P}^1$; $t = r +i \theta$ is the F.I. parameter of the underlying $U(1)$ gauge theory; $\omega^{\rm FS}$ is the K\"ahler form of the Fubini-Study metric on $\IC {\bf P}^1$; 
and $\Phi^* \left[{\omega^{\rm FS}}\right]$ is given by $-F_D$, where $F_D$ is the field strength of the underlying $U(1)$ gauge field. Fourth, recall that the surface operator in question is a supersymmetric configuration characterized by $F'_+= 0$.  Substituting this into the gauge kinetic terms of the 4d theory, we find that there is a contribution along $D$  proportional to 
\be
(\eta + \tau_{cl} \alpha)  \int_D {F_D \over 2\pi}.
\label{n+t}
\ee
By comparing (\ref{t}) and (\ref{n+t}), we see that we can naturally identify the surface operator parameter $(\eta + \tau_{cl} \alpha)$ with the F.I. parameter $t$, as claimed in~\cite{loop}. Hence, one just needs to know what $t$ corresponds to in the string-theoretic setup.

To this end, note that the fields, flavor symmetry and F.I. parameter of the underlying $U(1)$ gauge theory along $D$ can be interpreted in terms of the positions of the various branes and their motion etc. In particular,  the difference $\Delta x^6|_\ns$ in the positions of the $\ns$-brane and NS5-brane along the $x^6$-direction corresponds to the parameter $-r$, where for non-zero values of $\Delta x^6|_\ns$, the D2-brane and hence $D$, will be forced to lie along one of the two D4-branes. So we have an interpretation of $r$ as $-\Delta x^6|_\ns$. But what about $\theta$? According to~\cite{hori}, $\theta$ will correspond to the difference $\Delta x^{10}|_\ns$ in the positions of the $\ns$-brane and NS5-brane along the $x^{10}$-direction of the M-theory circle. Altogether, this means that  $(\eta + \tau_{cl} \alpha)$ can be identified as $(i\Delta x^{6} + \Delta x^{10})|_\ns$, modulo nonperturbative effects which will be clarified shortly. Meanwhile, notice that if we set the $\Theta$-angle to zero via an appropriate chiral rotation of the 4d fermions, we see that $\alpha$ is (for a particular 4d gauge coupling) proportional to $\Delta x^6$; this is depicted in fig.~1(a).

\bigskip\noindent{\it Topological And Non-Topological Surface Operators}

We now come to a very interesting observation.  Notice that $t$, being the F.I. parameter of the $U(1)$ theory on $D$, will get renormalized~\cite{MS} -- this means that  $(\eta + \tau_{cl} \alpha)$ will run as we vary the scale of the theory. Moreover, it is known~\cite{MS} that the exact expression for $t$  also involves instanton corrections that are not captured by $(i\Delta x^{6} + \Delta x^{10})|_\ns$ -- in other words, the exact expression for $(\eta + \tau_{cl} \alpha)$ will involve instanton corrections as well.  One can interpret this observation in two ways. 

Firstly, if we assume  $(\alpha, \eta)$ to be constant and scale-invariant, then the running of and the instanton corrections to $(\eta + \tau_{cl} \alpha)$  can be attributed to the renormalization of  and the instanton corrections to $\tau_{cl}$, respectively. This is a natural and reasonable interpretation, since it is known that $t$ can be identified with $\tau(a)$ of the $\CN=2$ gauge theory in $\bR$~\cite{hori}, and in this case, we indeed have $t \sim \tau(a)$ (after an appropriate chiral rotation of the 4d fermions to absorb a constant  shift of $\eta / \alpha$ in the $\Theta$-angle). Let us call the corresponding surface operators with constant $(\alpha, \eta)$ \emph{topological}  surface operators. For such surface operators, we have $(\alpha, \eta) = (\ae, \ne)$.  

Alternatively, one can assume $\tau_{cl}$ to be fixed while $(\alpha, \eta)$ gets corrected both perturbatively and nonperturbatively. Since $(\alpha, \eta)$ is neither constant nor scale-invariant in this case, we have a notion  of ``surface operators that flow'';  we shall call such surface operators \emph{non-topological} surface operators. As we will be able to corroborate many of the claims made in~\cite{loop} via this second interpretation, we shall continue to explore its implications. Moreover, in doing so, we will be led to a family of low-energy non-topological surface operators parametrized by the total space of the fibration of the SW curve over $\CM_q$. Let us elaborate on this now.  

\bigskip\noindent{\it The Low-Energy Non-Topological Surface Operator Parameters}

Firstly, note that the position of the D2-brane along the $x^4$--$x^5$ plane 
is characterized by a complex scalar $ \sigma \sim (x^4 + i x^5)|_{\rm D2}$,  where $\sigma$ is the (eigenvalue of the) scalar component of the twisted chiral multiplet of the $U(1)$ theory along $D$.  Secondly, note that a difference in the positions $(x^4 + i x^5)|_{\rm D4, u}$ and $(x^4 + i x^5)|_{\rm D4, l}$ of  the upper and lower D4-branes 
will result in non-zero twisted masses ${\widetilde m}_i$ for the two chiral superfields in the $U(1)$ multiplet of the theory along $D$. In turn, this will break the $SU(2)$ flavor symmetry of the $U(1)$ theory along $D$ to a $U(1)$ flavor symmetry -- indeed, the separation distance of the D4-branes  corresponds to the Coulomb branch vev $a$  of the background $\CN =2$, $SU(2)$ theory in $\bR$, and if $a$ is non-zero, its $SU(2)$ gauge symmetry will be spontaneously-broken to a $U(1)$ gauge symmetry (at a low-energy scale below $a$), which, in turn,  implies that the $SU(2)$ flavor symmetry must be broken to a $U(1)$ flavor symmetry in the $U(1)$ theory along $\br =D$. 
At any rate, note that a non-zero $a$ also halts the running of the 4d gauge coupling at the scale $\mu =a$, consistent with the vanishing of the $\beta$-function in the corresponding low-energy effective $U(1)$ theory in $\bR$. This too, is reflected in the halting of the running of the F.I. parameter $t$ when twisted masses are ``turned on''. With all the above understood, we can now write the \emph{scale-independent}, low-energy effective parameter of the non-topological surface operator  as (cf.~\cite{hori, MS})
\be
(\eta_{\rm eff} + \tau_{cl} \alpha_{\rm eff}) = (\eta + \tau_{cl} \alpha) +  \sum_{i=1}^2 {\rm ln} ({\sigma - {\widetilde m_i} \over \Lambda}) + \sum^\infty_{n=1} a_n\left[{(\sigma - {\widetilde m_1})(\sigma - {\widetilde m_2}) \over \Lambda^2} \right]^{-n}, 
\label{s.o. flow}
\ee
where $\Lambda$ is an RG-invariant scale parameter. 

The first and second terms in (\ref{s.o. flow}) correspond to the ``bare'' parameter and the one-loop corrections, respectively. This is captured by $(i\Delta x^{6} + \Delta x^{10})|_\ns$, as mentioned earlier. The third term however, is not captured by $(i\Delta x^{6} + \Delta x^{10})|_\ns$;   it arises due to the nonperturbative effects of worldsheet instantons from the sigma-model which are characterized by holomorphic maps $\Phi: D \to \IC {\bf P}^1$ -- in other words, the third term can be understood as a correction to $(\eta_{\rm eff} + \tau_{cl} \alpha_{\rm eff})$ due to 2d-instantons which are stretched between the two NS5-branes.

The relation between the ${\widetilde m}_i$'s and the Coulomb branch vev $a$ has been explained; thus, the dependence  of $(\ae, \ne)$ on $u(a)$ is manifest in (\ref{s.o. flow}). What about its $\sigma$-dependence?    

\bigskip\noindent{\it Lift To M-Theory And The SW Curve}

\def\bs{{\bf S}}
\def\ms{{{\rm M5}'}}

Notice that by incorporating the $x^{10}$-direction into our expression for $t$ earlier, we have implicitly assumed a lift to M-theory. This requires us to consider the large string coupling limit of the type IIA setup, if the circle of the $x^{10}$-direction is to be of a finite size, such that ${\Delta x^{10}}{|_\ns}$ is to have any meaning. As argued in~\cite{Witten-sol}, our above computation of  $(\eta_{\rm eff} + \tau_{cl} \alpha_{\rm eff})$ will not depend on the string coupling: one can rescale the string coupling and the distances $\Delta x^6|_{\ns}$ and $\Delta x^{10}|_{\ns}$ simultaneously, such that the relevant field theoretic quantities remain the same. Hence, there is no problem in going to M-theory.     

In the M-theory picture, the topology of spacetime will be given by ${\bf R}^{10} \times {\bf S}^1$, where ${\bf S}^1$ is the ``eleventh'' circle.  The type IIA fivebrane is just the M5-brane -- positioned at a certain point in $\bs^1$  -- which spans a six-dimensional subspace of ${\bf R}^{10}$. The type IIA fourbranes are just the M5-branes wrapped around $\bs^1$. The type IIA D2-brane is just an M2-brane. According to~\cite{Witten-sol}, the NS5-branes and D4-branes will merge into a \emph{single} M5-brane in the equivalent M-theory setup; its six-dimensional worldvolume will span the usual four spacetime dimensions with coordinates $(x^0, \dots, x^3)$,  as well as a complex curve in ${\bf Q}\cong {\bf R}^3 \times \bs^1$, where the holomorphic coordinates in $\bf Q$ are $v = x^4 + ix^5$ and $s = x^6 + i x^{10}$. The independent $\ns$-brane is now an independent ${\rm M5}'$-brane that is connected along the $x^7$-direction to the M2-brane embedded in the single M5-brane. This is illustrated in Fig.~\ref{Fig1}(b) above.   
Depending on which component of the complex curve the M5-brane wraps, it will correspond to either the NS5-branes or D4-branes in the type IIA setup: if the M5-brane wraps the component of the curve which is parameterized by \emph{constant} $s$ or $v$, it will correspond to  the NS5-branes or D4-branes, respectively. One can go further to show -- from the requirement that the degree and roots of the polynomial equation describing the complex curve in $\bf Q$, ought to be consistent with the numbers and positions of the NS5-branes and D4-branes in the type IIA setup -- that the complex curve is actually the SW curve ${\cal C}_{sw}$~\cite{Witten-sol}. 

\bigskip\noindent{\it A Family Of Low-Energy Non-Topologcial Surface Operators}

Now recall that $\sigma$ is the position of the M2-brane in the $x^4$--$x^5$ plane; thus, $\sigma$ will  correspond to some value of $v$. Since the M2-brane is embedded in the M5-brane, and since ${\cal C}_{sw}$ is parameterized by a polynomial equation involving $v$, $\sigma$ must correspond to a point in ${\cal C}_{sw}$.  Hence, via (\ref{s.o. flow}), one can see that the surface operator with IR parameters $(\ae, \ne)$ will be parameterized by a point $p \in {\cal C}_{sw}$, as claimed in~\cite{loop}. 

Referring back to (\ref{curve}) and  (\ref{s.o. flow}), it appears that for every non-vanishing point $u(a)$ on $\CM_q  \cong H^+ / \Gamma(2)$ that corresponds to some non-zero twisted masses $({\widetilde m}_1,{\widetilde m}_2)$, we have a ${\cal C}_{sw}$'s worth of effective IR parameters $(\eta_{\rm eff} + \tau_{cl} \alpha_{\rm eff})$ arising from all possible values of $p \in {\cal C}_{sw}$. Indeed, recall that for non-vanishing $r$, the D2-brane must lie along one of the D4-branes in the type IIA picture, and for a particular value of $u(a)$, one can move the pair of D4-branes freely along the $x^4$--$x^5$ plane  while keeping their separation distance fixed at $a$,  such that the position $p$ of the corresponding M2-brane takes all possible values over ${\cal C}_{sw}$. 
Therefore, over all physically-sensible values of $u(a)$, we have a family of low-energy non-topological surface operators parameterized by the total space of the ${\cal C}_{sw}$-fibration over $\CM_q$, as claimed.

\bigskip\noindent{\it The Effective Twisted Superpotential ${\cal W}_{\rm eff}$}

Notice that since $(\eta_{\rm eff} + \tau_{cl} \alpha_{\rm eff})$ can be identified with $\tau (a)$ in (\ref{period integrals}),  it can be solved in terms of solutions to the Picard-Fuchs equations given by linear combinations of hypergeometric functions; in particular, the values of the $a_n$'s in (\ref{s.o. flow}) can be determined exactly. In addition, one can also ascertain the effective twisted superpotential ${\cal W}_{\rm eff}$ of the 2d gauge theory as follows: from the relation $ t_{\rm eff} =  \partial{\cal W}_{\rm eff} / \partial \sigma$, the identification of $t_{\rm eff}$ as $\tau(a)$, and the relations $\tau = \partial_a a_d$ and that of (\ref{ad,a}), we find that we can, as claimed in~\cite{loop}, write\footnote{In the following computation, we have fixed one of the two D4-branes (that does not contain the D2-brane)  at the origin of the $x^4$--$x^5$ plane, such that one can identify the position $\sigma$ of the D2-brane (embedded in the other D4-brane) with the separation distance $a$ of the two D4-branes.}
\be
{\cal W}_{\rm eff} = \int^p_{p*} \lambda_{sw},
\ee
where $\lambda_{sw} \sim  dx \ ({\sqrt {x-p} / \sqrt{x^2 -p*}})$ is the SW-differential; $p$ represents a point in ${\cal C}_{sw}$ that parameterizes the surface operator; and $p*$ represents a reference point in ${\cal C}_{sw}$.

\bigskip\noindent{\it The Low-Energy Vacuum Structure In The Presence Of Surface Operators}

Note that the D2-brane, just like the D4-branes, will exert pulling forces on the five-branes that it is connected to. The pulling forces will act in the directions normal to the worldvolumes of the five-branes which are nevertheless parallel to the worldvolume of the D2-brane itself.  Just as the D4-branes exert a pulling force in the $x^6$-direction normal to the worldvolume of the NS5-branes -- thus resulting in a finite value for $\Delta x^6$ --  the D2-brane exerts a pulling force in the $x^7$-direction normal to the worldvolumes of the $\ns$-brane and NS5-branes -- thus resulting in a finite value for  $\Delta x^7|_\ns$. 

Notice that since the worldvolume of the D2-brane spans the $(x^0, x^1, x^7)$ directions, it will not contribute to a pulling force along the $x^6$-direction; hence, $\Delta x^6$ is unmodified by the presence of the D2-brane.  (The presence of the $\ns$-brane does not affect $\Delta x^6$ either, since its worldvolume does not span the $x^6$-direction, and it is a distance $\Delta x^7|_\ns$ away from the NS5/D4-branes configuration).    
Since the one-loop expression for $\tau(a)$ is proportional to $\Delta x^6$, we conclude that $\tau(a)$ is unmodified by the presence of a surface operator at the perturbative level. What about the nonperturbative instanton corrections to $\tau(a)$? From the identification of $t_{\rm eff}$ with $\tau(a)$, we learn that these corrections are captured by the 2d-instantons of the worldvolume theory of the  D2-brane itself. The only other thing that might influence the  worldvolume theory of the D2-brane (and hence the 2d-instantons), is the distance $\Delta x^7|_\ns$ which determines the strength of the gauge coupling in the 2d theory. However, since instanton effects are nonperturbative, that is, they are independent of the gauge coupling, the value of $\Delta x^7|_\ns$ is irrelevant in this regard.  Hence, the exact expression for $\tau(a)$ is unmodified by the presence of the D2-brane and $\ns$-brane. Therefore, we can conclude that the metric ${\rm Im} \, \tau(a)$ of $\CM_q$ is unchanged in the presence of a surface operator, consistent with our earlier field-theoretic analysis. 

One can also understand this from the viewpoint of the SW curve ${\cal C}_{sw}$; that is, ${\cal C}_{sw}$ remains unmodified in the presence of the M2-brane and ${\rm M5}'$-brane. Indeed, since the worldvolume of the M2-brane spans the $(x^0, x^1, x^7)$ directions, it will only exert a pulling force on the single M5-brane  -- whose worldvolume is a six-dimensional subspace of ${\bf R}^{3,1} \times \bf Q$ with coordinates $(x^0, \dots, x^6, x^{10})$ -- in the $x^7$-direction normal to it; in other words, the M2-brane will not exert any force along the $(x^4, x^5, x^6, x^{10})$ directions spanned by  $\bf Q$ in which the curve ${\cal C}_{sw}$ that the M5-brane wraps, lives; thus, the presence of the M2-brane will not deform the curve ${\cal C}_{sw}$. As for the ${\rm M5}'$-brane, it can only exert its influence on the M5-brane through the M2-brane along the $x^7$-direction, and this does not affect ${\cal C}_{sw}$ either. Therefore, consistent with our above conclusion about $\CM_q$, and our earlier field-theoretic analysis, the SW curve ${\cal C}_{sw}$ is unchanged in the presence of surface operators.    

\bigskip\noindent{\it About Curved Surface Operators In General $X$}

Notice that we have restricted our above discussion to completely flat surface operators with topology ${\bf R}^2$ that are trivially-embedded in flat ${\bf R}^4$. What happens if we consider a completely curved surface operator in ${\bf R}^4$? In other words, what if we consider the D2-brane to be completely curved? According to~\cite{vafa}, since the half-BPS D2-brane must preserve the original four global supersymmetries, regardless, it must be that the worldvolume theory of the curved D2-brane is a topological field theory, such that its supercharges, being scalars after the twist, continue to be globally well-defined even in curved spacetime. In fact, this is true as long as the D2-brane is at least partially-curved: in order for the global supercharges to be well-defined over \emph{all} regions in spacetime, flat or otherwise, the theory must always be twisted at the outset, even though the twist is trivial over flat space.

If the worldvolume theory of the D2-brane is topological, it will be scale-independent; in particular, there will not be a running of the F.I. parameter; consequently, the surface operator parameters $(\alpha, \eta)$ cannot flow. What this means is that an interpretation of a non-topological surface operator that is at least partially-curved, does not exist. In other words, \emph{surface operators that are at least partially-curved in ${\bf R}^4$ are necessarily topological}. Moreover, it is clear that this must be true regardless of their embeddings in ${\bf R}^4$. 

The above statement obviously holds if we move away from flat ${\bf R}^4$ to a general four-manifold $X$; in fact, if $X$ is curved, it would mean that the corresponding D4-brane is also curved, and according to the above rationale,  its  worldvolume theory would have to be topological as well.

To end this section, let us consider an example where $X$ is a ruled surface given by an ${\bf S}^2$-fibration over $\Sigma_g$, where $\Sigma_g$ is a Riemann surface of genus $g$. For $g \neq 1$, the curvature of $\Sigma_g$ is non-zero and constant. Hence, for all ruled surfaces with $g \neq 1$, where an admissible embedding $D$ is a linear sum of the ${\bf S}^2$-fiber and the $\Sigma_g$ base, as explained earlier, the corresponding surface operator is necessarily topological. For $g=1$, $\Sigma_g$ is equivalent to a flat torus. Nevertheless, the ${\bf S}^2$-fiber has constant positive curvature, and so, $D$ is at least partially-curved; according to our above discussion, the surface operator is again necessarily topological.  Moreover, for small enough  ${\bf S}^2$-fiber, the curvature of $X$ is non-zero for any $g$; consequently, the 4d theory must,  in this case,  be topological too.

\newsection{The Effective Theory Over A Generic Region In The $u$-Plane} 

As mentioned at the beginning of the previous section, we would like to express the topological observables of the twisted $SU(2)$ or $SO(3)$ gauge theory which correspond to the ``ramified'' Donaldson invariants, in terms of observables associated with the ``simpler'' abelian theory over the $u$-plane. Since the underlying $SU(2)$ or $SO(3)$ theory is fundamentally a microscopic theory defined in the UV, it is clear that the observables associated with the macroscopic abelian theory in the IR ought to be scale-invariant, at least. A little thought will then reveal that they must correspond to the correlation functions of ``descent'' operators in the \emph{topologically twisted} abelian theory over the $u$-plane;  these correlation functions are simply the low-energy counterparts of the correlation functions of the $I'_0(p)$ and $I'_2(S)$ operators that underlie the topological observables of the $SU(2)$ or $SO(3)$ theory.  

As the effective abelian theory at $u = \pm 1$ differs from that over a generic region in the $u$-plane where $u \neq \pm 1$, the ``ramified'' Donaldson invariants will be expressed as a sum of three parts; the first comes from the pure $U(1)$ theory over a generic region in the $u$-plane, the second comes from the $U(1)$ theory coupled to massless monopoles at $u=1$, and the third comes from the $U(1)$ theory coupled to massless dyons at $u=-1$. Specifically, we will have
\be
\label{Z = Z + Z}
Z'_{D} =   Z'_u + Z'_{SW}, 
\ee
where $Z'_{D}$ is the generating function of the ``ramified'' Donaldson invariants given in (\ref{Donaldson generating function - physical}),  $Z'_u$ is the corresponding low-energy counterpart of $Z'_{D}$ over a generic region in the $u$-plane, and $Z'_{SW} = Z'_{u=1} + Z'_{u=-1}$ is the SW contribution, where $Z'_{u \pm 1}$ are the low-energy counterparts of $Z'_{D}$ at $u = \pm 1$.  

In this and the next section, we will ascertain and analyze the contribution $Z'_u$ from the twisted $U(1)$ theory over a generic region in the $u$-plane. Then, in $\S$7, we will proceed to do likewise for the contributions $Z'_{u = \pm 1}$ from the twisted $U(1)$ theories coupled to massless monopoles and dyons at $u =\pm 1$ described by the ``ramified'' SW theory discussed in $\S$3.2.       

\newsubsection{A Topologically Twisted Pure $U(1)$ Theory}

We have seen that the low-energy effective theory over a generic region in the $u$-plane  is an $\CN=2$, $U(1)$ gauge theory in $\bR$ with (``electric'') action $S_{\rm eff}$. In order to twist this theory, one just needs to wick-rotate $S_{\rm eff}$ to Euclidean ${\bf R}^4$ space before applying the prescription described in $\S$2.3. In effect, the resulting topologically twisted $U(1)$ theory is just an abelian version of the topological $SU(2)$ theory defined earlier. In particular, the twisted $U(1)$ action is invariant under the nilpotent scalar supercharge $\cq$, whereby the $\cq$-transformations of the fields of the $U(1)$ multiplet are an abelian version of (\ref{susy tx - non-abelian}); they are given by\footnote{In the following, we have actually written the scalar field $\varphi$ as its vacuum expectation value $a$, so as to agree with the notation in~\cite{Moore-Witten}. Hopefully, this will not cause any confusion.}
\begin{equation}
 \label{susy tx-abelian}
\begin{matrix}
[\cq, A' ] =  \psi,
\quad & \quad
\{\cq, \psi \}   =  4 \sqrt{2} da, \\  \\
[\cq, a]   = 0,
\quad & \quad
[\cq,\bar a] =    \sqrt{2} i \zeta, \\  \\
\{\cq, \zeta\} = 0,
\quad & \qquad
 \{\cq, \chi_+\}   = i( F'_+  - K_+),  \\ \\
 \hspace{2.0cm} [\cq, K_+]   =   (d_{A'} \psi)_+,   \\
\end{matrix}
\end{equation}
where $A'$ is now understood as the $U(1)$ gauge field with field strength $F' = F'_L = F_L - 2 \pi \ae \delta_D$. The corresponding $G$-transformations of the fields are then 
\begin{equation}
 \label{G-action abelian}
\begin{matrix}
[G , a]   = {1 \over 4 \sqrt{2}} \psi,
\quad & \quad
[G, \bar a]   = 0, \cr \\
\{G , \psi \} = - 2(F'_-  + K_+ ),
 \quad & \quad
[G,  A']   = i \zeta - 2i \chi, \cr \\
\{G, \zeta\}   = -{i \sqrt{2} \over 2} d \bar a,
\quad & \quad
\{G, \chi_+ \}   = -{3i \sqrt{2} \over 4} * d \bar a, \cr \\
[G, K_+]  = -{3i \over 4} * d \zeta
+ {3 i \over 2} d \chi_+. \cr 
\end{matrix}
\end{equation}
Consequently, the Lagrange density of the $\cq$-invariant twisted $U(1)$ action will be given by\footnote{Note that in order to agree with the conventions in~\cite{Moore-Witten}, we have rescaled the physical Lagrange density by an overall factor of $1/4$ in writing $\mathscr L$. In addition, the sign of the topological $\ne$-term has also been switched; this just results in some sign changes in the transformations of $(\ae, \ne)$ under the $SL(2,\IZ)$ equivalence group, and does not alter the underlying physics at all. We will elaborate on these transformations at an appropriate point at the end of $\S$5.} 
\begin{eqnarray}
\label{Lagrangian twisted U(1)}
{\mathscr L} =  { i \over  16 \pi}   \bigl( \bar \tau F'_+ \wedge F'_+
+   \tau F'_- \wedge F'_-\bigr)
 +  {1 \over  8 \pi}   ({\rm Im}  \tau) da \wedge * d\bar a - {1 \over  8 \pi}
({\rm Im}  \tau) K \wedge *K \cr - {1 \over  16 \pi}
\tau  \psi \wedge * d \zeta
+ {1 \over  16 \pi} \bar \tau \zeta \wedge d * \psi
+ {1 \over  8 \pi}  \tau \psi
\wedge d \chi_+  - {1 \over  8 \pi} \bar \tau \chi_+ \wedge d \psi 
\nonumber \\
+ {i \sqrt{2}   \over  16 \pi } {d \bar \tau \over  d \bar a} \zeta \chi_+\wedge
(K_+ + F'_+ ) - {i \sqrt{2}   \over  2^7 \pi } {d \tau \over  da}
(\psi\wedge \psi) \wedge  ( F'_-  +  K_+)   +  {i \over 4} \ne \, \delta_D \wedge F' \\
 + {i \over  3 \cdot 2^{11} \pi  }      {d^2 \tau \over  da^2} \psi\wedge \psi
\wedge \psi\wedge \psi  - {\sqrt{2} i \over  3\cdot 2^5 \pi   }   \{ \cq , {d \bar \tau \over
d \bar a}\chi^+_{\mu\nu} \chi^{\nu\lambda}_+ \chi_\lambda^{+ \mu} \} \sqrt{\bar g}  d^4x. \nonumber
\end{eqnarray}
In addition, there are extra terms that take into account the coupling to gravity, and they are of the form  
\be
\label{coupling to gravity terms}
-b(u) \Tr R\wedge \widetilde R
 - c(u) \Tr R \wedge R  + {i \over  4} F' \wedge w_2(X),
\ee
where $b(u)$ and $c(u)$ are holomorphic functions of $u$; and $\Tr R\wedge \widetilde R$ and $\Tr R\wedge R$ are the densities of the Euler characteristic and signature of $X$, respectively. These terms have been determined explicitly in~\cite{mine}, and they are further discussed in $\S$5. 

\newsubsection{Observables And Contact Terms}

From our discussion in $\S$3.3 and (\ref{G-action abelian}), we have the following map from high to low energy observables:
\begin{eqnarray}
\label{map hi to low}
I_0 & \rightarrow & 2 u,   \nonumber \\
I_2(\Sigma) & \rightarrow & {\widetilde I}_2(\Sigma)  = {i \over  \pi\sqrt{2}}  \int_{\Sigma} G^2 u = {i \over  \pi\sqrt{2}} \int_{\Sigma}
\left[{1 \over 32} {d^2 u \over  d a^2} \psi  \wedge \psi  - {\sqrt{2}\over  4} {d u \over  d a} (F_-   +   K_+)  \right] + {i  \over 2} \int_D \ae {du \over da} \delta^-_{\Sigma}, \nonumber \\
\end{eqnarray}
where $\Sigma$ is an \emph{arbitrary} element of $H_2(X)$; and $\delta^-_{\Sigma}$ is the anti-self-dual component of the delta two-form operator that is Poincar\'e dual to $\Sigma$. The normalization constants in these formulas have been
fixed to match our results with those in the ordinary case without surface operators in the limit $(\ae, \ne) \to (0,0)$.\footnote{The reason that we have a factor of $2$ in the map $I_0 \rightarrow 2u$, is as follows. Firstly, note that a ``ramified'' generalization of the simple-type condition for $X$ would read $[{\partial^2 \over \partial p^2} - 4] Z'_D =0$; this is tantamount to an insertion of the operator $I_0^2 - 4$ in the correlator. Secondly, recall that the only time where one has $(u^2-1) =0$, is when massless monopoles and dyons appear. Thirdly, to reconcile the condition $(u^2 -1) =0$ with a simple-type condition $I^2_0 - 4 =0$ (as was shown to be useful in the ordinary case in~\cite{Moore-Witten}), we require a factor of 2.} Also, as we will henceforth be dealing with curved surface operators in a general $X$, the surface operators will necessarily be topological, as explained at the end of the last section; thus, we can set $(\ae, \ne) = (\alpha, \eta)$. 

\def\tD{{\widetilde D}}

According to  (\ref{map hi to low}), and the discussion in $\S$3.3, we then have the high-to-low energy map 
\begin{eqnarray}
{\rm exp} \hspace{0.0cm} [I_2(\Sigma)] & \rightarrow &  {\rm exp} \left[   {\widetilde {\cal I}}_2(\Sigma) + {\widetilde {\cal I}}_2(\tD) + \Sigma^2 T_{\Sigma} (u) + {\tD}^2T_{\tD}(u) + (\Sigma \cdot \tD) T_{\Sigma, \tD}(u)   \right],
\end{eqnarray}
 where $\tD =   i  \pi \alpha  D /2$,  $\Sigma \cdot \tD = \Sigma \cap \tD$, and
\be
\label{I-sigma}
{\widetilde {\cal I}}_2(\Sigma)  =  {i \over  \pi\sqrt{2}} \int_{\Sigma}\left[{1 \over 32} {d^2 u \over  d a^2} \psi  \wedge \psi  - {\sqrt{2}\over  4} {d u \over  d a} (F_-   +   K_+)  \right] ,
\ee
while
\be
\label{I-D}
{\widetilde {\cal I}}_2(\tD)  =  {  1 \over \pi} \int_{\tD}  {du \over da} \delta^-_{\Sigma} = {1 \over \pi} {du \over da} (\Sigma \cdot \tD)_-.
\ee
Here, the ``-'' subscript indicates that the intersection form in question -- namely $\Sigma \cdot \tD = \int_X \delta_\Sigma \wedge \delta_\tD$ -- involves only the anti-self-dual components of the delta two-forms $\delta_\Sigma$ and $\delta_\tD$.  

Nevertheless, recall that in the topological correlation functions of the microscopic theory that correspond to the ``ramified'' Donaldson invariants, $\Sigma = S$, where $S \in H_2(X \backslash D)$. In particular, this implies that $\Sigma \cdot \tD = 0$; therefore, the  map of interest is given by
\begin{eqnarray}
\label{exp map hi to low}
{\rm exp} [I_2(S)] & \rightarrow &  {\rm exp} \left[   {\widetilde {\cal I}}_2(S)  + S^2 T_{S} (u) + \tD^2T_\tD(u)  \right].
\end{eqnarray}
Notice that as required, the definition of ${\widetilde {\cal I}}_2(S)$ and therefore the low-energy observable on the RHS of (\ref{exp map hi to low}), does not depend on a choice of  extension of the $U(1)$ gauge bundle  over $D$; this is consistent with the definition of the high-energy observable $I_2(S)$ which appears on the LHS of (\ref{exp map hi to low}).

In order to ascertain the holomorphic functions $T_{S} (u)$ and $T_{\tD}(u)$, one can adopt the approach in~\cite{Moore-Witten},  and first express the low-energy observable on the RHS of (\ref{exp map hi to low}) purely in terms of physical fields. From (\ref{I-sigma}), it is clear that one can do this by integrating out the auxiliary field $K_+$. From $\mathscr L$ in (\ref{Lagrangian twisted U(1)}), we have the propagator $\langle K(x) K(y)\rangle \sim \delta^4(x-y) / {\rm Im} \, \tau$; utilizing this propagator to evaluate all pairwise ``interactions'' between the $K_+$ fields at different points $x$ and $y$ in spacetime, we arrive at the low-energy observable expressed purely in terms of physical fields:
\be
\label{explicit low observable}
{\rm exp} \left(-{i \over  4 \pi} \int_S
\left({du\over da}F_-\right) +S_+^2{(du/da)^2\over
8\pi {\rm Im} \, \tau} +  S^2 T_{S} (u) + \tD^2T_\tD(u)   + {\rm fermion \ terms}\right),
\ee
where $S_+^2 = \int_X \delta^+_S \wedge \delta^+_S$.  We will not need to concern ourselves with the fermion terms for now. 

Notice that even though (\ref{explicit low observable}) is $\cq$-invariant at the outset, due to the factor of $1 / {\rm Im} \, \tau$ in its $S_+^2$-term, it is not manifestly modular-invariant. Nevertheless, (ignoring the first term in (\ref{explicit low observable}) for now), if  $T_S(u)$ and $T_{\tD}(u)$ are physically consistent holomorphic functions, their inclusion should ensure that the net modular transformation of (\ref{explicit low observable}), is trivial. Since one can write $S^2 = S^2_+ + S^2_-$, it will mean that  $T_S(u)$ must transform under modular transformations in such a way as to keep the coefficient of the \emph{total} $S_+^2$-term
\be
F(u) = {(du/da)^2\over 8\pi {\rm Im} \, \tau} + T_S(u),
\ee
invariant.  

Now recall that an arbitrary $SL(2,\IZ)$ transformation is generated by the transformations ${\mathcal T} : \tau \to \tau +1$ and ${S}: \tau \to -1 / \tau$. Under $\mathcal T$, ${\rm Im} \, \tau \to {\rm Im} \, \tau$, $a \to a$ and $a_d \to  a_d + a$. Consequently, $F(u)$ is invariant under $\mathcal T$. However, under $S$, 
\be
{\rm Im} \, \tau \to {{\rm Im} \, \tau \over \bar \tau \tau}
\ee
and
\be
{du\over da}\to {du\over da_d}={du/da\over {da_d/da}}={1\over \tau}
{du\over da}.
\ee 
Thus, we find that under ${S}$, 
\be
F(u)\to F(u) -{i\over 4\pi \tau}\left({du\over da}\right)^2.
\ee
Therefore, we conclude that $T_S(u)$ must transform as 
\be
T_S(u)\to T_S(u) + {i\over 4\pi \tau}\left({du\over da}\right)^2, 
\label{shift in Ts(u)}
\ee
 under ${S}: \tau \to -1 / \tau$, so as to cancel the shift in $F(u)$. 

Another property that $T_S(u)$ should have is that it must be odd under $u \to -u$. This can be understood as follows. Recall that the microscopic $SU(2)$ theory has a $U(1)_R$ symmetry that is broken by ``ramified'' instantons to a ${\bf Z}_8$ symmetry. Recall also that this ${\bf Z}_8$ symmetry maps $u \to -u$, while on an operator of $R$-charge 2 such as ${\widetilde {\cal I}}_2(S)$, 
it will map ${\widetilde {\cal I}}_2(S) \to i {\widetilde {\cal I}}_2(S)$ and hence, $T_S(u) \to -T_S(u)$. Therefore, $T_S(-u) = - T_S(u)$, as claimed. Likewise, $T_D(u)$ must also have this property. 

As explained in $\S$3.3, since $T_S(u)$ arises due to quantum interactions, it must depend on the variable $u  /  \Lambda^2$ (of $R$-charge 0) which controls the strength of the effective gauge coupling. Moreover, $T_S(u)$ must also vanish in the semiclassical region where $u \to \infty$. Hence, since $T_S(u)$ must have $R$-charge 4, it will depend on $u$ as $T_S(u) = u f(\Lambda^2 / u)$, where $f(0) =0$, so that $T_S(u)$ will vanish in the semiclassical limit $\Lambda^2 / u \to 0$.  Again, the same must likewise be true of $T_\tD(u)$. 
  
All of the above-stated properties of $T_S(u)$ will be readily satisfied if (after coveniently setting $\Lambda$ to 1) 
\be
\label{TSu}
T_S(u) = -{1\over 24} \left(  E_2(\tau) \left({du \over da} \right)^2 - 8u  \right),
\ee   
where $E_2(\tau)$ is the Eisenstein series of weight 2 (whose transformation under $S$ is detailed in Appendix A).

On the other hand, $T_\tD(u)$, unlike $T_S(u)$,  is not required to satisfy a shift such as (\ref{shift in Ts(u)}) under $S$. However, as explained above, it must also take the form $T_\tD(u) = u f_\tD(\Lambda^2 / u)$, where $T_\tD(u) / u \to 0$ when $u \to \infty$, and $T_\tD(-u) = -T_\tD(u)$.  This implies that $f_\tD(x)$ can only be a polynomial of even-powers of $x$ that does not contain any constant terms. Since there are no other conditions imposed on $T_\tD(u)$, let us therefore consider the simplest possibility (after conveniently setting $\Lambda =1$) that 
\be
\label{TDu}
T_\tD(u) = {1 \over 4u},
\ee 
where the factor of $1/4$ is chosen to facilitate our later computations.   

Last but not least, note that $T_S$ and $T_\tD$ have no singularities near the special points in the $u$-plane; if one
works in the preferred local coordinate near $u=\pm 1$ (for instance,
$a_d$ at the monopole point, and $a - a_d$ at the dyon point) then $T_S$ and $T_\tD$ have
no singularities at $u=\pm 1$. For example, near the monopole point, $u$ will be replaced by $u_M$ (whose explicit expression in terms of $\tau_d  = \tau_M = - 1/ \tau$ is given in Appendix A), and the corresponding functions $T^M_S$ and $T^M_\tD$ can be seen to be well-behaved. This is because integrating out massless monopoles or dyons does not produce a singularity in $T_S$ and $T_\tD$: these functions arise due to local quantum interactions of the inserted operators and they do not depend on the part of the measure affected by integrating out massless modes. 

\newsubsection{Vanishing Of The Generic $u$-Plane Contribution For $b_2^+ > 1$}

Note that the generic $u$-plane contribution $Z'_u$ associated with the Lagrange density $\mathscr L$ vanishes  when $b_2^+(X) > 1$. The arguments leading to this claim are identical to those in~\cite{Moore-Witten} for the ordinary case without surface operators -- one just replaces the ordinary field strength in the analysis detailed in~\cite{Moore-Witten} with the ``ramified'' field strength $F'$.  As such, we shall, for brevity, not repeat everything here. Nevertheless, let us summarize the pertinent points required for a coherent understanding of the arguments behind this important claim. 

Firstly, recall that over a generic region in the $u$-plane, the pure $U(1)$ theory can be described in different equivalent frames related to one another via an $SL(2,\IZ)$ transformation. In particular, unlike near the special points $u=1, -1$ and $\infty$, there is no preferred frame over the rest of the $u$-plane; thus, a frame in which the theory appears strongly-coupled, is as equivalent to an $S$-dual frame in which the theory appears weakly-coupled. A consequence of this freedom is that any computational technique which requires a specialization of the gauge couplings to be either weakly or strongly-coupled, cannot be used to evaluate the $u$-plane contribution in a physically consistent manner. Hence, the topological field-theoretic method in $\S$2.5 and $\S$3.2 -- whereby one reduces to supersymmetric configurations in the weak-coupling limit  as a means to ascertain the contributions in question -- cannot be applied here.   Nevertheless, $\mathscr L$ is still a topological-twisted  Lagrange density; this means that the corresponding theory ought to be metric-independent. Consequently, one can define the metric on $X$ to be $t^2 {\bar g}_0$ for fixed $\bar g_0$,  look at the behavior of the theory as $t \to \infty$, and see which contributions do survive in this large volume limit.  

Secondly, note that because of supersymmetry, the one-loop determinants of the various fields cancel.  Almost all Feynman diagram contributions vanish because -- the theory on the $u$-plane being unrenormalizable and without marginal or relevant couplings in the renormalization group sense -- the vertices only scale as negative powers of $t$.  In determining which contributions  do actually survive, we will find that the generic $u$-plane contribution vanishes when $b^+_2 > 1$. 

Thirdly, note that since $\zeta$ is a spacetime scalar, it will have a single zero-mode with constant wave-function 1.  On the other hand, since $\psi$, $\chi_+$ and $F'$ are a spacetime one-form, self-dual two-form and two-form,  respectively, their corresponding zero-modes will correspond to harmonic one-forms, self-dual two-forms and two-forms in $X$ which are thus counted by the Betti numbers $b_1$, $b^+_2$ and $b_2$. Moreover, the zero-modes of $\zeta$, $\psi$, $\chi$ and $F'$, will have geometrical scaling dimensions  0, 1, 2 and 2, accordingly.  

Since we will be interested in $X$ with $b_1=0$ and $b_2^+ \geq 1$, let us start by explaining the case where $b_1 = 0$ and $b^+_2 =1$. In such examples, there are no $\psi$ zero-modes and one $\chi_+$ zero-mode. The single, ever-present $\zeta$ zero-mode and the single $\chi_+$ zero-mode can be soaked up using the interaction vertices $\int_X (d\bar \tau / d \bar a)\zeta \chi_+ \wedge K_+$ or $\int_X(d\bar \tau / d \bar a) \zeta\chi_+ \wedge F'_+$ from the low energy effective action (cf.~(\ref{Lagrangian twisted U(1)})). In doing this, one sets $d\bar \tau / d \bar a$ to its vacuum expectation value at the given point on the $u$-plane, and $\zeta$ to a constant; also, since in these vertices, $\zeta$ and $\chi_+$  are necessarily expressed in terms of their zero-modes, so must $F'_+$; this is because $\int_X \zeta\chi_+ \wedge {\widetilde F}'_+ =0$ if ${\widetilde F'}_+$ is a quantum fluctuation  of $F'_+$ that is hence orthogonal to the zero-modes. On the other hand, there are no zero-modes for the auxiliary field $K_+$. Nevertheless, one can integrate it out as we did earlier, and the vertex will be replaced by $\int_{\Sigma} (d\bar \tau / d \bar a)^2 \zeta \chi_+$. Since the interaction vertices $\int_{\Sigma} (d\bar \tau / d \bar a)^2 \zeta \chi_+$ and  $\int_X(d\bar \tau / d \bar a) \zeta\chi_+ \wedge F'_+$ are expressed purely in terms of zero-modes, from the geometrical scaling dimensions of their integrands, we find that  their $t$-dependences will be given by $t^{2 -0 -2} = t^0$ and $t^{4- 0 - 2 - 2} =t^0$, respectively; thus, they will survive in the $t \to \infty$ limit. By a careful but straightforward analysis, it can be shown that these are the only contributions that survive (cf.~\cite{Moore-Witten}). The idea is to make a judicious assignment of the scaling dimensions of the \emph{quantum} fields such that their kinetic energies are all scale-invariant, as required, and thereby considering all possible interaction vertices whose effective $t$-dependences are of the form $t^n$, where $n \geq 0$; this is done keeping in mind that there is no propagator $\langle \psi \psi \rangle$,  and therefore, all factors of $\psi$ must be set equal to their zero-modes, which, in this case,  there are none. In short, only tree diagrams contribute to $Z'_u$ when $b^+_2 =1$.

What if $b_1 =0$ and $b^+_2  > 1$? A relevant example would be the case of $b_1 =0$ and $b^+_2 =3$. Here, there is a single $\zeta$ zero-mode, as usual, no $\psi$ zero-modes, and three $\chi_+$ zero-modes.  Notice that from the $\cq$-exact term in $\mathscr L$, we have the terms $\zeta\chi_+^3$ and $\chi_+^2(F'_+ - K_+)$. The only way that one can try to soak up all the fermion zero-modes exactly is to use the $\zeta\chi_+^3$ term or the $\chi_+^2(F'_+ - K_+)$ term (along with the $\zeta\chi_+(K_+ + F'_+)$ term) in $\mathscr L$.  However, these additional terms scale as $t^{-2}$, and therefore vanish as $t \to \infty$.  Clearly, we will have such a situation whenever $b_2^+>1$: there is no way to absorb the $\chi_+$ zero-modes with vertices that have non-negative powers of $t$.  In short,  $Z'_u$ vanishes whenever $b^+_2 > 1$ and $b_1 =0$.

\newsection{The Explicit Expression For The ``Ramified'' $u$-Plane Integral} 

We would now like  to compute the exact expression for the ``ramified'' $u$-plane integral $Z'_u$, so that we can ascertain the generic contribution to the ``ramified'' Donaldson invariants when $b_1=0$ and  $b^+_2 =1$. In particular,  we would like to compute the $u$-plane counterpart of the ``ramified'' Donaldson generating function
\be
{\bf Z}'_{\xi, \bar g}(p,S) =  \langle e^{p I'_0 +  I'_2(S)} \rangle_{\xi},
\label{Donaldson generating function - physical - repeat}
\ee
with $SO(3)$ gauge bundle $E$, and fixed $\xi = w_2(E)$. (We have omitted the dependence on the ``ramified'' instanton number $k'$ in (\ref{Donaldson generating function - physical - repeat}), as this has already been summed over.) As noted earlier, the $SU(2)$ case is just a specialization to $\xi =0$. 

\newsubsection{Form Of The Integral For  $b_1=0$, $b_2^+=1$}

To ascertain the explicit form of the $u$-plane counterpart of (\ref{Donaldson generating function - physical - repeat}),  a number of factors need to be considered. They are:

\vspace{0.2cm}

{(i)} Interactions with gravity that vanish on flat ${\bf R}^4$ but are present in the twisted 

theory  on a curved $X$. 

\vspace{0.1cm}
{(ii)} The integration measure for the zero-modes.

\vspace{0.1cm}
{(iii)} The relevant low-energy observables.

\vspace{0.1cm}
{(iv)} The path-integral of the photons.

\vspace{0.1cm}
{(v)} The absorption of fermion zero-modes and elimination of the auxiliary
field $K_+$.

\vspace{0.2cm}

\hspace{-1.0cm}We consider these factors individually and then put them together.

\bigskip\noindent{\it Interactions With Gravity}

On the $u$-plane, there are couplings to gravity that do not
appear on flat ${\bf R}^4$ but do appear if one works on curved $X$.
For the topologically twisted pure $U(1)$ theory, these couplings were analyzed and determined in $\S$3 of~\cite{mine}, and they multiply
the measure by a factor
\be
\label{factor}
e^{b \chi + c \sigma} \sim A^{\chi} B^{\sigma}
\left((u^2-1){d\tau\over du}\right)^{\chi/4}\left(u^2-1\right)^{\sigma/8}
\ee
where $\chi $ and $\sigma$ are the Euler characteristic and the signature of
$X$, and $A$ and $B$ are universal constants (independent of $X$)
that were not determined in~\cite{mine}. These constants will be fixed later to match the definitions in the mathematical literature.  

For our results in the limit $(\alpha, \eta) \to (0,0)$ to agree with that in the ordinary case without surface operators, we must also multiply the above factor by 2.  Altogether, since $b_1=0$ and $b_2^+=1$ implies that $\chi+\sigma=4$,  the additional factor in the path-integral due to interactions with gravity will be given by
\be
\label{coupling to gravity}
F(\chi, \sigma) = 2 A^\chi B^\sigma (u^2-1){d \tau \over  du}
\left(  {({du \over  d \tau})^2 \over  u^2-1} \right)^{\sigma/8}.
\ee

\bigskip\noindent{\it Zero-Mode Integration Measure}

With $b_1=0$, there are no zero-modes for the gauge field $A'$. Hence, the only bosonic zero-modes come from the scalar field $a$, and they correspond to a particular point on the $u$-plane.  As seen earlier, the metric  on the $u$-plane is ${\rm Im} \, \tau \, |da|^2$;
so the zero-mode measure for $a$ is 
\be
{\rm Im} \, \tau \, da\,d\bar a.
\ee

As usual, there is a single $\zeta$ zero-mode, with constant wave-function 1.
Let us write $\zeta=\zeta_0+\zeta'$, where $\zeta_0$ is a constant anticommuting
$c$-number and $\zeta'$ is orthogonal to the constants.  For $b_2^+=1$, there is a single
$\chi_+$ zero-mode, corresponding to a normalized harmonic self-dual two-form
$\omega_+$ where $\int_X\omega_+ \wedge \omega_+ =1$.
Notice that the normalization defines $\omega_+$ up to a sign, the significance of which will be clear shortly. 
Also, let us write $\chi_+=\chi_0\omega_+ +\chi'_+$, where $\chi_0$ is an anticommuting constant
and $\chi'_+$ is orthogonal to $\omega_+$. 

Specifically in our theory, for any fermi field $\varrho$ with anticommuting constants $\varrho^i_0$ in its zero-modes, the fermionic integration measure for each zero-mode can be written as
\be
d(\sqrt {\textrm{Im} \,\tau}\ \varrho^i_0) = {d \varrho^i_0 \over \sqrt {\textrm{Im} \,\tau}},
\ee
that is, for every fermion zero-mode, there is a factor of $({\rm Im}\,\tau)^{-1/2}$ in the fermionic integration measure. Thus, the integration measure for the fermion zero-modes of $\zeta$ and $\chi_+$ is just
\be
{ {d\zeta_0\, d\chi_0\over {\rm Im}\,\tau}.}
\ee
Notice that a sign change in $\omega_+$ implies a sign change in $\chi_0$, which in turn effects a sign change in the above fermionic integration measure. This corresponds to the fact that defining the sign of the ``ramified'' Donaldson
invariants requires a choice of an orientation of $H^{2,+}(X)\oplus H^1(X)$~\cite{structure}.  For $b_1=0$ and $b_2^+=1$, a choice of an
orientation of $H^{2,+}(X)\oplus H^1(X)$ is a choice of $\omega_+$.  Our formulas will thus depend on a choice of
$\omega_+$, and will be odd under reversal of sign of $\omega_+$.

Combining the above, the zero-mode measure is simply
\be
\label{zero-modes measure}
{ \,da\,d\bar a \,d\zeta_0\,d\chi_0}
\ee
with no factors of ${\rm Im}\,\tau$.

\bigskip\noindent{\it Observables of the Low-Energy Theory}

The low-energy counterpart of ${\rm exp} (p I_0 + I_2(S))$ is, according to (\ref{map hi to low}) and (\ref{exp map hi to low}), given by   $ {\rm exp} [ 2 p u + {\widetilde {\cal I}}_2(S)  + S^2 T_{S} (u) + \tD^2T_\tD(u)]$.    For $b_1=0$ and $b_2^+=1$, the $\psi \wedge \psi$ term
in ${\widetilde {\cal I}}_2(S)$ can be dropped: since there is no $\langle \psi \psi \rangle$ propagator, $\psi$ must be replaced by its zero-modes -- which are counted by $b_1$ -- within correlation functions, and in this case, there are none.  The net result is that we have the following high-to-low energy map of observables:
\begin{eqnarray}
\label{exp map hi to low total}
{\rm exp} [pI_0 + I_2(S)] & \rightarrow &  {\rm exp} \left[  2pu  -{i \over  4 \pi} \int_{S} {d u \over  d a} (F_-   +   K_+)
 + S^2 T_{S} (u) + \tD^2T_\tD(u)  \right], 
\end{eqnarray}
where $T_S(u)$ and $T_\tD(u)$ are given in (\ref{TSu}) and (\ref{TDu}), respectively.  

\bigskip\noindent{\it Photon Path-Integral  }

The ``ramified'' $u$-plane integral contains an important factor given by the partition function of the
free photons. This factor is responsible for the expected modular invariance of the pure $U(1)$ theory over the $u$-plane.  Let us determine its explicit form now.   

For $b_1 =0$, there are no zero-modes of $A'$. Also, in computing the photon partition function, one must sum over all $U(1)$-bundles $L$ and include a prefactor of $({{\rm Im} \, \tau})^{-1/2}$~\cite{mine first}. For  $2 \lambda = -{c_1(L)}$, one can write $F  = 4 \pi \lambda$; for a lattice $\Gamma = H^2(X, \IZ)$, $\lambda$ is a half-integral class that is congruent to ${1\over 2}w_2(E)$ modulo $\Gamma$.\footnote{This is because for non-vanishing $w_2(E)$, we have $c_1(L) \in - w_2(E) + 2H^2(X, \IZ)$; this means that $\lambda \in {1 \over 2}w_2(E) + H^2(X, \IZ)$.} Thus, from (\ref{Lagrangian twisted U(1)}), we see that the elementary photon partition function can be written as
\be
{1 \over \sqrt {{\rm Im} \, \tau}} \sum_{\lambda \in {1 \over 2}w_2(E) + \Gamma} {\rm exp} \left[  -{i \pi \bar \tau} \left(\lambda_+ -  {\alpha \over 2} \, \delta^+_D \right)^2  - {i \pi \tau} \left(\lambda_- - {\alpha \over 2} \, \delta^-_D\right)^2 \right].
\ee
Here, $\lambda_+$ and $\lambda_-$ are the self-dual and anti-self-dual projections of $\lambda$ that obey $(\lambda_+)^2 > 0$ and $(\lambda_-)^2 < 0$, where for any two-forms  $\Omega$ and $\theta$ in $X$, $(\Omega)^2 = \int_X \Omega \wedge \Omega$ and $\Omega \cdot \theta = (\Omega, \theta) = \int_X \Omega \wedge \theta$.  Note that
the self-dual projection is explicitly $\lambda_+=\omega_+(\omega_+, \lambda)$,
with $\omega_+$ the normalized self-dual harmonic two-form introduced earlier.

There is also an important phase factor due to the third term in (\ref{coupling to gravity terms}), which arises when one integrates out the massive fermions that are the $SU(2)$ or $SO(3)$ partners of $\zeta$, $\psi$ and $\chi_+$. This phase factor is also responsible for the appearance of a ${\rm Spin}^c$-structure in the SW theory at $u = \pm 1$, as we shall see in the next section.  It may be described as follows. Pick an arbitrary but fixed class $\lambda_0 \in {1 \over 2} w_2(E) + \Gamma$. By including a sign factor of $(-1)^{(2 \lambda^2_0 + \lambda_0 \cdot w_2(X))}$ (where $\lambda^2_0 = (\lambda_0)^2$) as was done in~\cite{Moore-Witten}, so that our conventions in the limit $(\alpha, \eta) \to (0,0)$ agree with those in the mathematical literature for the ordinary case,  we can write this phase factor as
\be
(-1)^{(\lambda - {\alpha \over 2} \delta_D   - \lambda_0) \cdot w_2(X)} e^{2\pi i \lambda^2_0},
\label{phase factor}
\ee
where $w_2(X)$ is integer-lifted modulo 2.  

There is no preferred choice of $\lambda_0$ (unless $w_2(E)=0$, which then means that we can take $\lambda_0=0$).  If $\lambda_0$
is replaced by $\widetilde\lambda_0$, then (\ref{phase factor}) is multiplied
by
\be
(-1)^{\beta\cdot w_2(X)},
\label{o factor}
\ee
where $\beta$ is the integral class $\beta=\lambda_0-\tilde\lambda_0$.
Thus, with the factor  (\ref{phase factor}) included, the overall sign of the ``ramified'' Donaldson
invariants depends on a choice of $\lambda_0$. This fact is consistent with the results in~\cite{structure}:   if the integral lift $2\lambda_0$ of $w_2(E)$ is changed, the orientation of the moduli space of ``ramified'' instantons is multiplied
by the factor (\ref{o factor}), as argued in the appendix of~\cite{structure}. 
Thus, when the factor (\ref{phase factor}) is included, the ``ramified'' $u$-plane integral depends on the same choices,
and transforms in the same way when the choices are changed, as the orientation
of the moduli space of ``ramified'' instantons. Last but not least, notice that when $\lambda$ is changed by an element of $\Gamma$, the phase factor (\ref{phase factor}) changes by a factor of $\pm 1$. Thus, in its dependence on $\lambda$, the phase factor behaves as a sign factor. 

Putting all the factors together,  the total photon partition function on the $u$-plane can be written as
\be
\label{Z photon}
Z'_{\rm photon} = {e^{2 \pi i \lambda_0^2} \over \sqrt {{\rm Im} \, \tau}} \sum_{\lambda \in {1 \over 2}w_2(E) + \Gamma} \hspace{-0.2cm} (-1)^{(\lambda - {\alpha \over 2} \delta_D   - \lambda_0) \cdot w_2(X)} \, {\rm exp} \left[  -{i \pi \bar \tau} \left(\lambda_+ -  {\alpha \over 2} \, \delta^+_D \right)^2  - {i \pi \tau} \left(\lambda_- - {\alpha \over 2} \, \delta^-_D\right)^2 \right].
\ee

\bigskip\noindent{\it Absorption Of Fermion Zero-Modes And Elimination Of Auxiliary Field $K_+$}

For $b_1=0$ and $b_2^+=1$, there are no zero-modes for $\psi$ and thus, according to our earlier explanation, the interaction vertices associated with the $\psi$-terms in $\mathscr L$, do not contribute to the path-integral. On the other hand, there is a single zero-mode from $\zeta$, and a single zero-mode from $\chi_+$. We have already determined in $\S$4.3 which interaction vertex should be used
to absorb these two fermion zero-modes; this interaction vertex
corresponds to a factor in the path integrand that reads
\be
\label{vertex}
\exp\left[ - {i \sqrt{2} \over 16\pi}\int_X
{d\bar \tau\over d\bar a}\zeta \chi_+\wedge (F_+ - 2 \pi \alpha \delta^+_D + K_+)
\right].
\ee

Let us now express everything in terms of the physical fields by  integrating out the auxiliary field $K_+$. The only other $K_+$-dependent factor in the path-integral, other than (\ref{vertex}), comes from the inserted observable (\ref{exp map hi to low total}). Altogether, since only the first-order term in the expansion of (\ref{vertex}) will contribute non-vanishingly to the path-integral,  the total $K_+$-dependence of the path-integral is in a factor
\be
\label{K-terms}
{ - \exp\left(-{i\over  4 \pi} {du\over
da}\int_S K_+\right)\cdot\left({i \sqrt{2} \over 16\pi} \int_X
{d\bar \tau\over d\bar a}\zeta\chi_+ \wedge (F_+ - 2 \pi \alpha \delta^+_D  + K_+)\right).}
\ee
To integrate $K_+$ out, recall that $K_+$ is a Gaussian field with propagator $\langle K_+(x) K_+(y) \rangle \sim \delta^4(x-y) / {\rm Im} \, \tau$. Upon integrating out $K_+$, (\ref{K-terms}) becomes 
\be
\label{K-terms integrated}
- \exp\left(S_+^2\left({ (du/da)^2\over 8 \pi  {\rm
Im}\,\tau}\right)\right)
\cdot \left({  \sqrt{2} \over 16\pi} \int_X {d\bar\tau\over d\bar a}
\zeta \chi_+ \wedge (F_+ - 2 \pi \alpha \delta^+_D + i { (du/ da)\over {\rm Im}\tau}
                       \, \delta_S^+)\right).
\ee

To simplify this further, notice that since $\zeta$, $\chi_+$ and $F_+$ must be expressed purely in terms of their zero-modes in (\ref{K-terms integrated}), one can replace $\zeta$ by 1, $\chi_+$ by $\omega_+$, and $F_+$ by the fixed cohomology class $4 \pi \lambda_+$. Hence, the resulting factor will be given by
\be
\label{5.16}
-{\sqrt{2} \over  4}
{d\bar \tau\over  d\bar a}
 \cdot
\exp\left((S^2 - S_-^2)\left({  (du/da)^2\over 8 \pi y} \right)\right)\cdot
\left(  (\omega_+, \lambda - {\alpha \over 2} \delta_D) + {i \over  4 \pi y}  {du\over  da} (\omega_+,S)
 \right),
 \ee
where $\tau = x + i y$.  Here, $(\omega_+,S) = \int_X \omega_+ \wedge \delta_S = \int_S \, \omega_+$. 
  
Let us now define the ``ramified'' lattice sum factor:
\begin{eqnarray}
\label{Psi'}
\Psi' & =  &  \exp\left[  - { 1 \over  8 \pi y}({d   u \over  d  a})^2 S_-^2 \right]
 e^{2\pi i \lambda_0^2} \sum_{\lambda\in {1\over 2} w_2(E) +  \Gamma}  (-1)^{(\lambda - {\alpha \over 2} \delta_D   - \lambda_0) \cdot w_2(X)}  \left[ (\lambda - {\alpha \over 2} \delta_D, \omega_+) +  {i \over  4 \pi y} {d u\over  d a} (S, \omega_+)\right]   \nonumber \\
&&  
\hspace{-0.4cm}\cdot \exp\left[ - i \pi \bar\tau (\lambda_+ - {\alpha \over 2} \delta^+_D)^2  - i \pi   \tau(\lambda_-   - {\alpha \over 2} \delta^-_D)^2 -i
{ d    u \over  d   a} (S_-, \lambda  - {\alpha \over 2} \delta_D) - 2 \pi i (\lambda - {\alpha \over 2} \delta_D, {\eta \over 2} \delta_D) \right]
\end{eqnarray}
which, modulo the $\eta$-phase factor ${\rm exp} [- 2 \pi i (\lambda - {\alpha \over 2} \delta_D, {\eta \over 2} \delta_D)]$, coincides with the ordinary lattice sum factor $\Psi$ in (3.18) of~\cite{Moore-Witten}, albeit with $\lambda$ replaced by $\lambda - {\alpha \over 2} \delta_D$, and a restriction of the homology two-cycle $S$ to $X \backslash D$. 

Combining that which appears in (\ref{Z photon}) and (\ref{5.16}), and including the $\eta$-term and the $(du / da)$-dependent term of the inserted observable in (\ref{exp map hi to low total}) whilst noting that $S \cap D =0$, we have 
\be
Z' = - {\sqrt 2 \over 4} 
{d\bar \tau\over  d\bar a}
y^{-1/2}\exp\left(S^2{(d u/ d a)^2\over 8 \pi y
} \right)\cdot \Psi'.
\ee

\bigskip\noindent{\it Putting Them All Together}

Incorporating the zero-mode measure in (\ref{zero-modes measure}), the interactions with gravity in (\ref{coupling to gravity}), the rest of the inserted observable in (\ref{exp map hi to low total}),  as well as integrating over the space of  \emph{all}  vacua associated with $a$, we finally arrive at the ``ramified'' $u$-plane integral
\be
\label{Z'u}
Z'_u = \int_{\CM_q}
{dx dy \over  y^{1/2}} \mu(\tau) e^{2 p u + S^2 \hat T_S(u) + \tD^2 T_\tD (u)} \Psi',
\ee 
where $\hat T_S = T_S +{1 \over  8 \pi y} \bigl({du \over  da}\bigr)^2$, and the measure
factor is:
\be
\label{measure factor}
\mu(\tau)   = - {\sqrt{2} \over  4} {da \over  d \tau} F(\chi, \sigma) =-4 \sqrt{2}i (u^2-1){da \over  du}
\left( { ( {2i \over  \pi} {du \over  d \tau} )^2 \over  u^2-1 } \right)^{\sigma/8},
\ee
where the normalization factors $A$ and $B$ have been fixed so as to
agree with the known results for the ordinary Donaldson invariants in~\cite{Moore-Witten} when $(\alpha, \eta) \to (0,0)$, whence  $\Psi' \to \Psi$ and $\tD  \to 0$ in (\ref{Z'u}).\footnote{Recall that $\tD =   i \pi \alpha  D/2$, which vanishes as $\alpha \to 0$.} We have also set $\chi+\sigma=4$,
since this is so for $X$ of $b_1=0,\,b_2^+=1$.

Note that the integration measure is really $d\tau d\bar \tau$; making use of the relation $\tau=x+ i y$, we have actually rewritten it as $dxdy$ in $Z'_u$. The point is that the integration measure is manifestly compatible with the the integration region $\CM_q$ -- the modular curve of $\Gamma(2)$. At any rate, note that the group $\Gamma^0(4)$ is conjugate in $GL(2, \mathbb Q)$ to the subgroup $\Gamma(2)$ of $SL(2,\mathbb Z)$. What this means is that $\CM_q$ could equally well be identified as the modular curve of $\Gamma^0(4)$, where the corresponding family of SW curves differ from the original family by a two-isogeny.  Following~\cite{Moore-Witten}, we shall henceforth use this identification of $\CM_q$, as this convention will facilitate comparisons with established mathematical results  at least in the ordinary limit. Thus, in the notation of~\cite{Moore-Witten},  $\CM_q \cong \Gamma^0(4) \backslash {\cal H}$, where $\cal H$ is the upper half of the complex plane. The translation between the two descriptions is given in  Appendix A.   

From the form of the integrand in (\ref{Z'u}), one can see that $Z'_u$ is manifestly homotopy-invariant: for a particular surface operator with parameters $(\alpha, \eta)$, $Z'_u$ for a simply-connected $X$ is completely determined -- via the ``ramified'' lattice sum $\Psi'$ --  by the lattice $\Gamma = H^2(X, \IZ)$ with its intersection pairing, and the self-intersection number of $D$.
In fact, following~\cite{Moore-Witten}, one can proceed to explicitly show that the ``ramified'' $u$-plane integral is indeed metric-independent as expected of a twisted theory, at least within a chamber in the space of self-dual two-forms on $X$. The point is to prove that the integral is a constant function of $\omega_+$  with wall-crossing. Modulo the manifestly topological $\eta$-term, notice that $\Psi'$ coincides with $\Psi$ up to a constant shift in $\lambda$; since the proof of the topological invariance in $\S$11.3 of~\cite{Moore-Witten} makes no reference to the lattice in which $\lambda$ lives, this constant shift will be irrelevant, at least in the present context. Hence, one can read across from their results and show explicitly  that the ``ramified'' $u$-plane integral is indeed topologically-invariant. For brevity, we shall not present the details here. Nonetheless, as we shall see, the parameters $(\alpha, \eta)$ will play an important role in determining the wall-crossing behavior of $Z'_u$ as we move from one chamber to another in the space of self-dual two-forms on $X$.        

\newsubsection{Verification Of Modular Invariance}

\def\half{{1\over 2}}

As explained earlier, the effective low-energy $U(1)$ theory which underlies the ``ramified'' $u$-plane integral can be expressed in different  but physically equivalent  frames that are related to one another via modular transformations. This means that its corresponding observables -- in this case, the ``ramified'' $u$-plane integral -- must be single-valued and therefore invariant under such modular transformations. Hence, let us now verify  the modular invariance of the ``ramified'' $u$-plane integral as a consistency check on our calculations obtained hitherto.   

Modular invariance is most readily checked by relating $\Psi'$ in (\ref{Psi'}) to the standard
Siegel-Narain theta functions $\Theta_{\Gamma}$ which transform
simply under modular transformations. These functions can be described as follows. 

Let $\Lambda$ be a lattice of signature $(b_+, b_-) = (b^+_2,b^-_2)$.  Let $P$
be a decomposition
of $\Lambda\otimes \IR$ as a sum of orthogonal subspaces  of definite
signature:
\be{ P:\Lambda \otimes \IR \cong
\IR^{b_+,0} \perp \IR^{0,b_-}.
}\ee
Let $P_\pm(\lambda)= \lambda_\pm$ denote the projections onto the two factors.
We also write $\lambda = \lambda_+ + \lambda_-$. With these conventions, $(P_-(\lambda))^2 \leq 0$.

Let  $\Lambda+ \gamma $ denote a translate of the lattice
$\Lambda$.
The Siegel-Narain theta function can then be written as (cf.~\cite{Borcherds})
\begin{eqnarray}
& \hspace{-0.0cm} \Theta_{\Lambda + \gamma} (\tau, \varpi,\beta; P, \xi) 
 \equiv  e^{i \pi (\beta,\varpi)}
 \exp[{ \pi \over  2 y} ( \xi_+^2 - \xi_-^2) ] \,  \cdot  &  \nonumber \\ 
&  \sum_{\lambda\in \Lambda + \gamma}
\exp \left[ i \pi \tau (\lambda+ \beta)_+^2 +
i \pi \bar \tau (\lambda+ \beta)_-^2
+ 2 \pi i (\lambda+\beta, \xi) - 2 \pi i
(\lambda+  \beta, \varpi) \right].&
\end{eqnarray}
Its main transformation law is:
\be{
\Theta_{\Lambda'} (-1/\tau, \varpi, \beta; P, {\xi_+ \over  \tau} +
{\xi_- \over  \bar \tau} )
= \sqrt{\vert \Lambda' \vert \over  \vert \Lambda \vert}
(-i \tau)^{b_+/2} (i \bar \tau)^{b_-/2}
\Theta_{\Lambda } ( \tau, \beta, - \varpi; P, \xi ),
}\label{tx law}\ee
where $\Lambda'$ is the dual lattice.
If there is a characteristic vector, say $v_2$, such that
\be
\label{char}{
(\lambda,\lambda) = (\lambda, v_2)~ \mod ~2
}\ee
for all $\lambda$,
then we have in addition:
\be{
\Theta_{\Lambda } (\tau+1,\varpi, \beta; P, \xi)
= e^{-i \pi(\beta,v_2)/2}
\Theta_{\Lambda } ( \tau, \varpi - \beta - \half v_2,\beta ; P, \xi ).
}\label{tx law 2}\ee
 
However, what we will need is the  \emph{complex-conjugate}  of the Siegel-Narain theta function:
\begin{eqnarray}
\label{conjugate theta function}
& \hspace{-0.0cm} {\overline \Theta}_{\Lambda + \gamma} (\tau,  \varpi,\beta; P, \xi) 
 \equiv  e^{-i \pi (\beta,\varpi)}
 \exp[{ \pi \over  2 y} ( {\bar\xi}_+^2 - {\bar \xi}_-^2) ] \,  \cdot  &  \nonumber \\ 
&  \sum_{\lambda\in \Lambda + \gamma}
\exp \left[-i \pi \bar\tau (\lambda+ \beta)_+^2 
-i \pi  \tau (\lambda+ \beta)_-^2
- 2 \pi i (\lambda+\beta, \bar\xi) + 2 \pi i
(\lambda+  \beta, \varpi) \right].&
\end{eqnarray}
Using (\ref{tx law}), one can show, after a small calculation, that its main transformation law is:
\be
\label{tx law - conjugate}
{\overline\Theta}_{\Lambda'} (-1/\tau, \beta,-\varpi; P, -{\xi_+ \over  \tau} -
{\xi_- \over  \bar \tau} )
= \sqrt{\vert \Lambda' \vert \over  \vert \Lambda \vert}
(-i \tau)^{b_-/2} (i \bar \tau)^{b_+/2}
{\overline \Theta}_{\Lambda } ( \tau, \varpi, \beta; P, \xi ).
\ee
And the complex-conjugate counterpart of (\ref{tx law 2}) is:
\be{
{\overline\Theta}_{\Lambda } (\tau+1,\varpi, \beta; P, \xi)
= e^{i \pi(\beta,v_2)/2}
{\overline \Theta}_{\Lambda } ( \tau, \varpi - \beta - \half v_2,\beta ; P, \xi).
}\label{tx law 2 - conjugate}
\ee

Let us define the generalized theta function:
\be
\label{gen theta}
{\overline\Theta}_{\Gamma; \bar\kappa}  ={\bar\kappa}^{-4(\beta, \varpi)} \, {\overline\Theta}_{\Gamma}\left(\tau,  \varpi, \beta; P_{\omega_+}, \xi \right),
\ee
where $\bar \kappa = e^{-2\pi i/8}$.   Now, let
\be
\label{beta}
\beta = \half (w_2(E) - {\alpha} \delta_D),
\ee
\be
\label{varpi}
\varpi =  \half (w_2(X) - \eta \delta_D),
\ee 
and
\be
\xi = \xi_+ + \xi_- = \rho y {d\bar a \over d \bar u} \omega_+ +  {1 \over  2 \pi}  {d \bar u \over d \bar a} S_-.
\ee
Let us also define
\be
\hat f (p, S, \tD, \tau)  \equiv {i \CN \over 4\pi} 
\left((u^2-1){d \tau \over  d u} \right)^{\chi/4} (u^2-1)^{\sigma/8} {du \over
d \tau} \exp\left[ 2p u+S^2 \hat T_S(u) + \tD^2 T_\tD(u)
\right]
\ee   
for an appropriate normalization constant $\cal N$. Last but not least, 
let us introduce the auxiliary integral ${\cal G}(\rho)$
\be
\label{grho}{
{\cal G}(\rho)  \equiv
\int_{\Gamma^0(4) \backslash {\cal H}} {dx dy \over  y^{3/2}}
 \, \hat f(p, S, \tD, \tau) \, {\overline \Theta}_{\Gamma; \bar \kappa} (\rho).
}\ee
Then, by a careful comparison of (\ref{Psi'}), (\ref{Z'u}), (\ref{measure factor}), (\ref{conjugate theta function}) and (\ref{gen theta})-(\ref{grho}), we find that the ``ramified'' $u$-plane integral can be written as\footnote{In writing the following expression for $Z_u$, we have dropped the sign factor $(-1)^{(2 \lambda^2_0 + \lambda_0 \cdot w_2(X))}$ which was included earlier to demonstrate an agreement with certain observations found in the mathematical literature. Indeed, what we would like to verify is the modular invariance of the physically formulated theory, and dropping this factor is  hence inconsequential.}
\be
 Z'_u = (S,\omega_+)  {\cal G}(\rho) \biggr\vert_{\rho=0}
 + 2
 {d {\cal G}  \over  d   \rho}   \biggr\vert_{\rho=0}.
\ee

\def\gof{{\Gamma^0(4)}}

Notice that one can write the integrand of (\ref{grho})  as ${dx dy \over  y^2} \CJ$, where $\CJ = \hat f \cdot y^{1/2} \overline\Theta_{\Gamma; \bar \kappa}$.  
Note also that the integration region in (\ref{grho}) is given by   
\be\label{fnddom}{
\gof \backslash\CH \cong
\left[ \CF \cup T\cdot \CF \cup T^2\cdot\CF
\cup T^3 \cdot \CF \right] \cup S \cdot \CF \cup T^2 S \cdot \CF
},
\ee
where $\CF$ is a fundamental domain for $PSL(2,\IZ)$ defined by
\be
\CF = \{\tau \in \CH: -\half \leq {\rm Re} \tau \leq \half, |\tau| \geq 1\},
\ee
while $T: \tau +1$ and $S: \tau \to -1 /\tau$.  The first four domains give the region of the cusp at
$\tau \rightarrow i \infty$ and correspond to the semiclassical
region. The domain $S \cdot \CF$ gives the region of the
cusp near $\tau=0$ and corresponds to the
monopole cusp. The domain $T^2 S \cdot \CF$ gives the region of the cusp near $\tau=2$
and corresponds to the dyon cusp.

In order to ascertain how $Z'_u$ will behave under modular transformations, it is clear that we must map ${dx dy \over  y^2} \CJ$ in these six regions to $\CF$. Since $\CJ$ manifestly depends on $\tau$ through $\hat f$ and $\overline \Theta_{\Gamma; \bar \kappa}$, one can see from the actions of $S$ and $T$ on $\CF$ in (\ref{fnddom}) that $\CJ$, when mapped to $\CF$,  will be given by the following functions:
\begin{eqnarray}
\label{six functions}
\CJ_{(\infty,0)}(\tau )    &\equiv & \CJ(\tau),  \nonumber \\
\CJ_{(\infty,1)}(\tau )  &\equiv & \CJ(\tau+1),\nonumber \\
\CJ_{(\infty,2)}(\tau )   &\equiv & \CJ(\tau+2), \nonumber \\
\CJ_{(\infty,3)}(\tau )   &\equiv & \CJ(\tau+3), \\
\CJ_{M}(\tau ) &   \equiv & \CJ(-1/\tau), \nonumber \\
\CJ_D (\tau ) &  \equiv & \CJ(2 -1/\tau), \nonumber
\end{eqnarray}  
each associated with the indicated region in $\gof \backslash \CH$. The subscript $M$ and $D$ refer to the monopole and dyon cusps, respectively. Denoting 
\be
I=(\infty,0), (\infty,1),(\infty,2),(\infty,3), M, D,
\ee  
we can then write
\be
\CG(\rho)  =
\int_{\CF} {dx dy \over  y^{3/2}}
\sum_I
\hat f_{I}(p, S, \tD, \tau) \overline \Theta_{\Gamma; \bar\kappa; I} (\rho),
\ee
where 
\be
 \overline \Theta_{\Gamma; \kappa; I} (\rho) = e^{-i \phi_I} \overline {\Theta}_\Gamma(\tau, \varpi_I, \beta_I; \xi_I)
\ee
are the transforms of the generalized theta functions implied by (\ref{six functions}), and the $\phi_I$'s are the appropriate phases. 

Let us assume that $2\varpi$ is a characteristic vector, such that we can identify it with $v_2$ in (\ref{char}). Then, one can check via (\ref{six functions}),    (\ref{tx law 2 - conjugate}) and (\ref{tx law - conjugate}) that
\begin{eqnarray}
e^{-i \phi(\infty, n)} & = & {\bar\kappa}^{-(n+1) (w_2(X) - \eta \delta_D, w_2(E) - \alpha \delta_D)},  \quad n =0,1,2,3, \nonumber \\
 e^{-i\phi_M} & = & e^{-i \phi_D} = {\bar \kappa}^{-(w_2(X) - \eta \delta_D, w_2(E) - \alpha \delta_D)}, 
\end{eqnarray}
and that 
\begin{eqnarray}
\label{w tx}
\varpi_{(\infty, 0)} =  \varpi, & \qquad & \beta_{(\infty, 0)}   =  \beta, \nonumber\\
\varpi_{(\infty, 1)}  =  - \beta, &  \qquad & \beta_{(\infty, 1)} = \beta, \nonumber\\
\varpi_{(\infty, 2)} = - 2 \beta - \varpi, & \qquad & \beta_{(\infty, 2)} = \beta, \nonumber\\
\varpi_{(\infty, 3)} =  -3 \beta - \varpi, & \qquad & \beta_{(\infty, 3)} = \beta, \\
\varpi_{M} = \beta, & \qquad & \beta_{M} = - \varpi, \nonumber\\
\varpi_{D} = \beta,  & \qquad  &\beta_{D} = 2 \beta + \varpi, \nonumber
\end{eqnarray}
where $\varpi$ and $\beta$ are as given in (\ref{varpi}) and (\ref{beta}). 

As for the $\hat f_I$'s, note that  $\hat f$ differs from its ordinary counterpart by the factor ${\rm exp} [\tD^2 T_\tD(u)]$. Since this factor is holomorphic in $u$ and -- in contrast to the other factor ${\rm exp} [S^2 \hat T_S(u)]$ -- is also independent of $\tau$ and $du/da$, it is modular-invariant, as emphasized earlier. What this means is that it is a ``spectator'' factor under modular transformations of $\hat f$; in other words, the modular transformations of $\hat f$ are given by those of its ordinary counterpart. Consequently, we have (cf.~\cite{Marcos}) 
\begin{eqnarray}
{\hat f}_{(\infty, 0)} (\tau +1) = {\hat f}_{(\infty, 1)}(\tau), & \qquad & {\hat f}_{(\infty, 0)} (-1 /\tau) = (-i \tau)^{\sigma/2} {\hat f}_M(\tau), \nonumber \\
{\hat f}_{(\infty, 1)} (\tau +1) = {\hat f}_{(\infty, 2)}(\tau), & \qquad & {\hat f}_{(\infty, 1)} (-1 /\tau) = (-i \tau)^{\sigma/2} {\hat f}_{(\infty,3)}(\tau), \nonumber \\
{\hat f}_{(\infty, 2)} (\tau +1) = {\hat f}_{(\infty, 3)}(\tau), & \qquad & {\hat f}_{(\infty, 2)} (-1 /\tau) = (-i \tau)^{\sigma/2} {\hat f}_D(\tau), \nonumber \\
{\hat f}_{(\infty, 3)} (\tau +1) = {\hat f}_{(\infty, 0)}(\tau), & \qquad & {\hat f}_{(\infty, 3)} (-1 /\tau) = (-i \tau)^{\sigma/2} {\hat f}_{(\infty, 1)}(\tau), \nonumber \\
{\hat f}_{M} (\tau +1) = \kappa^{\sigma} {\hat f}_{M}(\tau), & \qquad & {\hat f}_{M} (-1 /\tau) = (-i \tau)^{\sigma/2} {\hat f}_{(\infty, 0)}(\tau), \nonumber \\
{\hat f}_{D} (\tau +1) = \kappa^{\sigma} {\hat f}_{D}(\tau), & \qquad & {\hat f}_{D} (-1 /\tau) = (-i \tau)^{\sigma/2} {\hat f}_{(\infty, 2)}(\tau). \nonumber \\
\end{eqnarray}

With the further assumption that $4 \varpi^2 \equiv \sigma \, {\rm mod} \, 8$, one can go on to show that ${\overline\Theta}_{\Gamma; k; I}$ and $\hat f_I$ are in conjugate (unitary) representations of $SL(2,\IZ)$; in other words, ${\overline\Theta}_{\Gamma; k; I}$ and  $\hat f_I$ transform with   \emph{opposite}   phases under modular transformations.  Hence, $\CG(\rho)$ is modular-invariant, and therefore, so is $Z'_u$, as required.  

Nevertheless, to prove that $Z'_u$ is modular-invariant, we have had to make two important assumptions. The first assumption can be stated as
\be
\label{ass1}
(\lambda,\lambda) ~ \mod ~2 \equiv (\lambda, w_2(X) - \eta \delta_D)
\ee
for all $\lambda \in \Gamma$, while the second assumption can be stated as
\be
\label{ass2}
( w_2(X) - \eta \delta_D,  w_2(X) - \eta \delta_D) \equiv \sigma \ {\rm mod} \ 8.
\ee

Note that in four-dimensions, one will always have $(\gamma, w_2(X)) \equiv (\gamma, \gamma)  \ {\rm mod} \ 2$ for any $\gamma \in \Gamma$. Thus, the first assumption implies that we must have $\eta (\lambda, \delta_D) \in 2 \IZ$.  Since $\lambda \in H^2(X, \IZ)$,  it can be expanded in the basis of two-forms which is Poincar\'e dual to the basis of homology two cycles $\{U_i\}_{i =1, \dots, b_2(X)}$ that has a purely diagonal, unimodular intersection matrix, and from which $D$ -- whereby $D \cap D \neq 0$ -- can be conveniently chosen to be one of the $U_i$'s. This means that we must have $\eta q (\delta_D, \delta_D) \in 2 \IZ$ for some non-zero integer $q$.  In other words,  if $D \cap D \neq 0$, we must have $\eta \in 2 \IZ$ in order for the first assumption to hold, regardless.   

Note also that one will always have $(w_2(X), w_2(X)) \equiv \sigma  \ {\rm mod} \ 8$. Consequently, since $w_2(X)$ is integer-lifted -- which means that it can be expanded in the same integral basis of two-forms as $\lambda$ -- the second assumption implies that we must have $(D \cap D) (\eta^2 - 2 \eta p) \in 8\IZ$ for some non-zero integer $p$. Therefore, if $D \cap D \neq 0$, we must have $\eta \in 4 \IZ$. 

Altogether, this means that when $D \cap D \neq 0$, the two assumptions will be guaranteed to hold if and only if $\eta \in 4 \IZ$.  This observation is in complete agreement with the analysis in $\S3.1$, which shows (after taking into account the fact that the twisted and untwisted theories differ  by an overall factor of $1/4$) that when $D \cap D \neq 0$,  the  ``electric'' and ``magnetic''  descriptions of the underlying physical theory are only truly equivalent if and only if $\eta \in 4 \IZ$!  In hindsight, this agreement is expected, since the modular invariance of $Z'_u$ will imply that the ``electric'' and ``magnetic'' descriptions of the underlying $U(1)$ theory \emph{are} necessarily physically equivalent. Nevertheless, it is still satisfying to know that we have obtained the same condition on $\eta$ via two \emph{a priori} different formalisms -- one involving a duality transformation in superspace~\cite{mine}, and the other involving the modular properties of the Siegel-Narain theta function.

Last but not least, note that $\alpha_{(\infty, 0)} =  \alpha$ and  $\eta_{(\infty, 0)} = \eta$.
Hence, from (\ref{w tx}),  we find that corresponding to $\tau_M = - 1/ \tau$, we have
\be\label{alphaM}
\alpha_M = - \eta, \quad {\rm and} \quad \eta_M =  \alpha,
\ee
and corresponding to $\tau_{(\infty, 2)} = \tau + 2$ (since $w_2(X) \neq 0$), we have
\be
\alpha_{(\infty, 2)} = \alpha, \quad  {\rm and} \quad \eta_{(\infty, 2)} = -\eta - 2\alpha. 
\ee
On the other hand,  the transformation $(\alpha, \eta) \to (\eta, - \alpha)$  under $S: \tau \to -1/\tau$, and  the transformation $(\alpha, \eta) \to (\alpha, \eta - 2\alpha)$ under $T^2 : \tau + 2$, which leave the underlying physical theory with action $S_{\rm eff}$ in (\ref{Seff}) invariant, imply that $\alpha_M = \eta$, $\eta_M = - \alpha$,  $\alpha_{(\infty, 2)} = \alpha$ and $\eta_{(\infty, 2)} = \eta - 2\alpha$. However, as noted in footnote~18,  this discrepancy is just due to a trivial sign difference in the definitions of the $\eta$-term in $\mathscr L$ and $S_{\rm eff}$.

\newsection{Wall-Crossing Formulas Of The ``Ramified'' $u$-Plane Integral}

As mentioned earlier, it is known that the ``ramified'' Donaldson invariants are chamber-dependent when $b^+_2 = 1$~\cite{KM2} -- in other words, they jump as one moves across ``walls'' that divide the space of self-dual two-forms on $X$ into chambers.  If we denote $\Delta_{D}$ as the jump in the ``ramified'' Donaldson invariants across a ``wall'', then from the relation (\ref{Z = Z + Z}), and from the fact that such discontinuities are typically associated with singular points in moduli space such as $u = \infty, 1 ,-1$, we can write
\be
\label{deltas}
\Delta_{D}  = \Delta_{u = \infty} + \Delta_{u=1} + \Delta_{u =-1}  + \Delta_{SW}.
\ee
Note that the proof of the topological invariance of the correlation functions that correspond to the ``ramified'' Donaldson invariants can only fail if a particular total derivative in the path-integral does not vanish at infinity, that is, if the field space is non-compact; such is the case at $u = \infty$ only. This means that $\Delta_D$ will be given by $\Delta_{u = \infty}$, or
\be
\label{delta D = delta u}
\Delta_D = \Delta_{u = \infty}.
\ee 
Indeed, as we shall see shortly, the formula of $\Delta_{u = \infty}$ will coincide with the formula of $\Delta_D$ from the existing mathematical literature when we take the ordinary limit $(\alpha, \eta) \to (0,0)$.  ($\Delta_D$ has yet to be computed by mathematicians, and we shall provide its explicit formula below.) In turn, this means that $\Delta_{u=1} + \Delta_{u=-1}$ must cancel  $\Delta_{SW}$ in (\ref{deltas}), or
\be
\label{delta u = delta SW}
\Delta_{u=1} + \Delta_{u =-1}  = \Delta_{SW}
\ee
up to a sign. In other words, one can match the wall-crossing formulas of the ``ramified'' $u$-plane integral at $u= \pm 1$ with the wall-crossing formulas of the ``ramified'' SW theory.  

\newsubsection{About The Definition And Nature Of The ``Ramified'' $u$-Plane Integral}

Notice that the ``ramified'' $u$-plane integral (\ref{Z'u})  does not actually converge; this due to its 
bad behavior near $u=\infty$ and in some cases also near $u=\pm 1$.  Clearly, if one expands (\ref{Z'u}) in powers of $p$ in order to compute the ``ramified'' Donaldson invariants of increasing order, then as $u$ diverges at infinity, one will eventually run into a divergent integral.  Likewise, near $u=\pm 1$, if $\sigma$ is sufficiently negative, we will also run into a problem. 

The way around this problem is to cut off these divergences as follows. First, we expand (\ref{Z'u}) out to a given order in $p$ and $S$ to obtain an integral that should give a ``ramified'' Donaldson invariant of some given order.  Then, for $\tau = x + iy$, we perform the integral for $y<y_0$, for some cutoff $y_0$, and then take the limit as $y_0\to\infty$ only at the end. A similar procedure can be employed near the cusps at $u=\pm 1$; for example, at the monopole cusp where $\tau_M = \tau_d$, we first integrate over ${\rm Im}\,\tau_d <y_0$, before taking the limit as $y_0\to\infty$.

\def\quart{{1 \over 8}}

\def\lama{{\lambda'}}

\def\lamap{{\lambda'_+}}

\def\lamam{{\lambda'_-}}

This procedure eliminates the divergences for the following reason.  Set $q=
\exp(2\pi i \tau)$.   Then the term in (\ref{Z'u}) that is of any given order in
$p$ and $S$  is a sum of terms, each of which is some power of $y$ times a factor $\mu(\tau) \cdot F(u, du/da, \alpha, \eta, \tD^2)$ times a sum of the form
\be
\label{sumexp}
\sum_{\nu,\mu}q^\nu \bar q^\mu,
\ee
where $\nu = - \quart (2\lambda - \alpha \delta_D)_-^2$ and $\mu = \quart (2\lambda - \alpha \delta_D)^2_+$ are not integers, but obey $\mu - \nu \in {1 \over  8} \IZ$.\footnote{Note that henceforth, we will revert back to the original definition where $\lambda \in \half w_2(E) + \Gamma$. This means that $2\lambda \in 2 \Gamma + w_2(E)$, where $w_2(E)$ is integer-lifted, and for a physically consistent choice of $D$, we also have $\alpha \delta_D \in \Gamma$. Therefore, $(2\lambda - \alpha \delta_D)^2 \in \IZ$.} As with $\lambda_+$, one can write $(\lambda - {\alpha \over 2}\delta_D)_+ = \lambda'_+ = \omega_+ (\omega_+, \lama)$, where $(\lamap)^2 \geq 0$ and $(\lamam)^2 \leq 0$. Consequently,  after we take into account the factor $\mu(\tau) \cdot F(u, du/da, \alpha, \eta, \tD^2)$, we find that $\mu$ is bounded from below by zero but not $\nu$. This important observation is true because the factor $\mu(\tau) \cdot F(u, du/da, \alpha, \eta,\tD^2)$ is purely $q$-dependent via the functions $\mu(\tau)$, $u$ and $du/da$; in particular, there are contributions in  negative powers of $q$ from $u$ and $(d\tau / du)^{- \sigma /4}$ but not in $\bar q$; in other words,  the factor $\mu(\tau) \cdot F(u, du/da, \alpha, \eta, \tD^2)$ will only contribute to a negative shift in the value of $\nu$ in (\ref{sumexp})  but not $\mu$.  

Keeping the lower bound of $\mu$ in mind, now consider an integral of the following form:
\be
\lim_{y_0\to\infty}\int_{y_1}^{y_0}{dy\over y^c} \int_0^kdx
\sum_{\nu,\mu}q^\nu \bar q^\mu,
\ee
where $y_1$ is an irrelevant lower cutoff that is included so as
to study one cusp while keeping away from others.  The question here is whether the integral
converges for $y_0\to \infty$.
The $x$ integral runs from 0 to $k$  where (for $\Gamma^0(4)$) $k=4$ for the
cusp at
infinity, and $k=1$ for the other cusps. A detailed examination of (\ref{Z'u}) and the ``ramified'' lattice sum $\Psi'$ reveals that in all cases, either $c>1$ or if $c < 1$, there are, for a generic metric on $X$ (the metric enters in the definition of $ \Psi'$), no terms with
$\nu=\mu=0$. As we shall see explicitly below, integrating first over $x$ projects the sum in (\ref{Z'u}) onto terms with $\nu=\mu$; since $\mu \geq 0$, the sum is projected onto terms ${\rm exp}(-4 \pi \mu y)$ (where $y \geq 0$) that consequently vanish exponentially or, if $\nu=\mu=0$, are constant at infinity.  Therefore, for a generic metric on $X$, the $y$ integral converges as  $y_0\to\infty$, since all terms that will survive the $x$-integral either have $c>1$, or have $c <1$ but with $\nu,\mu>0$. In summary, this cutoff procedure leads to the integral being a well-defined formal power series in $p$ and $S$ for a generic metric on $X$.

Nevertheless, at special points in the space of metrics on $X$ where one encounters ``ramified'' abelian instantons, there are terms with $c=1/2$ (that is, $c <1$) and $\nu=\mu=0$.  These points are where $Z'_u$ is discontinuous and thus, where wall-crossings can occur.  Let us elaborate on this now. 

\bigskip\noindent{\it A Discontinuity In $Z'_u$}

As explained above,  any ill-behavior and thus discontinuity in $Z'_u$ should occur where $c < 1$ and $\nu = \mu =0$. Hence, the relevant terms that might contribute to a discontinuity in $Z'_u$ ought to be given by 
\be
\label{I(w)}
I(\omega_+) \equiv \int_{\CF} {dx dy \over  y^{1/2}} c(d) e^{2 \pi i x d - 2 \pi yd}
e^{-i \pi x \left [(\lamap)^2 + (\lamam)^2 \right]} e^{- \pi y \left[(\lamap)^2 - (\lamam)^2 \right]}  (\omega_+, \lama) 
\ee 
for some number $d$ and some $\lama = \lambda - {\alpha \over 2} \delta_D$. (The integrand of $Z'_u$ also contains additional terms proportional to $y^{-3/2}$ and $y^{-5/2}$ instead of $y^{-1/2}$, that do not have the factor $(\omega_+, \lama)$; these terms will not contribute to a discontinuity in $Z'_u$; they will be suppressed, as we shall explain shortly). In (\ref{I(w)}), $c(d)$ is the  coefficient of $q^d$ in the $q$-expansion of a nearly holomorphic modular form ${\cal C} (q)$ whose details will be specified briefly. In any case, it is useful to note at this point that $c(d)$ is a function of $p$, $(S_-, \lamam)$ and $(S_+,  \lamap)$, among other things. 

What we really wish to do is to analyze $Z'_u$ for discontinuities as the decomposition $\lama = \lambda'_+ + \lambda'_-$ varies while $\lama$ stays fixed. Since $\lamap = \omega_+ (\omega_+, \lama)$, the decomposition will depend on $\omega_+$; hence, we should study $I(\omega_+)$ for discontinuities as we vary $\omega_+$. As before, let us do the integration over the $x$-variable first. In doing so, the remaining integral is projected onto $2d = (\lama)^2$. For this value of $d$, the remaining $y$-integral  looks like
\be
\int_{y_1}^\infty {dy\over y^{1/2}} \, c\left((\lama)^2/2\right) \, e^{-2\pi y
(\lamap)^2} (\omega_+, \lama).
\ee 
The above is an elementary integral (if one replaces $y_1$ by 0) and converges for all non-zero $\lamap$. However, notice that in the vicinity of $\lamap = 0$,  the integral is discontinuous.  The discontinuity comes from the large $y$ part of the integral, and so is independent of the lower cutoff $y_1$. The discontinuity in $I(\omega_+)$
as $\omega_+$ crosses  from
$(\omega_+, \lama) = 0^-$ to
$(\omega_+,\lama) = 0^+$
is computed to be
\be
\label{the jump}
I^+(\omega_+) - I^-(\omega_+) = \sqrt{2} c(d)  = \sqrt{2} \left[ q^{-(\lama)^2/2} {\cal C}(q) \right]_{q^0},
\ee
where  $[ \cdot ]_{q^0}$ indicates the constant term in a Laurent expansion in powers of $q$. It may also be expressed as a residue. Since $\lamap=0$ at the discontinuity, we may make the following replacements in ${\cal C}(q)$:  $(S_+,  \lamap) =0$ and   $(S_-, \lamam) = (S, \lama)$.  
In addition, since the discontinuity comes from the large $y$ part of the integral, we can set the second term in $\hat T_S = T_S +{1 \over  8 \pi y} \bigl({du \over  da}\bigr)^2$  to zero in (\ref{the jump}); likewise, the terms involving $1 /y$ in (\ref{Psi'}) will also be suppressed. Last but not least, since $(\lama)^2 = (\lamap)^2 + (\lamam)^2$, and since $(\lamam)^2 < 0$, the condition $\lamap = 0$ will imply that $(\lama)^2 < 0$. 

Recall that there are three cusps at $\tau = i \infty$, $\tau =0$ and $\tau =2$ in $Z'_u$ that correspond to the points $u = \infty$, $u =1$ and $u=-1$, respectively. According to (\ref{delta D = delta u}), the first cusp leads to the ``ramified'' Donaldson wall-crossing formulas, while according to (\ref{delta u = delta SW}),  the second and third cusps lead to the ``ramified'' SW wall-crossing formulas. 
As we shall see below, the discontinuities in $Z'_u$ at the three cusps do not cancel one another. Hence, one can conclude that $Z'_u$ is \emph{not} a bona-fide topological invariant. Indeed, the conditions $(\lama)^2 < 0$ and $\lamap =0$ define a ``ramified'' abelian instanton; since the macroscopic $U(1)$ gauge group is a ``part of'' the microscopic $SU(2)$ or $SO(3)$ gauge group, such a ``ramified'' abelian instanton will correspond to a singular point in the moduli space  $\CM'$ of ``ramified'' $SU(2)$ or $SO(3)$ instantons; that singular points in $\CM'$ ought to correspond to jumps in $Z'_u$ and thus, the ``ramified'' Donaldson invariants, is certainly not an unexpected phenomenon, at least from our experience with the ordinary Donaldson invariants.

At any rate,  the conditions $(\lama)^2<0$ and $\lamap=0$  will define  chambers in the forward light cone
$V_+= \{\omega_+ \in H^{2,+}(X;\IR): (\omega_+)^2 > 0 \}$ :
for any $\lama = \lambda - {\alpha \over 2} \delta_D$ with $(\lama)^2<0$, one can define a
wall bounding a chamber in $V_+$   by
\be
\label{wall}{
W_{\lambda; \alpha} \equiv \{ \omega_+: (\omega_+, \lambda) = {\alpha \over 2} (\omega_+, D)  \},
}\ee
such that when $\omega_+$ crosses any one of these ``walls'', $Z'_u$ will jump as given in (\ref{the jump}) for some ${\cal C}(q)$.

\newsubsection{Wall-Crossing Formulas Of The ``Ramified'' Donaldson Invariants}

\def\lamaf{{\lambda - {\alpha \over 2} \delta_D}}

Let us now ascertain the explicit wall-crossing formula of $Z'_u$ at the cusp $\tau = i \infty$, which should correspond to a ``ramified'' Donaldson wall-crossing formula that has yet to be computed in the mathematical literature. In this case, note that one has to multiply (\ref{the jump}) by a factor of four: in computing (\ref{the jump}), we performed the $x$-integral from $x=0$ to $x=1$ only, and to cover the cusp at $\tau = i \infty$, one needs to perform the $x$-integral from $x=0$ to $x=4$.  Note also that at this cusp, we have $\beta_{I}  =  \half (w_2(E) - \alpha \delta_D)$ for all $I =1, \dots, 4$ in (\ref{gen theta}). Thus, according to (\ref{the jump}), the quantity $\Delta_{u =\infty}$ in (\ref{delta D = delta u}) will be given (up to a sign) by
\begin{eqnarray}
\label{wallcrossing Don}
& Z'_{u,+}  - Z'_{u,-}  =  -   32i (-1)^{ \{(\lambda-\lambda_0)\cdot w_2(X)\}} \, e^{(2\pi i\lambda_0^2  + w_2(X)[\tD])} & \nonumber \\
&\hspace{0.0cm}\cdot \left[q^{- (\lamaf)^2/2}  (u^2-1) h(\tau)  \left( { ( {2i \over  \pi} {du \over  d \tau} )^2 \over  u^2-1}
\right)^{\sigma/8} \exp \left \{ 2p u  +  S^2 T_S(u) + \tD^2 T_\tD (u) - i (\lambda, S) / h \right \} 
\right]_{q^0}. & \nonumber \\
\end{eqnarray}
Here, $h \equiv {da \over du}$ and $\tD = i \pi \alpha D /2$. Note that  in computing the above formula, we have made use of the fact that $S \cap D = 0$,  and that for a nontrivially-embedded surface operator where $\tD^2 \sim \alpha^2D\cap D \neq 0$, one will have the conditions $\eta \in 4 \IZ$ and $\alpha D \cap D \in \IZ$. Consequently, we have taken the liberty to set $\eta$ to zero via an integral shift, since it is defined modulo 4 in $\mathscr L$  in (\ref{Lagrangian twisted U(1)}). 

In Appendix A, we supply the explicit formulas of the various modular forms in (\ref{wallcrossing Don}) in terms of Jacobi $\vartheta$-functions. By using these expressions, one can simplify (\ref{wallcrossing Don}); in doing so, we finally obtain (up to a sign) the wall-crossing formula of the generating function of the ``ramified'' Donaldson invariants of $X$ -- with a  \emph{nontrivially-embedded }two-surface  $D$ -- as:
\begin{eqnarray}
\label{wallcrossing Don simplified}
& \hspace{-0.0cm} \Delta_D  =  -   {i \over 2}  (-1)^{ \{(\lambda-\lambda_0) \cdot  w_2(X) \}} \, e^{(2\pi i\lambda_0^2 + w_2(X)[\tD])} & \nonumber \\
&\hspace{0.0cm}\cdot \left[q^{- (\lamaf)^2/2} \, {\vartheta_4^{8 + \sigma} \over h(\tau)^3} \, \exp \left \{ 2p u  +  S^2 T_S(u) + \tD^2 T_\tD (u) - i (\lambda, S) / h \right \} 
\right]_{q^0}, & 
\end{eqnarray}
with $h= \half \vartheta_2 \vartheta_3$ and $u   = \half { \vartheta_2^4 + \vartheta_3^4 \over  (\vartheta_2 \vartheta_3)^2}$.

On the other hand, if one has a trivially-embedded surface operator such that $\tD^2 =0$, then the wall-crossing formula  for the generating function of the ``ramified'' Donaldson invariants of $X$ -- with a \emph{trivially-embedded} two-surface $D$ --  will be given (up to a sign) by
\begin{eqnarray}
\label{wallcrossing Don trivial-embed}
& \hspace{-0.0cm} \tilde \Delta_D  =  -   {i \over 2} (-1)^{ \{(\lambda-\lambda_0) \cdot   w_2(X)  + {\eta \over 2} l \}} \, e^{(2\pi i\lambda_0^2 + w_2(X)[\tD])} & \nonumber \\
&\hspace{0.0cm}\cdot \left[q^{- (\lamaf)^2/2} \, {\vartheta_4^{8 + \sigma} \over h(\tau)^3} \, \exp \left \{ 2p u  +  S^2 T_S(u)  - i (\lambda, S) / h \right \} 
\right]_{q^0}. & \nonumber \\
\end{eqnarray}
Here, $l = 2\int_D \lambda = 2 \lambda[D]$, where $l$ is the monopole number.

Notice that (\ref{wallcrossing Don simplified}) and (\ref{wallcrossing Don trivial-embed}) depend on the surface operator parameters $(\alpha, \eta)$, the monopole number $l$,  the ``ramified'' (abelian) instanton number $k'_L = -2(\lamaf)^2$, and the self-intersection number $D \cap D$ of the surface operator, among other things. Moreover, as the notation on  $W_{\lambda; \alpha}$ implies, the ``wall'' defined in (\ref{wall}) -- across which the jump (\ref{wallcrossing Don simplified}) or (\ref{wallcrossing Don trivial-embed}) occurs -- manifestly depends on $\alpha$, as anticipated by KM in~\cite{KM2}.

As a consistency check on our computations, note that in the ordinary limit $(\alpha, \eta) \to (0,0)$, both the wall-crossing formulas of the generating function of the ``ramified'' Donaldson invariants in (\ref{wallcrossing Don simplified}) and (\ref{wallcrossing Don trivial-embed})  reduce to eqn.~(4.6) of~\cite{Moore-Witten}, which in turn coincides with the wall-crossing formula given in~\cite{MW 15, MW 10} for the generating function of the ordinary Donaldson invariants of four-manifolds with $b_1 =0, b^+_2 =1$, as anticipated.

\newsubsection{Wall-Crossings Formulas At The Seiberg-Witten Cusps}

\def\lsq{{{L'}^{\otimes 2}}}

Let us now ascertain the explicit wall-crossing formula of $Z'_u$ at the cusp $\tau =0$, which should correspond to a ``ramified'' SW monopole wall-crossing formula -- a connection that will be clarified in the next section. At this cusp, we have $\varpi_M = \half (w_2(E) - \alpha \delta_D)$ and $\beta_M = -\half (w_2(X) - \eta \delta_D)$ in (\ref{gen theta}). 

As discussed in~\cite{Appendix 1}, the ``monopoles'' that appear near $u=1$ are sections of a rank-two complex vector bundle $S_+ \otimes L'$, where for $w_2(X) \neq 0$, $L'$ does not exist as a line bundle on its own but ${L'}^{\otimes 2}$ does; consequently, we have the integral class $c_1(\lsq) \in -w_2(X) + 2 \Gamma$. Let us define $\lambda' = \half c_1(\lsq)$; then $\lambda' \in -\half w_2(X) + \Gamma$. Notice that this definition of $\lambda'$ is consistent with $\beta_M = -\half (w_2(X) - \eta \delta_D)$ at this cusp, as it should. 
As before, the ``walls'' are defined by 
\be
(\omega_+, \lambda)  =  {\alpha_M \over 2} (\omega_+, D) \quad {\rm and} \quad  (\lama)^2  <  0,
\ee
where $\alpha_M = - \eta$. 

\def\lamafd{{\lambda - {\alpha_M \over 2}}}

Therefore, according to (\ref{the jump}) specialized to the monopole cusp at $\tau =0$, the quantity $\Delta_{u=1}$ in (\ref{delta u = delta SW}) for a \emph{nontrivially-embedded}  surface operator is (up to a sign)
\begin{eqnarray}
\label{wallcrossing mono nontrivial}
& \Delta_{u=1}  =  -   {i \over 8} e^{2\pi i( \lambda_0 \cdot \lambda + \lambda_0^2)} e^{2 \lambda[\tD]}  & \nonumber \\
&\hspace{0.0cm}\cdot \left[q_M^{- \lambda^2/2} \, {\vartheta_4^{8 + \sigma} \over (h_M)^3} \, \exp \left \{ 2p u_M  +  S^2 T^M_S(u_M)  - i (\lambda, S) / h_M \right \} 
\right]_{q_M^0}, & 
\end{eqnarray}
where $q_M = {\rm exp} (2 \pi i \tau_M)$; $\lambda \in -\half w_2(X) + \Gamma$; and 
\begin{eqnarray}
h_M  & = & \half \vartheta_2 \vartheta_3, \nonumber \\
u_M  & = &\half { \vartheta_2^4 + \vartheta_3^4 \over  (\vartheta_2 \vartheta_3)^2},  \\
T^M_S  & = & - {1 \over  24}\left[   {  E_2\over  h_M(\tau)^2}  - 8 u_M\right]. \nonumber
\end{eqnarray} 
Note that in computing (\ref{wallcrossing mono nontrivial}), we have made use of the fact that when $D \cap D \neq 0$,  one must have $\eta \in 4 \IZ$; as such, we have taken the liberty to set $\alpha_M$ to zero via an integral shift.   

On the other hand, for a \emph{trivially-embedded} surface operator where $D \cap D =0$, the quantity $\Delta_{u=1}$ in (\ref{delta u = delta SW}) will be given (up to a sign) by
\begin{eqnarray}
\label{wallcrossing mono trivial}
& \tilde \Delta_{u=1}  =  -   {i \over 8} e^{2\pi i( \lambda_0 \cdot \lambda +\lambda_0^2)} (-1)^{ \{ {\alpha_M \over 2} w_2(E)[D] \}} e^{2 \lambda[\tD]} & \nonumber \\
&\hspace{0.0cm}\cdot \left[q_M^{- (\lambda - {\alpha_M \over 2}\delta_D)^2/2} \, {\vartheta_4^{8 + \sigma} \over (h_M)^3} \, \exp \left \{ 2p u_M  +  S^2 T^M_S(u_M)  - i (\lambda, S) / h_M \right \} 
\right]_{q_M^0}. & \nonumber \\
\end{eqnarray}

As another consistency check, note that in the limit $(\alpha, \eta) = (\eta_M, -\alpha_M) \to (0,0)$, both the wall-crossing formulas (\ref{wallcrossing mono nontrivial}) and (\ref{wallcrossing mono trivial}) reduce to the ordinary SW wall-crossing formula in eqn.~(4.8) of~\cite{Moore-Witten}, as they should. With (\ref{wallcrossing mono nontrivial}) and (\ref{wallcrossing mono trivial}), we shall be able to determine some factors in the SW contribution $Z'_{u=1}$ to the generating function  $Z'_D$ of the ``ramified'' Donaldson invariants in the next section. There are similar wall-crossing formulas for the dyon cusp too; however, for brevity, we shall not compute them: as we shall see later, the sought-after contribution $Z'_{u=-1}$ due to the massless dyons can be obtained from $Z'_{u=1}$ via a discrete symmetry transformation; unlike $Z'_{u=1}$, the determination of $Z'_{u = -1}$ does not require an explicit knowledge of the wall-crossing formulas at the dyon cusp.

At the monopole cusp, $q_M \ll 1$. Consequently, for (\ref{wallcrossing mono nontrivial}) and (\ref{wallcrossing mono trivial}) to be non-vanishing, there must be negative powers of $q_M$. In this respect, the only relevant term in the $q_M$-expansion of (\ref{wallcrossing mono nontrivial}) is $q^{- \half d_\lambda}_M$, where (for $b^+_2 =1$)
\be
d_{\lambda} = {1\over 4} \left[ (2 \lambda)^2 - (9 - b^-_2)  \right].
\ee 
Thus, we must have 
\be
\label{more 1}
d_{\lambda} \geq 0
\ee
for there to be wall-crossings in $Z'_u$ at the monopole cusp, in the presence of a \emph{nontrivially-embedded} surface operator. 

Likewise, the only relevant term in the $q_M$-expansion of (\ref{wallcrossing mono trivial}) is $q^{- \half d_{\lama}}_M$, where (for $b^+_2 =1$)
\be
d_{\lambda'} = {1\over 4} \left[ (2 \lambda')^2 - (9 - b^-_2)  \right].
\ee 
Here, $\lambda' = \lambda - {\alpha_M \over 2} \delta_D$. In other words, we must have 
\be
\label{more 2}
d_{\lambda'} \geq 0
\ee
for there to be  wall-crossings in $Z'_u$ at the monopole cusp, in the presence of a \emph{trivially-embedded} surface operator. 

Since wall-crossings occur where there are abelian instantons corresponding to $\lambda_+ =  0$ or $\lambda'_+ =0$, as explained, it will mean that  $(\lambda)^2 < 0$ or $(\lambda')^2 < 0$, respectively.  Therefore, from (\ref{more 1}) and (\ref{more 2}), it will mean that the unique condition for there to be wall-crossings in $Z'_u$  at the monopole cusp is 
\be
\label{wall crossing condition}
8 + \sigma = 9 -b^-_2 < 0. 
\ee
As we shall see in the next section, (\ref{wall crossing condition})  is in perfect correspondence with SW theory in the presence of surface operators, where $d_\lambda$ and $d_{\lambda'}$ are the virtual dimensions of the \emph{ordinary} and ``ramified'' SW moduli spaces, respectively.

\newsection{The Complete Formulas Of The Generating Function Of The ``Ramified'' Donaldson Invariants}

In this section, we will derive the complete formulas of the generating function $Z'_D$ of the ``ramified'' Donaldson invariants of  $X$ with $b^+_2 \geq 1$ and $b_1 =0$.  To this end, we will first determine the SW contributions $Z'_{u = \pm 1}$ to $Z'_D$ of (\ref{Z = Z + Z}). Thereafter, we will use the results for $Z'_u$  in the previous section to write down the complete formulas of $Z'_D$ when the surface operator is trivially \emph{and} nontrivially-embedded.  In considering $X$ to be of ``ramified'' SW simple-type, ``ramified'' generalizations of Witten's ``magic formula''~\cite{monopoles}, will also be obtained. 

\newsubsection{The ``Ramified'' Seiberg-Witten Contributions}
 
We shall now determine the SW contributions $Z'_{u = \pm 1}$ to $Z'_D$. To do this, we will first explain how the ``ramified'' SW invariants introduced at the end of $\S$3.2, and discussed in~\cite{Appendix 1}, play a part in the computation of  $Z'_{u =  1}$.  Then, by using the wall-crossing formulas  derived in the previous section for $Z'_u$ at the monopole cusp, we will proceed to determine some universal functions that will lead us to the precise formulas of $Z'_{u=1}$ when the surface operator is trivially and nontrivially-embedded. Finally, by utilizing a discrete symmetry on the $u$-plane, we will obtain the corresponding precise formulas of $Z'_{u=-1}$ at the dyonic point.   

\vspace{0.4cm}
\noindent{\it The Supersymmetric Localization Of $Z'_{u=1}$ And The ``Ramified'' SW Invariants}
 
 \def\dsw{{ d^\lamad_{\rm sw}}}
 
  \def\dswu{{ d^\lama_{\rm sw}}}
 
 \def\lamad{{\lambda'_d}}
 
 Recall that $Z'_u$ is the low-energy counterpart of $Z'_D$ at the point $u=1$; in other words, it will be given by the corresponding correlation function in the ``ramified'' SW theory with topological action $S_{u=1}$ in (\ref{Su=1})  (rescaled by $1/4$ with a sign change in the $\eta$-term, and with couplings  to gravity of the sort in (\ref{coupling to gravity terms}) included). 
 
 As explained in $\S$3.2, there is a preferred ``magnetic'' frame at $u=1$; as such, the effective gauge coupling is weak, and the correlation function $Z'_{u=1}$ can be taken to localize onto supersymmetric configurations that minimize the action. (Note that unlike the earlier case of $Z'_u$ -- which is defined over a generic region in the $u$-plane where a universal equivalence of all frames related to one another via modular transformations forbids a preferred coupling regime -- there is no physical inconsistency in our present assumption, as emphasized earlier.)  Consequently, methods from topological field theory are applicable, and as explained in $\S$3.2, any non-vanishing correlation function will reduce to an integral of a top-form in the corresponding SW moduli space $\CM^{\lamad}_{\rm sw}$ (where $\lamad$ is proportional to $x'$ in $\S$3.2 when $X$ is Spin). For $b_1 =0$, the only operator that can be interpreted as a differential form in $\CM^{\lamad}_{\rm sw}$ is, according to (\ref{a_d}), $J^d_0 (p) =  a_d$, and it corresponds to a two-form in  $\CM^{\lamad}_{\rm sw}$. In other words, any non-vanishing correlation function in the (topological) ``ramified'' SW theory -- in particular, $Z'_{u=1}$ --  will depend on the ``ramified'' SW invariant
 \be
 \label{SW inv}
 SW_{\lamad}  = \int_{\CM^{\lamad}_{\rm sw}} (a_d)^{d^\lamad_{\rm sw}/2},
 \ee 
where $\dsw$ is the (virtual) dimension of the ``ramified'' SW moduli space given by
\be
\dsw = - {{2\chi + 3 \sigma} \over 4} + (\lamad)^2.
\ee 
(See~\cite{Appendix 1} for its derivation). Note that $a_d$ in (\ref{SW inv}) is really the vacuum expectation value of the scalar field $a_d$, and it is expressed purely in terms of zero-modes, as emphasized in $\S$3.2.  

\bigskip\noindent{\it A ${\rm Spin}^c$-Structure}

\def\spin{{{\rm Spin}^c}}

As reviewed in~\cite{Appendix 1}, and demonstrated through an \emph{a priori} path-integral computation in~\cite{mine}, when $w_2(X) \neq 0$, the ``magnetic'' $U(1)$ gauge theory with ``ramified'' field strength ${F^d_L}'$ involves a ${\rm Spin}^c$-structure which can be thought of as a choice of the complex vector bundle $S_+ \otimes L'_d$, where $L'_d$ is the corresponding effective $U(1)$ bundle of the gauge theory; the monopoles $M$ are charged under the gauge field associated with $L'_d$ and are in fact sections of $S_+ \otimes L'_d$. As briefly mentioned in the previous section, a pertinent feature of such an involvement of a $\spin$-structure is that $\lamad = \half c_1({L'_d}^{\otimes 2})$ is now a half-integral class; specifically, $\lamad = -\half w_2(X) + \Gamma$, where $\Gamma = H^2(X, \IZ)$ as usual. We shall call $\lama$ the first Chern class of the $\spin$-structure. 

To avoid cluttering the formulas, we will henceforth drop the above subscript ``$d$''; all relevant quantities in this subsection are understood to be in the ``magnetic'' frame. Likewise, the above ``$L$'' subscript will also be dropped, with the understanding that all discussion in this subsection pertains to a $U(1)$ theory.     

\bigskip\noindent{\it Wall-Crossing} 

As explained in~\cite{Appendix 1},  jumps in the `ramified'' SW invariants can occur when one moves in the space of metrics on $X$ with $b_1 =0, \, b^+_2 =1$,  and encounters a ``ramified'' abelian instanton $\lambda'_+ =0$. Notice from (\ref{SW1}) that such a condition is tantamount to setting $M=0$, that is, there are no (massless) monopoles where the invariants jump by $\pm 1$ in the ``ramified'' SW theory at $u=1$. Since $Z'_u$ does not vanish when $b_1=0, \, b^+_2 =1$,  and since jumps in $Z'_u$ also occur when one meets a ``ramified'' abelian instanton $\lambda'_+ =0$, it will mean that the contributions from $Z'_{u=1}$ and $Z'_u$ are indistinguishable where wall-crossings of the ``ramified'' invariants  at $u=1$ are concerned. This is consistent with our earlier assertion in (\ref{delta u = delta SW}) that $\Delta_{SW} = \Delta_{u=1} + \Delta_{u=-1}$ (since a similar story holds for the dyonic theory at $u=-1$). In particular, one can equate the wall-crossing formulas of $Z'_u$ at the monopole cusp $\tau =0$ to the wall-crossing formulas of $Z'_{u=1}$ that we will be presenting later in this subsection.  

In the case that $b_1 =0, \, b^+_2 =1$, we will have 
\be
\dswu = - 2 - {\sigma \over 4} + (\lama)^2. 
\ee
For $\dswu < 0$, $\CM^{\lama}_{\rm sw}$ is generically empty and  $SW_{\lama}$ will vanish; in other words, we must have $\dswu \geq 0$ for $Z'_{u =1}$ to be non-zero. Hence, together with the fact that $(\lama)^2 < 0$ at a ``wall'', the condition for wall-crossings to occur can be stated as
\be
8 + \sigma = 9 - b^-_2 < 0. 
\ee 
This condition is exactly the condition (\ref{wall crossing condition}) for wall-crossings to occur in $Z'_u$ at the monopole cusp, consistent with our above discussion. 

\bigskip\noindent{\it Effective Interactions With Gravity}

From $\S$3.2, we find that the action of the ``ramified'' SW theory at $u=1$ -- with the effective couplings to gravity included -- ought to be given by
\begin{eqnarray}
S_{u=1} & = & { \{ \cq_{\textrm{sw}} , W \} \over e_d^2}   +\int_X\left (c(u)F'\wedge F' +p(u)\Tr R \wedge R + l(u) \Tr R\wedge \widetilde R + \cdots \right), 
\label{SW lagrangian}
\end{eqnarray}
where $c(u)$, $p(u)$ and $l(u)$ are holomorphic functions in $u$. The additional couplings to gravity involving $c, \, p$ and $l$ are analogous to the terms in (\ref{coupling to gravity terms}) for the pure $U(1)$ theory defined over a generic region in the $u$-plane. Nevertheless, it should be stressed that although these terms in $c, \, p$ and $l$ also result in a factor of the form (\ref{factor}) after exponentiation, they are \emph{not} the same in $Z'_{u}$ and $Z'_{u=1}$. This is because the couplings to gravity here are those of a $U(1)$ theory with massless monopoles, while the factor in (\ref{factor}) arises from couplings to gravity of the $U(1)$ theory over a generic region in the $u$-plane in which the massive monopoles are integrated out. Moreover, the term $\int_X c(u)F'\wedge F'$ due to the appearance of massless monopoles is also absent in (\ref{coupling to gravity terms}). On the other hand, the term $ {i \over 4} \int_X F' \wedge w_2(X)$ in (\ref{coupling to gravity terms}), which seems to be absent in (\ref{SW lagrangian}), actually manifests itself as a $\spin$-structure  in the ``ramified'' SW theory; together with the $\eta$-term, it will generate an important topological term within the above ellipses, which measures the first Chern class of the $\spin$-structure over $\tD$~\cite{mine}. We will elaborate on this topological term shortly, but for now, it suffices for us to note that the couplings involving $c,  \, p$ and $l$ will result in the following factor in $Z'_{u=1}$:
\be
C(u)^{(\lama)^2/2}P(u)^{\sigma/8} L(u)^{\chi/4} ,
\ee 
where the functions $C, \, P$ and $L$ are essentially proportional to the exponentials of $c,\, p$ and
$l$.

\bigskip\noindent{\it The Low-Energy Observable}

We now come to the low-energy SW observable that one should insert in the path-integral in order to define $Z'_{u=1}$. Since the canonical descent procedure is scale and duality-invariant, we can use it to obtain the SW observable in exactly the same way the observable of the pure $U(1)$ theory over a generic region in the $u$-plane was obtained in $\S$4.2. 

Recall that what we would like to ascertain is the low-energy SW counterpart to the microscopic correlation function ${\bf Z}'_{\xi, \bar g}(p,S) =  \langle e^{p I'_0 +  I'_2(S)} \rangle_{\xi}$,  with $SO(3)$ gauge bundle $E$, and fixed $\xi = w_2(E)$. The microscopic operator ${\rm exp}({p I'_0})$ maps to the operator ${\rm exp} (2 p u)$, regardless. For $b_1 =0$,  the microscopic operator ${\rm exp}({p I'_2(S)})$ maps to $ {\rm exp} \left[   {\widetilde {\cal I}}_2(S)  + S^2 T^d_{S} (u) + \tD_d^2 T^d_\tD(u)  \right]$, where ${\widetilde {\cal I}}_2(S)  =  {i \over 4 \pi} \int_{S}\left[{d u \over  d a_d} (F'_-   +   K_+)  \right]$,\footnote{Notice that there is a sign difference in the expressions for ${\widetilde {\cal I}}_2(S)$ here and in (\ref{I-sigma}). This is due to the subtle fact that $h$ is a modular form of weight 1, and thus, in the ``magnetic'' frame, one has $du_M / da_d = - 1 / h_M( \tau_d)$, while in the ``electric'' frame, one has $du / da = 1 / h(\tau)$.} and $\tD_d  = i \pi \alpha_d  D /2$. Once again, we would like to remind the reader that all quantities  here are expressed in the ``magnetic'' frame whether or not a ``$d$'' label has been appended. 

Note that from the supersymmetric variations of the fields in the ``ramified'' SW theory, we have the relation $\delta \chi = i(F'_+ - K_+)$. Since we can localize onto supersymmetric configurations in this theory, we can substitute the condition implied by $\delta \chi =0$ into the expression for ${\widetilde {\cal I}}_2(S)$. Then, since $\alpha_d$ is equivalent to zero (via an integral shift) when $D \cap D\neq 0$, we have, in the presence of a nontrivially-embedded surface operator, the map from high to low energy observables: 
\be
\label{7.7}
{\rm exp} ( p I'_0 +  I'_2(S) ) \to {\rm exp} \left( 2 p u +  {i \over 4 \pi} \int_{S} {d u \over  d a_d} F + S^2 T^d_S(u) \right).
\ee  
On the other hand,  if the surface operator is trivially-embedded, that is, $D \cap D =0$, then $\alpha_d = - \eta$ cannot be set to zero.  Nevertheless, the high-to-low energy map of observables would still be given by (\ref{7.7}): this is so because $S \cap D =0$, and therefore, the extra contribution involving the non-zero quantity $\alpha_d \delta_D$ in the second term on the RHS of the map, vanishes. However, it should be emphasized that unlike in the nontrivially-embedded case, the field strength $F$ \emph{does} contain a singularity proportional to $\alpha_d \delta_D$ along $D$ in the trivially-embedded case.

An important distinction between the above observables and those in the pure $U(1)$ theory over a generic region in the $u$-plane, is that the observables here \emph{can} be expressed purely in terms of zero-modes. This is because one can employ a semiclassical approximation to ``integrate out'' the fluctuating modes  (as was done for the non-abelian microscopic theory in $\S$2.5) since we are in a unique and \emph{weakly-coupled} (``magnetic'') frame. It is through these observables that $Z'_{u=1}$ will depend on the (``ramified'') SW invariants of (\ref{SW inv}), as mentioned earlier.

\bigskip\noindent{\it The Form Of $Z'_{u=1}$} 

We are now ready to determine the explicit formula of $Z'_{u=1}$, which, for a \emph{nontrivially-embedded} surface operator, is given by the correlation function
\be
Z'_{u=1} = \sum_{\lambda} \,    \langle   e^{\{2 p u +  {i \over 4 \pi} \int_{S} {d u \over  d a_d} F + S^2 T^d_S(u) \}}  \rangle_{\xi,  \eta_d, \lambda},   
\ee
where $\lambda$ is an \emph{ordinary} (first Chern class of the) $\spin$-structure (since  $\alpha_d$ is equivalent to zero). As required of any path-integral computation, we have summed over all possible $U(1)$-bundles and hence, $\lambda$'s, in the above expression. Nevertheless, we have held $\xi$ and $\eta_d = \alpha$ to be fixed. This correlation function is to be evaluated with respect to the action $S_{u=1}$ in (\ref{SW lagrangian}). (The computation in the case of a trivially-embedded surface operator is similar; as such, it would suffice for us to focus on the nontrivially-embedded case. We will however, present the formulas of all cases at the end.) 

After putting all the factors together, and incorporating the various terms in the ellipses of (\ref{SW lagrangian}) that were determined via a path-integral computation in~\cite{mine}, we get 
\begin{eqnarray}
\label{7.9}
\langle   e^{\{2 p u +  {i \over 4 \pi} \int_{S} {d u \over  d a_d} F + S^2 T^d_S(u) \}}  \rangle_{\xi, \eta_d, \lambda} & = & 
\int_{\CM^{\lambda}_{\rm sw}} 2e^{2i\pi(\lambda_0\cdot\lambda+\lambda_0^2)} e^{2 \lambda[\tD]} \exp\left(
2pu + {i\over  4\pi} \int_S {du\over da_d}F
+S^2 T^d_S(u)\right) \nonumber \\ 
&& \cdot C(u)^{\lambda^2/2}P(u)^{\sigma/8} L(u)^{\chi/4}.
\end{eqnarray}
The factor of 2 appears here because we would eventually like to  compare our results with those found in the mathematical literature where it is not customary to divide by the order of the gauge group  (as is done physically as part of the Fadde'ev-Popov gauge fixing procedure). Also, the $\eta_d$-dependence is in $\tD = i \pi \eta_d D /2$.  

Note that the non-vanishing contributions to the above correlation function come only from terms in the integrand which contain the top-form $(a_d)^{d^{\lambda}_{\rm sw}/2}$. Thus, in order to determine the correlation function, we need to single out these terms. To do this, one must expand the integrand in powers of $a_d$ near $a_d=0$, and extract the coefficient of
$a_d^{n}$, where $n=d^\lambda_{\rm sw}/2=-(2\chi+3\sigma)/8+\lambda^2/2$. In carrying out the integration over $\CM^{\lambda}_{\rm sw}$, one will then be left with the extracted coefficient multiplied by the \emph{ordinary}  SW invariant $\int_{\CM^{\lambda}_{\rm sw}} a_d^n = SW_{\lambda}$. Consequently, we can write the correlation function as a residue:
\begin{eqnarray}
\label{SW nontrivial}
\langle   e^{\{2 p u +  {i \over 4 \pi} \int_{S} {d u \over  d a_d} F + S^2 T^d_S(u) \}}  \rangle_{\xi,  \eta_d, \lambda}  &= & SW_\lambda \cdot
{\rm Res}_{a_d=0}
{da_d\over a_d^{1-(2\chi+3\sigma)/8+\lambda^2/2}}\cdot
2e^{2i\pi(\lambda_0\cdot\lambda+\lambda_0^2)}  e^{2 \lambda[\tD]} \nonumber \\
& & \exp\left(2pu + {i\over  4\pi} \int_S{du\over da_d}F+S^2T^d_S(u)\right) \cdot
 C(u)^{\lambda^2/2}P(u)^{\sigma/8} L(u)^{\chi/4}. \nonumber \\
\end{eqnarray}
Notice that we have yet to normalize the operator $a_d$ topologically;
it is not necessary to do so at this point as any change in the normalization can be absorbed
in a rescaling of the yet-undetermined functions $C,P$, and  $L$.

\bigskip\noindent{\it Determination Of $C$, $P$ and $L$}

Let us now determine the functions $C$, $P$ and $L$. As mentioned earlier, one can do so by comparing the wall-crossing formula obtained from (\ref{SW nontrivial}) at $b^+_2 =1$, with that obtained from $Z'_u$ at the monopole cusp in (\ref{wallcrossing mono nontrivial}). As was first established in~\cite{monopoles}, for $n \geq 0$ and $b_1 =0$, the ordinary SW invariants $SW_\lambda$ jump by $\pm 1$ when crossing a ``wall'' where an abelian instanton $\lambda_+ =0$ can be found.  The sign depends on which direction one crosses the ``wall'', but it will not be necessary to keep track of this if we just want to determine the functions. 

Thus, since $\chi + \sigma =4$ in this case, in crossing a ``wall'', the change in (\ref{SW nontrivial}) will be given by
\begin{eqnarray}
\label{SW nontrivial wall-cross}
\Delta\langle   e^{\{2 p u +  {i \over 4 \pi} \int_{S} {d u \over  d a_d} F + S^2 T^d_S(u) \}}  \rangle_{\xi, \eta_d, \lambda}  &= & \pm
{\rm Res}_{a_d=0}
{da_d\over a_d^{-\sigma/8+\lambda^2/2}}\cdot
2e^{2i\pi(\lambda_0\cdot\lambda+\lambda_0^2)}  e^{2 \lambda[\tD]} \nonumber \\
& & \exp\left(2pu + {i\over  4\pi} \int_S{du\over da_d}F+S^2T^d_S(u)\right) \cdot
 C(u)^{\lambda^2/2}P(u)^{\sigma/8} L(u)^{\chi/4}. \nonumber \\
\end{eqnarray}

In order to compare the formulas, we have to take note of the following. Firstly,  $da_d/du = -h_M(\tau_d)={i \over  2} \vartheta_3\vartheta_4(\tau_d)$, as explained in footnote~24. Secondly,  $T^d_S$ is  the $q_M = {\rm exp} (2 \pi i \tau_M)$ expansion of $T^M_S$ at the monopole cusp $\tau =0$ -- in other words, $T^d_S$ coincides with $T^M_S$ from (\ref{wallcrossing mono nontrivial}). Thirdly, the ``$d$'' label in this section can be interpreted as the ``$M$'' label in (\ref{wallcrossing mono nontrivial}), and vice-versa. Therefore, in order for the formulas to agree, we must have:  
\begin{eqnarray}
\label{functions}
C & = & {a_d \over  q_d},    \nonumber \\
 P & = & - {4 \vartheta_2(\tau_d)^8 \over    h_M^4}  a_d^{-1}, \\
 L & = & -{2i  \over h_M^2}, \nonumber 
 \end{eqnarray}
where we have used the identity
\be
\label{7.13}
q_d{da_d \over  d q_d} = {1 \over  16i} {\vartheta_2^8 \over \vartheta_3
\vartheta_4}.
\ee

\bigskip\noindent{\it The Complete Formulas Of $Z'_{u =1}$}

Substituting (\ref{functions}) in (\ref{SW nontrivial}), we obtain -- in the presence of a \emph{nontrivially-embedded } surface operator -- the complete formula of $Z'_{u=1}$ as:
\begin{eqnarray}
\label{Zu=1 nontrivial}
Z'_{u=1} & = &  \sum_{\lambda} \, {SW_\lambda \over 16} \cdot e^{2i\pi(\lambda_0\cdot\lambda+\lambda_0^2)}  e^{2 \lambda[\tD]} \nonumber \\
&& \cdot {\rm Res}_{q_d=0} \left[{dq_d\over q_d} q_d^{-\lambda^2/2} {\vartheta_2^{8+\sigma} \over  a_d h_M } \left(2i {a_d \over  h_M^2 }\right)^{(\chi+\sigma)/4} \exp\left[ 2 p u_M - i(\lambda,S)/h_M + S^2 T^M_S \right] \right]. \nonumber \\
\end{eqnarray}

One can repeat the computations behind (\ref{7.9})-(\ref{7.13}) for the case of a trivially-embedded surface operator, where the SW theory is now ``ramified'' with a $\spin$-structure  $\lama = \lambda - {\alpha_d \over 2} \delta_D$. Nonetheless, as established in~\cite{Appendix 1}, for $d^\lama_{\rm sw} \geq 0$ and $b_1 =0$, the ``ramified'' SW invariants also jump by $\pm 1$  when crossing a ``wall'' where an abelian ``ramified'' instanton $\lambda'_+ =0$ can be found.  Consequently, we obtain -- in the presence of a  \emph{trivially-embedded} surface operator -- the complete formula of $Z'_{u=1}$ as:
\begin{eqnarray}
\label{Zu=1 trivial}
Z'_{u=1} & = &  \sum_{\lambda} \, {SW_{\lambda'} \over 16} \cdot e^{2i\pi(\lambda_0\cdot\lambda+\lambda_0^2)} \, (-1)^{ \{{\alpha_d \over 2} w_2(E) [D] \} } \, e^{2 \lambda[\tD]} \nonumber \\
&& \cdot {\rm Res}_{q_d=0} \left[{dq_d\over q_d} q_d^{-(\lambda - {\alpha_d \over 2} \delta_D)^2/2} {\vartheta_2^{8+\sigma} \over  a_d h_M } \left(2i {a_d \over  h_M^2 }\right)^{(\chi+\sigma)/4} \exp\left[ 2 p u_M - i(\lambda,S)/h_M + S^2 T^M_S \right] \right]. \nonumber \\
\end{eqnarray}
Here, $\alpha_d = - \eta \neq 0$, and $SW_{\lama}$ is the ``ramified'' SW invariant.  

In order to compute the above formulas explicitly, we must expand $a_d$ in terms of $q_d$. This expansion is known to be given by~\cite{Moore-Witten}
\be
a_d = -{i \over  6} \left({2 {E_2}-\vartheta_3^4-\vartheta_4^4 \over
\vartheta_3\vartheta_4} \right).
\ee
 
As yet another consistency check on our above computations, note that in the limit $(\alpha, \eta) \to (0,0)$, the formulas of $Z'_{u=1}$ in (\ref{Zu=1 nontrivial}) and (\ref{Zu=1 trivial}) will reduce to eqn.~(7.17) of~\cite{Moore-Witten} obtained by Moore and Witten for the ordinary SW contribution to the generating function of the ordinary Donaldson invariants. It is interesting however, to note that even though $\alpha_d =0$ in the nontrivially-embedded case,  the formula in (\ref{Zu=1 nontrivial}) does \emph{not} coincide with the ordinary formula obtained by Moore and Witten; it contains an extra phase factor ${\rm exp} (2 \lambda [\tD])$, which is not equal to 1 as long as $\alpha \neq 0$ (as then $\tD \neq 0$), that is, when the (microscopic) Donaldson theory is ``ramified''. This extra phase factor is a crucial ingredient in our physical proof -- to be furnished in the next section -- of KM's remarkable universal formula which expresses the ``ramified'' Donaldson invariants purely in terms of the ordinary Donaldson invariants.


 \bigskip\noindent{\it The Dyon Contribution $Z'_{u=-1}$}

 As claimed earlier, the dyon contribution $Z'_{u =-1}$ to $Z'_D$ can be obtained from $Z'_{u=1}$ via a discrete global symmetry on the $u$-plane. This discrete global symmetry in question is the (spontaneously-broken) $\bz_8$ symmetry discussed in $\S$3.1; as required, it maps $u \to -u$, and hence, the contribution at $u=1$ to that at $u=-1$. Recall that this symmetry also maps $T_S(u) \to -T_S(u)$; this just reflects the fact that the appropriate variables, parameters and modular forms at the monopole and dyon cusps are related by (cf.~(\ref{w tx}) and \cite{Marcos})
\be 
\label{relations}
u_D = -u_M,    \quad h_D = i h_M, \quad T^D_S = - T^M_S, \quad T^D_{\tD} = - T^M_{\tD}, \quad \alpha_D = - \alpha_M + 2  \eta_M, \quad \eta_D = \eta_M,
\ee 
where $(\alpha_M, \eta_M) = (-\eta, \alpha)$. 

On a flat manifold such as ${\bf R}^4$,  one can obtain $Z'_{u=-1}$  from $Z'_{u=1}$ by making the above substitutions; in other words, $Z'_{u=-1}(p, S, \tD, \alpha_D, \eta_D) = Z'_{u=1}(-p, - iS, \tD, \eta + 2  \alpha, \alpha)$. However, on a curved four-manifold, there is further breaking of the discrete global symmetry by gravitational instantons. Consequently,  we will have $Z'_{u=-1}(p, S, \tD, \alpha_D, \eta_D) = f \cdot Z'_{u=1}(-p, - iS, \tD, \eta + 2  \alpha, \alpha)$, where $f$ is an additional factor that is not equal to 1 on a curved four-manifold.  

Let us determine what $f$ is. Firstly, note that the left and right-moving massless fermions $\varrho$ and $\bar \varrho$ in the underlying effective $U(1)$ theory have $R$-charge $1$ and $-1$, respectively; hence, under the $\bz_8$ symmetry, $\varrho \to e^{i\pi / 4} \lambda$ and $\bar \varrho \to e^{-{i\pi / 4}} \bar \lambda$. Secondly, note that the net number of $\varrho$ minus $\bar \varrho$ zero-modes is (after twisting) given by $- (\chi + \sigma) / 2$; consequently, the path-integral measure must transform with a factor $[e^{i \pi / 4}]^{(\chi + \sigma) / 2} = i^{(\chi + \sigma) / 4}$ since any non-zero correlation function ought to be invariant under the discrete global symmetry. Thirdly, since this additional factor arises purely out of gravitational effects, it does not depend on whether the field strength is ``ramified'' or not; in particular, it arises in both the nontrivially and trivially-embedded cases. Finally, note that for our results to agree with standard mathematical conventions in Donaldson theory in the ordinary limit $(\alpha, \eta) \to (0,0)$, we need to include an additional factor of $e^{-2 \pi i \lambda^2_0}$. With all these four points understood, we find that the factor $f$ must be such that 
\be
\label{relation to dyon Z}
Z'_{u= -1} (p, S, \tD, \alpha_D, \eta_D) =  i^{\{(\chi + \sigma) / 4 - w_2(E)^2 \} } Z'_{u=1} (-p, - iS, \tD, \eta + 2  \alpha, \alpha).
\ee

\newsubsection{The Complete Expressions For The Generating Function Of The ``Ramified'' Donaldson Invariants}

We are now finally ready to present the complete expression for $Z'_D$ -- the generating function of the ``ramified'' Donaldson invariants of a smooth, compact, oriented four-manifold $X$ with an embedded two-surface $D$ of genus $g$.  

\bigskip\noindent{\it The Formula Of $Z'_D$ In The Case Of A Nontrivially-Embedded Surface Operator}

Consider a nontrivially-embedded surface operator $D$ of genus $g$ and non-zero self-intersection number $D\cap D$ in $X$. Let $X$ be such that $b_1=0$ and $b^+_2 =1$. Then, 
\begin{eqnarray}
\label{zd1-a}
Z'_D & =  &  \int_{\Gamma^0(4) \backslash {\cal H} } {dx dy \over  y^{1/2}} \mu(\tau) e^{2 p u + S^2 \hat T_S(u) + \tD^2 T_\tD (u)} \Psi' + Z'_{u=1} (q_M, p, S, \tD, 0, \alpha)   \nonumber \\
&& \qquad +  i^{\{(\chi + \sigma) / 4 - w_2(E)^2 \} }Z'_{u=1} (q_M, -p, -iS, \tD, 2\alpha, \alpha).
\end{eqnarray} 
Here, $\hat T_S = T_S +{1 \over  8 \pi y} \bigl({du \over  da}\bigr)^2$; and $\mu(\tau)$, $T_S(u)$, $T_{\tD}$, $\Psi'$ and $Z'_{u=1}$, are given in (\ref{measure factor}), (\ref{TSu}), (\ref{TDu}), (\ref{Psi'}) and (\ref{Zu=1 nontrivial}), respectively. (Note also that $ q_M = q_d$, as pointed out earlier.)

For $X$ with $b_1 =0$ and $b^+_2 > 1$, the ``ramified'' $u$-plane integral $Z_u$ vanishes, and we have
\begin{eqnarray}
\label{zd2-a}
Z'_D & =  &  Z'_{u=1} (q_M, p, S, \tD, 0, \alpha)  +  i^{\{(\chi + \sigma) / 4 - w_2(E)^2 \} }Z'_{u=1} (q_M, -p, -iS, \tD, 2\alpha, \alpha).
\end{eqnarray} 
As clearly indicated in the above formulas of $Z'_D$ in (\ref{zd1-a}) and (\ref{zd2-a}), the ``dyonic'' field strength  at $u=-1$ is ``ramified'' while the ``magnetic'' field strength at $u=1$ is not: $\alpha_M =0$ while $\alpha_D = 2 \alpha$.  Nevertheless, as shown in~\cite{Appendix 1}, the orientations of the ``ramified'' and ordinary SW moduli spaces both depend on a choice of orientation of $H^1(X, \IR) \oplus H^{2,+}(X, \IR)$, which, in turn, defines the sign of the integration measure of the ``ramified'' $u$-plane integral and the ``ramified'' Donaldson invariants generated by $Z'_D$. This observation is reflected in the above formulas of $Z'_D$ which are therefore physically and mathematically consistent.

\bigskip\noindent{\it The Formula Of $Z'_D$ In The Case Of A Trivially-Embedded Surface Operator}

Consider a trivially-embedded surface operator $D$ of genus $g$, where $D\cap D = 0$. Let $X$ be such that $b_1=0$ and $b^+_2 =1$. Then, 
\begin{eqnarray}
\label{zd1-b}
Z'_D & =  &  \int_{\Gamma^0(4) \backslash {\cal H} } {dx dy \over  y^{1/2}} \mu(\tau) e^{2 p u + S^2 \hat T_S(u)} \Psi' + Z'_{u=1} (q_M, p, S, \tD, -\eta, \alpha)   \nonumber \\
&& \qquad +  i^{\{(\chi + \sigma) / 4 - w_2(E)^2 \} }Z'_{u=1} (q_M, -p, -iS, \tD, \eta +2\alpha, \alpha).
\end{eqnarray} 
Here, $Z'_{u=1}$ is given in (\ref{Zu=1 trivial}).    

For $X$ with $b_1 =0$ and $b^+_2 > 1$, we have
\begin{eqnarray}
\label{zd2-b}
Z'_D & =  &  Z'_{u=1} (q_M, p, S, \tD, -\eta, \alpha)  +  i^{\{(\chi + \sigma) / 4 - w_2(E)^2 \} }Z'_{u=1} (q_M, -p, -iS, \tD, \eta + 2\alpha, \alpha). \nonumber \\
\end{eqnarray} 
Note also from the above formulas of $Z'_D$ in (\ref{zd1-b}) and (\ref{zd2-b}), that \emph{both} the ``magnetic'' field strength at $u=1$ and the ``dyonic'' field strength  at $u=-1$ are ``ramified'': $\alpha_M =-\eta$ while $\alpha_D = \eta + 2 \alpha$.   

Last but not least, it should also be pointed out that in the monopole contribution $Z'_{u=1}$, one ought to sum over $\lambda \in -w_2(X) /2 + \Gamma$, while in the dyon contribution $Z'_{u=-1}$, one ought to sum over $\lambda \in w_2(X) /2 + w_2(E) + \Gamma$: this is because at $u=1$ and $u=-1$,  the theory is coupled to the ``magnetic'' and ``dyonic'' field strengths, where the appropriate (local) variables are $a_d$ and $a - a_d$, respectively.

\newsubsection{``Ramified'' Generalizations Of Witten's ``Magic Formula''}

Let us now consider $X$ with $b_1=0, \, b^+_2  > 1$, to be of ``ramified'' SW simple-type; that is, for all  the finite number of \emph{basic classes} $\lama$ such that $SW_{\lama} = SW(\lama) \neq 0$,  we have $\dswu =0$~\cite{Appendix 1}.  Then, $Z'_{u=1}$ in (\ref{Zu=1 trivial}) simplifies considerably, since we can replace $(\lama)^2/ 2$  with $(\chi + \sigma) /4 + \sigma / 8$ in the residue, whence only the leading terms in the $q_M$-expansion contribute non-vanishingly. Moreover, since the ordinary limit can also be understood as having a non-zero $\alpha$ or $\eta$ which takes values in $4\IZ$, a four-manifold $X$ of ``ramified'' SW simple-type is necessarily of ordinary SW simple-type also. In other words, the simplification applies to all cases.    

\smallskip\noindent{\it The Formula For A Nontrivially-Embedded Surface Operator}

Using $u_M = 1 + \dots$, $T^M_S = 1/2 + \dots$, $T^M_\tD =  1/4 + \dots$,  $h_M = 1/ (2i) + \dots$ and $a_d = 16 i q_d + \dots$, and noting the relations in (\ref{relations}), we find that for a nontrivially-embedded surface operator,  (\ref{zd2-a}) can be written as 
\begin{eqnarray}
\label{magic 1}
{\bf Z}'_{\xi, \bar g}(p,S) & = & 2^{1 + {7\chi \over 4} + {11 \sigma \over 4} } e^{-\tD^2 /2} \left \{ \sum_{\lambda \in -\half w_2(X) + \Gamma}  SW(\lambda)  e^{2i\pi(\lambda_0\cdot\lambda+\lambda_0^2)}  e^{2 p  + (S + \tD)^2 /2} e^{2(S + \tD, \lambda)}  \right.  \nonumber \\
&& \left.  \hspace{-0.5cm} + i^{\{(\chi + \sigma) / 4 - w_2(E)^2 \} } \hspace{-1.0cm} \sum_{\lambda \in \half w_2(X) + w_2(E) + \Gamma} \hspace{-0.5cm} {SW\left(\lambda - \alpha\delta_D\right)}  e^{2i\pi(\lambda_0\cdot\lambda+\lambda_0^2 + \alpha^2 D^2 /2)}  e^{-2 p  - (S +  \tD)^2 /2} e^{-2i(S + i \tD, \lambda)} \right \}, \nonumber \\
\end{eqnarray}
where $\lambda$ in the first term on the RHS of (\ref{magic 1}) is an \emph{ordinary} (first Chern class of the) $\spin$-structure. In computing the above formula, we have noted that in the dyon contribution, there is a non-vanishing observable term ${\rm exp}(\tD^2_DT^D_{\tD})$ in the integrand, where $\tD_D = i \pi \alpha_D D /2$; and from (\ref{relations}), we have $\alpha_D = 2 \alpha$ (since $\alpha_M =0$ for a nontrivially-embedded surface operator, as explained earlier). We have also appealed to the fact that $\alpha (D \cap D)$ is actually gauge-equivalent to an (even) integer as required of the condition (\ref{intersection number}).

\bigskip\noindent{\it The Formula For A Trivially-Embedded Surface Operator}

On the other hand, for a trivially-embedded surface operator, we find that (\ref{zd2-b}) can be written as 
\begin{eqnarray}
\label{magic 2}
{\bf Z}'_{\xi, \bar g}(p,S) & = & 2^{1 + {7\chi \over 4} + {11 \sigma \over 4} } \left \{ \sum_{\lambda \in -\half w_2(X) + \Gamma}  {SW({\lambda + {\eta \over 2} \delta_D})}  e^{2i\pi(\lambda_0\cdot\lambda+\lambda_0^2)} (-1)^{ \{{\eta \over 2} w_2(E) [D] \} }   e^{2 p  + S^2 /2} e^{2(S + \tD, \lambda)}  \right.  \nonumber \\
&& \left.  \hspace{-2.5cm} + i^{\{(\chi + \sigma) / 4 - w_2(E)^2 \} } \hspace{-1.0cm} \sum_{\lambda \in \half w_2(X) + w_2(E) + \Gamma} \hspace{-0.5cm} {SW\left(\lambda - \left({\eta \over 2} + \alpha\right) \delta_D\right)}  e^{2i\pi(\lambda_0\cdot\lambda+\lambda_0^2)}  (-1)^{ \{( \alpha +{\eta \over 2}) w_2(E) [D] \} }  e^{-2 p  - S^2 /2} e^{-2i(S + i \tD, \lambda)} \right \}. \nonumber \\
\end{eqnarray}
Eqns.~(\ref{magic 1}) and (\ref{magic 2}) above are just ``ramified'', $G =SO(3)$ generalizations of Witten's celebrated ``magic formula'' for the ordinary case with  $G = SU(2)$ of~\cite{monopoles}.  Note that in the ordinary limit $(\alpha, \eta) \to 0$, and for $G=SU(2)$, that is, for $w_2(E) =0$,  (\ref{magic 1}) and (\ref{magic 2}) both reduce to
\begin{eqnarray}
\label{witten's magic}
{\bf Z}_{\xi, \bar g}(p,S) & = & 2^{1 + {7\chi \over 4} + {11 \sigma \over 4} } \hspace{-0.5cm} \sum_{\lambda \in \half w_2(X) + \Gamma}  e^{2i\pi(\lambda_0\cdot\lambda+\lambda_0^2)} \left[ e^{2 p  + S^2 /2} e^{2(S, \lambda)}  + i^{\Delta} e^{-2 p  - S^2 /2} e^{-2i(S, \lambda)} \right] {SW(\lambda)}, \nonumber \\
\end{eqnarray}
where $\Delta = (\chi + \sigma) / 4$. By identifying $\lambda$, $x$ and $n_x$ in eqn.~(2.14) of~\cite{monopoles} with $p$, $2\lambda$ and $SW(\lambda)$ above, respectively, we find that (\ref{witten's magic}) coincides \emph{exactly} with Witten's ``magic formula'', as it should. 

\bigskip\noindent{\it A Generalized Simple-Type Condition}

Let us consider $X$ to have $b_1=0$ and $b^+_2 > 1$, so that $Z'_D = Z'_{u=1} + Z'_{u=-1}$. Recall from footnote~19 that we have a correspondence between the operator $({\partial \over \partial p^2} -4)$ and $(u^2 -1)$ with regard to their action on $Z'_D$.  Note also that the formulas (\ref{Zu=1 nontrivial})-(\ref{Zu=1 trivial}), and their ``dyonic'' counterparts at $u=-1$, entail expansions in powers of $(u-1)$ (or $a_M$) and $(u+1)$ (or $a_D$), in which one only ``sees'' terms of order at most $(u-1)^n$ and $(u+1)^m$, respectively, where $m$ and $n$ are certain half-integers. This means that $(u^2 -1)^r = (u-1)^r(u+1)^r$ act by zero on $Z'_D$, if $r$ is greater than ${\rm max} \,\{m,n\}$. In other words, we have the condition:
\be
({\partial^2 \over \partial p^2} -4)^r Z'_D = 0,
\ee 
for $r$ greater than ${\rm max} \, \{m,n\}$. Hence, the manifold is of generalized simple-type in the sense formulated by KM for the ordinary invariants. 

In the case of a \emph{nontrivially-embedded }surface operator, we have 
\be
m = {\lambda^2 \over 2} - {(2 \chi + 3 \sigma) \over 8}, \qquad n = {(\lambda - \alpha \delta_D)^2 \over 2}  - {(2 \chi + 3 \sigma) \over 8}.
\ee
On the other hand, in the case of a  \emph{trivially-embedded}  surface operator, we have 
\be
m = {(\lambda + {\eta \over 2}\delta_D)^2 \over 2} - {(2 \chi + 3 \sigma) \over 8}, \qquad  n = {(\lambda - ({\eta \over 2} + \alpha) \delta_D)^2 \over 2}  - {(2 \chi + 3 \sigma) \over 8}.
\ee

\newsection{Physical Proofs Of Various Seminal Results In Four-Dimensional Geometric Topology} 

As a check and an application of the formulas computed in the previous section, we will now present some elegant physical proofs of various seminal results in four-dimensional geometric topology obtained by KM in a series of important papers~\cite{structure, structure 1, KM1, KM2}.

\newsubsection{The ``Ramified'' Donaldson Invariants Are The Ordinary Donaldson Invariants}

Let us consider $X$ to be of SW simple-type with $b_1=0$ and odd $b^+_2 \geq 3$, such that $(\chi + \sigma) /2 = b^+_2 - b_1 +1$ is an even integer. Let us take $D$ to be a \emph{nontrivially-embedded} two-surface in $X$ of genus $g \geq 2$ and self-intersection number $D^2 > 0$. In this case, we have, from (\ref{magic 1}), the following relation: 
\begin{eqnarray}
\label{magic 1a}
{\bf Z}'_{\xi, \bar g}(p,S) & = &  2^g \cdot 2^{1 + {7\chi \over 4} + {11 \sigma \over 4} }\left \{ \sum_{\lambda \in -\half w_2(X) + \Gamma}  SW(\lambda)  e^{2i\pi(\lambda_0\cdot\lambda+\lambda_0^2)}  e^{2 p  + (S + \tD)^2 /2} e^{2(S + \tD, \lambda)}  \right.  \nonumber \\
&& \left.  \hspace{0.0cm} + i^{\{(\chi + \sigma) / 4 - w_2(E)^2 \} } \hspace{-1.0cm} \sum_{\lambda \in \half w_2(X) + w_2(E) + \Gamma} \hspace{-0.5cm} {SW\left(\lambda - \alpha\delta_D\right)}  e^{2i\pi(\lambda_0\cdot\lambda+\lambda_0^2 + \alpha^2D^2/2)}  e^{-2 p  - (S +  \tD)^2 /2} e^{-2i(S + i \tD, \lambda)} \right \}, \nonumber \\
\end{eqnarray}
where due to the term $e^{- \tD^2 /2}$ in (\ref{magic 1}), a factor of $2^g$ appears in (\ref{magic 1a}) after one uses the relation $2g -2 \geq D^2$ proved in (\ref{min genus}), and absorbs a field-independent real $c$-number in the overall normalization of the correlation functions. 

\bigskip\noindent{\it The ``Ramified'' Donaldson Series} 

To facilitate the comparison of our results with those in the existing mathematical literature, it will be useful to define the ``ramified'' Donaldson series.  
As in the ordinary case, the ``ramified'' Donaldson series can be written as
\be
\label{series}
{\mathscr D}' _\xi (S) = {\bf Z}'_{\xi, \bar g}(p,S) \vert_{p=0} + {1\over 2} {\partial \over \partial p} {\bf Z}'_{\xi, \bar g}(p,S) \vert_{p=0}, 
\ee
where $\xi$ is an integral lift of $w_2(E)$.  

From (\ref{magic 1a}), we thus find that 
\be
\label{proof 1}
{\mathscr D}' _\xi (S) = 2^g  \left \{ \sum_{\lambda \in \half w_2(X) + \Gamma}  a_\lambda  e^{(S + \tD)^2 /2} e^{2(S + \tD, \lambda)}  \right \},  
\ee
where 
\be
\label{proof 2}
a_\lambda = 2^{2 + {7\chi \over 4} + {11 \sigma \over 4} } e^{2i\pi(\lambda_0\cdot\lambda+\lambda_0^2)}  SW(\lambda),
\ee
in which $SW(\lambda)$ is an \emph{ordinary} SW invariant.

\bigskip\noindent{\it The ``Ramified'' Donaldson Invariants As The Ordinary Donaldson Invariants}

Let ${\mathscr D}_\xi(\Sigma)$ be the \emph{ordinary}  Donaldson series, where $\Sigma$ is any two-cycle in $H_2(X)$.  Then, from (\ref{proof 1}), (\ref{proof 2}), and (9.36) of~\cite{Marcos}, we have
\be
\label{series equal}
{\mathscr D}' _\xi (S)  = 2^g \cdot {\mathscr D}_\xi (S + \tD). 
\ee
In other words, we have an equivalence between a ``ramified'' and ordinary Donaldson series!  

Note that from our discussion in $\S$2.5, we find that the (``ramified'') Donaldson series can also be written  as
\be
{\mathscr D} _\xi (\Sigma) =   \langle \, e^{I_2(\Sigma)} \, \rangle_{\xi} + \half  \langle \, I_0  \cdot e^{I_2(\Sigma)} \, \rangle_{\xi}.
\ee
Thus, from (\ref{series equal}), we have 
\be
\label{series equal 2}
 \langle \, e^{I'_2(S)} \, \rangle_{\xi} + \half  \langle \, I'_0 \cdot e^{I'_2(S)} \, \rangle_{\xi} =  2^g  \left \{ \langle \, e^{I_2(S)} \cdot e^{I_2(\tD)} \, \rangle_{\xi} + \half  \langle \, I_0 \cdot e^{I_2(S)} \cdot e^{I_2(\tD)} \, \rangle_{\xi} \right \},
\ee
where the unprimed operators are just the operators in the ordinary theory.

Recall that the operators $I_0$ and $I'_0$ are evaluated at the \emph{same} point $p$ in $X \backslash D$, and since  their expressions are both independent of the field strengths or $D$, they actually coincide. Moreover, recall that the fermi fields appearing in (\ref{<phi>}) are expressed purely in terms of their zero-modes; that is, they are actually independent of the spacetime points $x$ and $y$.  Consequently, one can ``extract'' $I'_0$ and $I_0$ from the correlation functions in (\ref{series equal 2}) after integrating over the fermi zero-modes.  Keeping these points in mind, let us expand the exponential terms in  (\ref{series equal 2}); in doing so,  we find that (\ref{series equal 2}) can be expressed as
\be
\label{exp expand}
\hspace{-0.5cm} \sum^\infty_{ d =0}  \langle \, \{I'_2(S) \}^d \, \rangle_{\xi} / d! +  {{\mathscr I}(p)} \cdot  \langle \,  \{ I'_2(S) \}^d \, \rangle^\ast_{\xi} / d! = 2^g \sum^\infty_{ d =0}  \langle \, {\cal I} (\tD) \cdot \{I_2(S) \}^d \, \rangle_{\xi} / d! +  {{\mathscr I}(p)}  \cdot  \langle \,  {\cal I} (\tD) \cdot \{ I_2(S) \}^d \, \rangle^\ast_{\xi} / d!
\ee
where the ``$\ast$'' denotes that there are four less fermi zero-modes in the path integral measure of the correlation functions in question;
\be
 {{\mathscr I}(p)} = c \left(\int_X d^4y \, G(p-y) \right)^2
\ee 
for some real constant $c$ and Green's function $G(p-y)$; and
\be
\label{sum p}
{\cal I} (\tD) = \sum_{m =0}^{\infty}  \{I_2(\tD) \}^m / m!.
\ee
Via a term-by-term comparison of both sides of (\ref{exp expand}),  we find that
\be
\label{exp expand 1}
\langle \, \{I'_2(S) \}^{d} \, \rangle_{\xi}  = 2^g \,  \langle \, {\cal I} (\tD) \cdot \{I_2(S) \}^d \, \rangle_{\xi},
\ee
for some integer $d$.

Let us now take note of some relevant facts before proceeding further.  Let $2s$ be the dimension of the moduli space $\CM'$ of ``ramified'' $SO(3)$-instantons for a particular surface operator with parameters $(\alpha, \eta)$ and genus $g$. As explained earlier, $I'_2(S)$ has $R$-charge 2, and the total $R$-charge that the operators in the correlation function on the LHS of (\ref{exp expand 1}) must have in order for the correlation function to be non-vanishing, is \emph{exactly} $2s$; in other words,  $d$ is necessarily equal to $s$ in (\ref{exp expand 1}). On the other hand, let $2m = 2s + 2p$ be the dimension of the moduli space $\CM$ of ordinary $SO(3)$-instantons.   As explained earlier, $I_2(S)$ and $I_2(\tD)$ both  have $R$-charge 2, and the total $R$-charge that the operators in the correlation function on the RHS of (\ref{exp expand 1}) must have in order for the correlation function to be non-vanishing, is \emph{exactly} $2m$. Altogether, this means that (\ref{exp expand 1}) can actually be written as
\be
\label{proof 1-final}
\langle \, \{I'_2(S) \}^{s} \, \rangle_{\xi}  = 2^g \,  \langle \, {\cal I}_p (\tD) \cdot \{I_2(S) \}^s \, \rangle_{\xi},
\ee
where the non-local operator 
\be
\label{proof 2-final}
{\cal I}_p (\tD)  =  \{I_2(\tD) \}^p / p! 
\ee
is of $R$-charge $2p$. Thus, we have an expression relating the ``ramified'' and ordinary Donaldson invariants in (\ref{proof 1-final}). 

Via the non-negativity of $p$ as implied by (\ref{sum p}),  the formulas $2s = 8k - {3 \over 2}(\chi + \sigma) + 4 {l} - 2( g-1)$ and $2m = 8k - {3 \over 2}(\chi + \sigma)$, and the fact that $ l = - c_1(L) [D] =  w_2(E)[D] + 2 \Gamma [D]$, we have
\be
\label{proof 3-final}
p = (g-1) - w'  [D]  \,\, \, ({\rm mod} \, 2)  \quad \geq 0,
\ee
where $w' = 2w_2(E)$.  

Now recall from our discussion in $\S$2.5 that one can interpret the $R$-charge of the operators as the degree in the sense of Donaldson theory. Also recall that we have $S \in H_2(X \backslash D, \IR)$ and (in KM's mathematical convention) $\tD \in H_2(D, \IR)$.\footnote{Notice that $\tD = i \pi \alpha D /2$, and hence, $\tD \in H_2(D, \IC)$. However, recall from $\S$2.1 that we have defined our connection to be valued in the \emph{real} Lie algebra, while KM have defined theirs to be valued in the \emph{complex} Lie algebra; consequently, our expression for the connection in (\ref{connection}) differs from theirs in eqn.~(1.4) of~\cite{KM1} by a factor of $i$. Hence, in order to cast our results in KM's mathematical conventions, we must replace $\alpha$ with $i\alpha$ (and $\eta$ with $i \eta$); in doing so, we will have $\tD \in H_2(D, \IR)$, as stated. We shall do likewise when writing down all subsequent formulas which are mathematically relevant.} Hence, since we have $b_1=0$ and $b^+_2 \geq 3$ odd (so that $X$ is ``admissible'' in the sense of~\cite{structure}), and since $D^2 > 0$ and $g \geq 2$,  we find that (\ref{proof 1-final}),  supported by the relations (\ref{proof 2-final}) and (\ref{proof 3-final}), is \emph{exactly} the universal formula proved by Kronheimer and Mrowka in Theorem 5.10 of~\cite{structure}!  Moreover, notice that the coefficient of $\{I_2(\tD) \}^p$ in  ${\cal I}_p (\tD)$ is $1/p!$; this agrees \emph{perfectly} with proposition 6.3 of~\cite{structure}. In fact, since $p$ is a point in $X \backslash D$, and since $b_1=0$, the relation (\ref{proof 2-final}) also matches proposition 6.2 of~\cite{structure} \emph{precisely}: $x$ in proposition 6.2 of~\cite{structure} -- which corresponds to $p$  -- can be set to zero, as it is defined in $H_0(D)$ while $p \in H_0(X \backslash D)$; likewise, since $b_1 =0$ and we have no one-cycles in $X$,   
 $\Gamma$ in proposition 6.2 of~\cite{structure} -- which consists of products of one-cycles in $D$ -- can also be set to zero. Our physical proof of the KM universal formula is thus complete.

\newsubsection{The Generalized Thom Conjecture}

We will now proceed to give a physical proof of a refined minimal genus formula for embedded two-surfaces in $X$ -- which generalizes the celebrated Thom conjecture regarding smooth holomorphic  curves of degree $d$ being genus-minimizing in their  homology  class in ${\bf P}^2$ -- that was obtained by KM as part of their structure theorem (Theorem 1.7 of~\cite{structure}). Firstly,  for $X$ with $b^+_2 > 1$, $b^+_2 - b_1$ odd, and $b_1 =0$,  the formula can be stated as 
\begin{equation}
2 g-2  \geq D\cap D + K_r [D]
\label{theorem result 2}
\end{equation}
for all $r=1, \dots, n$, where $K_r$ is one of the $n$  ordinary basic classes for the non-vanishing Donaldson invariants, and $g$ is the genus of $D$, where $g \geq 1$ and $D \cap D \geq 0$. 

Let us comment on the above formula before proceeding further. Firstly,  notice that this is a refined version of the minimal genus formula in (\ref{inequality theorem}). Secondly, the above topological constraints on $X$ can be restated as odd $b^+_2 \geq 3$. Thirdly, we have an even value of $b^+_2 - b_1 +1 = {(\chi + \sigma) \over 2}  \geq 4$. These are the same conditions behind (\ref{inequality theorem}), whose physical proof was already given  in $\S$2.6. As such, we can read across from our earlier results, and conclude that $4l$ must lie in the range
\be
 2 D \cap D - (2g -2) \leq  4l \leq (2g -2).
 \label{inequality}
\ee
Since $g \geq 1$ and therefore $(2g -2) \geq 0$, we can infer from (\ref{inequality}) that
\be
  (2g -2) \geq D \cap D   - 2 l,
 \label{KM's result}
\ee
where ${l} = \int_D F_L / 2 \pi$. 

The structure theorem of KM involves the ordinary Donaldson invariants of $X$ of simple-type. As such, to connect with their results, we shall take $(\alpha, \eta) \to (0,0)$.\footnote{Note that the relation (\ref{inequality}), and thus (\ref{KM's result}), is still applicable in this ordinary limit: recall that in $\S$2.6, (\ref{inequality}) was derived  by considering the boundary values of $\alpha$ in the $SO(3)$ theory.} 
Consequently, our formulas (\ref{magic 1}) and (\ref{magic 2}) which relate the generating function of the ``ramified'' Donaldson invariants to the (``ramified'') SW invariants, just reduce to the ordinary one in~\cite{Moore-Witten}.  In particular, via the relation (\ref{series}), we find that the Donaldson series in this ordinary limit ought to be given by
\be
\label{structure eqn}
{\mathscr D}_w(S)  =  {\rm exp}\left({S^2 \over 2}\right)  \sum_{\lambda} \,   (-1)^{(w^2 + K_\lambda \cdot w) /2} \,  \beta_\lambda  \,  e^{(S, K_\lambda)},  
\ee
where $K_\lambda = 2 \lambda$ is an integral lift of $w_2(X)$; $w = 2\lambda_0$ is an integral lift of $w_2(E)$; and
\be
\label{beta lambda}
\beta_\lambda = 2^{2 + {7\chi \over 4} + {11 \sigma \over 4} }  SW(\lambda)
\ee 
are\emph{ non-zero} rational numbers, that is, $\lambda$ is a SW basic class. Note that (\ref{structure eqn}) coincides exactly with KM's structure theorem (Theorem 1.7 of~\cite{structure}). In particular, the $K_\lambda$'s -- which are proportional to the \emph{SW basic classes} $\lambda$  -- are also  the\emph{ Donaldson basic classes} in the sense of KM; in other words,  they are the $K_r$'s in (\ref{theorem result 2}). 

Coming back to the physics, note that the spontaneous-breaking of the microscopic $SO(3)$ gauge group to its $U(1)$ subgroup at low-energy means that the (ordinary) $SO(3)$ gauge bundle $E$ is reducible;  hence, it can be decomposed as
\be
E = {\bf R} \oplus L,
\ee
where $\bf R$ is a trivial rank-one real bundle over $X$, and $L$ is the  (ordinary) $U(1)$ gauge bundle at low-energy with field strength $F_L$. Note that $F_L$ is physically $\it equivalent$ to the field strength $F^d_L$ of the abelian SW theory, which in turn corresponds to one-half of the curvature $- 2 \pi c_1(L^2_d)$ of the determinant line bundle $L^2_d$ of the $\textrm{Spin}^c$-structure associated with a choice of the complex vector bundle $S_+ \otimes L_d$. In other words, for each (first Chern class of the) $\spin$-structure $\lambda = \half c_1(L^2_d)$,  we can identify  $-2l$  with $ 2\int_D  \lambda = 2 \lambda[D]$.  Consequently, since $K_\lambda = 2\lambda$, (\ref{KM's result}) can then be written as  
\be
(2g -2) \geq D \cap D   + K_\lambda[D]
\label{result now}
\ee
for all $\lambda$, which is just (\ref{theorem result 2}). This completes our proof.

\newsection{New Mathematical Results From The Physics}

Now that we have physically re-derived the above mathematically established theorems by KM, one might wonder if the physics can, in turn, offer any new and interesting mathematical results. Indeed it can, as we shall now show.

\newsubsection{``Ramified'' Generalizations Of The Kronheimer-Mrowka Structure Theorem}

Using the physical formulas obtained hitherto, we shall now present ``ramified'' generalizations of KM's prescient structure theorem~\cite{structure, structure 1} which led Witten to his original  ``magic formula'' for the ordinary invariants etc. These generalizations also make explicit the connection between the ``ramified'' Donaldson series and the ``ramified'' SW invariants. As before, let us consider $X$ to be admissible, that is, $b_1=0$ and $b^+_2 \geq 3$ is odd. 

\bigskip\noindent{\it The Formula For A Nontrivially-Embedded Surface Operator}

Note that (\ref{proof 1}) and (\ref{proof 2}) --  which led to our proof of KM's universal formula in (\ref{proof 1-final}) -- define the following ``ramified'' version of KM's structure theorem for when the surface operator is nontrivially-embedded with $D \cap D  > 0$ and $g \geq 2$: 
\be
\label{st1}
{\mathscr D}' _w (S) = {\rm exp}\left({S^2 \over 2}\right)  \sum_{\lambda} \, (-1)^{(w^2 + K_\lambda \cdot w) /2} \, {\beta}'_{\lambda; \alpha; D^2; g}    \, e^{(S, K_\lambda)},  
\ee
where 
\be
\label{st2}
{\beta}'_{\lambda; \alpha; D^2; g}  = 2^{2 + {7\chi \over 4} + {11 \sigma \over 4}} SW'_{\alpha; D^2; g} (\lambda),
\ee
and 
\be
\label{st3}
SW'_{\alpha; D^2; g}(\lambda) = 2^g \cdot e^{{\pi \alpha \over 8}(\pi \alpha D^2  - 4K_\lambda[D])} SW(\lambda).
\ee
Here, $SW(\lambda)$ is an \emph{ordinary} SW invariant for the ordinary $\spin$-structure $\lambda$.

By comparing (\ref{st1})-(\ref{st3}) with the ordinary structure theorem defined by (\ref{structure eqn}) and (\ref{beta lambda}), we see that we can \emph{interpret} $SW'_{\alpha; D^2; g}(\lambda)$ as a ``ramified'' SW invariant for when $D \cap D > 0$ and $g \geq 2$. Moreover, as a ``ramified'' invariant ought to, $SW'_{\alpha; D^2; g}(\lambda)$ depends on all the parameters of the surface operator, namely, $g$, $D \cap D$ and most importantly, $\alpha$.  


\bigskip\noindent{\it The Formula For A Trivially-Embedded Surface Operator}

Now consider a trivially-embedded surface operator with $D \cap D =0$ and any $g$. Then, from (\ref{magic 2}), we find that in this case, the ``ramified'' version of KM's structure theorem will be given by:
\be
\label{st1a}
  {\mathscr D}'_w(S)  =   {\rm exp}\left( {S^2 \over 2}\right) \sum_{\lambda}  \, (-1)^{(w^2 + K_\lambda \cdot w) /2}  \, b'_{\lambda; \alpha; \eta; g; w}  \,  e^{(S, K_\lambda)},   
\ee
where
\be
\label{b'}
b'_{\lambda; \alpha; \eta; g; w} = 2^{2 + {7\chi \over 4} + {11 \sigma \over 4} } \cdot  e^{ {\pi \over 2}( \eta w -  \alpha K_\lambda) [D]}  \,  SW(\lambda + {\eta \over 2} \delta_D).
\ee
Here, $\lambda$ is a ``\emph{ramified}'' $\spin$-structure, and $SW(\lambda + {\eta \over 2} \delta_D)$ is a bona-fide ``\emph{ramified}'' SW invariant with holonomy parameter $\eta$. 

The $g$-dependence of $b'_{\lambda; \alpha; \eta; g; w}$ can be traced to the ${\rm exp} (- \pi \alpha  K_\lambda[D] /2 )$ term: from the physical identification of $-2l$ with $K_\lambda[D]$  (as explained earlier), as well as the relation (\ref{KM's result}), we find that $K_\lambda[D] /2 = g + c$, for some real constant $c$.  Nevertheless, since we cannot say precisely what $c$ must be, we will leave the equations as presented above.    

\newsubsection{A Generalization Of The Kronheimer-Mrowka Universal Formula}

Let us now set the ``quantum'' $\eta$-parameter to zero in (\ref{st1a}) and (\ref{b'}); then, electric-magnetic duality of the underlying macroscopic $U(1)$-theory -- which in this case gives $(\alpha_M, \eta_M) = (0, \alpha)$ -- will imply that $\lambda$ in (\ref{st1a}) is now an ordinary $\spin$-structure -- which we shall  henceforth denote as $\bar \lambda$ -- such that the corresponding \emph{ordinary}  SW invariant will be given by
$
SW(\bar \lambda).
$
However, since we did not set the ``classical'' $\alpha$-parameter to zero, the Donaldson invariants stemming from the microscopic $SO(3)$ theory would remain ``ramified''.

In this case, since $D^2 =0$ and $(S, D) =0$, one can rewrite (\ref{st1a}) as
\be
 {\mathscr D}'_w(S)  = 2^{2 + {7\chi \over 4} + {11 \sigma \over 4} } \sum_{\bar \lambda}  \, e^{2  i \pi (\lambda_0 \cdot \bar\lambda + \lambda^2_0)}  \, e^{(S+ \tD)^2 /2} \,  e^{2(S + \tD, \bar\lambda)} \, SW(\bar \lambda). 
 \label{UF}
\ee
Using (9.36) of~\cite{Marcos}, one finds from (\ref{UF}) that  
\be
 {\mathscr D}'_w(S) =  {\mathscr D}_w(S + \tD),
\ee
where ${\mathscr D}_w(\Sigma)$ is the ordinary Donaldson series for any $\Sigma \in H_2(X)$. Thus, we have an equivalence between a ``ramified'' and ordinary Donaldson series, albeit for a trivially-embedded surface operator! 

Repeating the arguments in $\S$8.1, we conclude that (\ref{proof 1-final}) (less the factor of $2^g$), (\ref{proof 2-final}) and (\ref{proof 3-final}) must also hold in this current setting. In other words, electric-magnetic duality of a trivially-embedded ``classical'' surface operator with parameter $\alpha$ suggests that KM's universal formula ought to hold even when the two-surface is \emph{trivially-embedded} in an admissible four-manifold $X$ with $b_1=0$.

\def\tl{{\widetilde\lambda}}

\newsubsection{The ``Ramified'' Seiberg-Witten Invariants Are The Ordinary Seiberg-Witten Invariants}

 Let us now set  the ``classical'' $\alpha$-parameter to zero in (\ref{st1a}) and (\ref{b'}); then, the Donaldson invariants stemming from the microscopic $SO(3)$ theory will \emph{not} be ``ramified''.  On the other hand, if we do not set the ``quantum'' $\eta$-parameter to zero, electric-magnetic duality of the underlying macroscopic $U(1)$-theory -- which in this case gives $(\alpha_M, \eta_M) = (-\eta, 0)$  -- will imply that $\lambda$  in  (\ref{st1a}) ought to persist as a ``ramified'' $\spin$-structure -- which we shall henceforth denote as $\tl$ -- such that the corresponding \emph{bona-fide} ``ramified'' SW invariant will be given by
$
SW(\tl + {\eta \over 2} \delta_D).
$

According to our discussion on pg.~34, the integrality requirement of the class $2\tl$ is met by the fact that one can invoke a ``ramification''-preserving twisted $U(1)$-valued gauge transformation of the kind in~(\ref{twisted gauge tx}) which shifts $\alpha_M$ by a non-integer $u_M$, so that one can regard $\alpha_M (U \cap D)$  as an integer for any integral homology two-cycle $U \subset X$. Since $\alpha_M = -\eta$, it will mean that via such a gauge transformation, one can set the phase (\ref{eta term}) which appears in the path-integrals on the LHS of (\ref{st1a}) to 1, whilst leaving the gauge-invariant field strength $\tl + {\eta \over 2}\delta_D$ \emph{unchanged}; consequently, one can express the LHS of (\ref{st1a}) as the ordinary Donaldson series ${\mathscr D}_w(S)$ of (\ref{structure eqn}), so that from (\ref{st1a}) and (\ref{b'}), we have 
\be
\label{sum SW}
 \sum_{\lambda \in  \half w_2(X) + \Gamma} \,   (-1)^{(K_\lambda \cdot w) /2} \,  SW(\lambda)  \,  e^{(S, K_\lambda)} =  \sum_{\tl \in  \half w_2(X) + \Gamma}  \, (-1)^{(K_\tl - i \eta \delta_D) \cdot w /2}  \, SW(\tl + {\eta \over 2}\delta_D)  \,  e^{(S, K_\tl)}.
\ee    

Note that there is a one-to-one correspondence between $\lambda$ and $\tl$, since a ``ramified'' field strength is just the original, ordinary field strength with an extra singular component along $D$ that is fixed for a chosen effective value of $\eta$.  Therefore, we can infer from (\ref{sum SW}) that
\be
\label{SW dual}
 SW(\tl + {\eta \over 2}\delta_D)  =  (-1)^{i \eta w [D]} \, SW(\lambda),
\ee
where we have made use of the fact that for a corresponding pair $\{\tl, \lambda\}$, ${\rm exp}(S, K_{\tl}) = {\rm exp}(S, K_\lambda)$,\footnote{To see this, note that $(S, K_\tl) = 2(S, \tl) = 2(S, \lambda) - \eta(S, D) = (S,K_\lambda)$,  since $(S,D) =0$.} and  $K_\lambda - K_\tl = i \eta \delta_D$. 


Thus, from (\ref{SW dual}), we conclude that electric-magnetic duality of a trivially-embedded ``quantum'' surface operator  with parameter $\eta$ will imply that the ``ramified'' SW invariants are (up to a constant) just the ordinary SW invariants!  Hence, (\ref{SW dual}) can be viewed as a SW analog of KM's universal formula proved in (\ref{proof 1-final}), albeit for embedded two-surfaces with \emph{vanishing} self-intersection number. 

\bigskip\noindent{\it An Agreement With An Independent Mathematical Computation}

Last but not least, note that since $w$ is supposed to be an integral cohomology class, it will mean that we can express its Poincar\'e dual as an integer linear combination of  the integral homology 2-cycles $U_{i=1, \dots, b_2(X)}$ in $H_2(X, \mathbb Z)$ (which we have assumed to be torsion-free). Consequently, the above-mentioned fact that we can regard $\alpha_M (U \cap D)$ as an integer for any integral homology two-cycle $U \subset X$, means that we can regard $i \eta w [D]$ as an integer.\footnote{To arrive at this claim, we have taken note of the fact that  due to footnote~26, one has to replace $i\eta$ with $\eta$ in the present discussion; this is because the statement that $\alpha_M (U \cap D)$ can be regarded as an integer is based on the relevant Lie algebra being real.}  Therefore, via (\ref{SW dual}), we learn that if the two-surface is  trivially-embedded in a four-manifold $X$ with $b_1 =0$ and $b_2^+ \geq 3$, the ``ramified'' and ordinary SW invariants are\emph{ equal up to a sign}. Moreover, if the microscopic gauge group is $SU(2)$ -- that is, $w \in 2 \Gamma$ and hence $i \eta w [D]$ can be regarded as an \emph{even} integer in (\ref{SW dual}) -- the equivalence is actually \emph{exact}. This observation agrees with and generalizes an independent mathematical computation performed in $\S$3.3 of~\cite{Appendix 1} which demonstrates that the ``ramified'' and ordinary SW invariants of $K3$ -- which is a four-manifold with $b_1=0$ and $b^+_2 =3$ -- coincide exactly, up to a sign.

\newsection{Vanishing Of The ``Ramified'' Donaldson Invariants In Certain Chambers}

For $X$ with $b_1=0, \, b^+_2 =1$, the wall-crossing formulas for $Z'_D$ have been computed in $\S$6.2. If one can establish that $Z'_D$ vanishes in certain chambers of the forward light cone $V_+$, one can ascertain the \emph{exact }values of $Z'_D$ using the wall-crossing formula (\ref{wallcrossing Don simplified}) or (\ref{wallcrossing Don trivial-embed}), depending on whether the surface operator is nontrivially or trivially-embedded, respectively. 

\def\P{{\bf P}}

Let us, for ease of illustration, consider  $X = \P^1 \times \P^1$ and $X = \IF_1$, where $\IF_1$ is a Hirzebruch surface given by a \emph{nontrivial} $\P^1$ fibration of $\P^1$. Let us also consider $SO(3)$-bundles $E$ where $(w_2(E), \CF) \neq 0$; here, $\CF$ is the (non)trivial $\P^1$ fiber in $X$ with base $\CB = \P^1$. We will now argue that $Z'_D$ indeed vanishes in certain chambers for such \emph{rational ruled-surfaces}.  
   
\newsubsection{Vanishing Of Invariants Of $\P^1 \times \P^1$ In Certain Chambers}

\def\CB{{\cal B}} 
\def\CF{{\cal F}}

We shall now establish that $Z'_D = Z'_u + Z'_{SW}$ vanishes in certain chambers of $V_+$ when $X = \P^1 \times \P^1$. To this end, let us consider a chamber whereby the fiber $\P^1$ in $X$ has an area much smaller than the base $\P^1$. In this case, the scalar curvature of $X$ is guaranteed to be positive, and as shown in~\cite{monopoles} and~\cite{Appendix 1}, both the regular and ``ramified''  SW invariants $SW_\lambda$ and $SW_\lama$, vanish. Consequently, $Z'_{SW} = Z'_{u=1} + Z_{u=-1}$ will be zero in such a chamber (cf.~(\ref{Zu=1 nontrivial}), (\ref{Zu=1 trivial}) and (\ref{relation to dyon Z})), and the vanishing of $Z'_u$ will then imply the vanishing of $Z'_D$.   

To demonstrate the vanishing of $Z'_u$, it suffices to analyze the ``ramified'' lattice sum factor $\Psi'$ in this chamber. Note that if $\P^1 \times \P^1 = \CB \times \CF$ is given a product metric, then a harmonic two-form on $\CB \times \CF$ such as $F'$, is the sum of a pullback from $\CB$ and a pullback from $\CF$. Then, the gauge kinetic term in $\Psi'$ can be expressed as:
\be{
\int_{\CB \times \CF}  F' \wedge * F' =
{\vol(\CB) \over  \vol(\CF)} \biggl( \int_\CF F'\biggr )^2 +
{\vol(\CF) \over  \vol(\CB)} \biggl( \int_\CB F'\biggr )^2.
}\ee
Notice that in the limit  ${\vol(\CB) \over  \vol(\CF)} \rightarrow \infty$, as long as the magnetic flux $\int_\CF F'$ through $\CF$ is non-zero, the action would be singular and the resulting path-integral would vanish. This is indeed the case since $(w_2(E),\CF)\not= 0$.\footnote{Note that due to the spontaneous breaking of the microscopic $SO(3)$ gauge symmetry to a $U(1)$ gauge symmetry at low-energy, we have $c_1(L) = - F/2 \pi = w_2(E) \,\, {\rm mod} \,\, 2$, where $L$ is the $U(1)$-bundle with (singular) curvature $F$. This means that $F' / 2\pi = F /2\pi - \alpha \delta_D =  - w_2(E) \,\, {\rm mod} \,\, 1$, since $\alpha \delta_D \in \Gamma$. Thus, $(w_2(E), \CF) \neq 0$ implies that $\int_\CF F' \neq 0$.} Thus, $Z'_D$ on $X = \P^1 \times \P^1$ vanishes in such a chamber where $\vol(\CF) \to 0$.

\newsubsection{Vanishing Of Invariants Of $\IF_1$ In Certain Chambers}

\def\CH{{\cal H}}

We now turn to the case where $X = \IF_1$, which can be regarded as a blowup  $Bl_P(\P^2)$ of complex projective space at a single point, such that  the blowup produces an exceptional divisor that can be identified with $\CB$.  $X$ is obviously not a simple product space; nevertheless, the same basic idea that $Z'_u$ ought to vanish when $\vol (\CF) \to 0$, holds. (Again, since $X$ is guaranteed to have positive scalar curvature in the limit $\vol (\CF) \to 0$, $Z'_{SW}$ will vanish. In fact, the vanishing of the (``ramified'') SW invariants of  $X = \IF_1$ was explicitly demonstrated in~\cite{Appendix 1}.)

There are two natural bases for $H^2(X;\IZ)$.  One basis consists of the pair
$\langle \CF,\CB \rangle$.  In this basis, the intersection form is:
\[
\left(
\begin{array}{ccc}
   0  & \, &   1  \\
  1   & \, & -1  \\   
\end{array}
\right).
\]
Alternatively, we can introduce $\CH=\CB+\CF$, the pullback of a hyperplane class on $\P^2$.  In the basis $\langle \CH, \CB\rangle$, the intersection
form is
\[
\left(
\begin{array}{ccc}
 1 & \,  & 0  \\
 0 & \,  &  -1  \\
  \end{array}
\right).
\]
Let us choose an integral lift of $w_2(X)$ by setting $w_2(X) = \CF$. The Kahler cone is $ \{ x \CB + y \CH$: $x\leq 0 , x+y\geq 0\}$. Any Kahler metric of unit volume has a Kahler class of the form
\be
\omega_+ = \cosh \theta \CH - \sinh \theta \CB \qquad 0 \leq \theta < \infty.
\ee
Recall that the self-dual harmonic two-form $\omega_+$ is normalized so that $(\omega_+,\omega_+)=1$.
Define $ \epsilon \equiv e^{-\theta}$. We are interested
in the limit
\begin{align}
\label{limits}
\omega_+ \cdot \CF & = \epsilon \rightarrow 0 \cr
\omega_+ \cdot \CB & = \sinh \theta \rightarrow \infty
\end{align}
in which the area of $\CF$ becomes small.  This is the limit in which we expect a vanishing of $Z'_u$ to occur.

Let us take $w_2(E) = \CB$ (since $(w_2(E), \CF) \neq 0$  then). Also, there are two admissible choices of $\delta_D$ in this case -- $\CH$ or $\CB$. Let us take $\delta_D = \CH$, whereby $D \cap D > 0$. (The analysis for when $\delta_D = \CB$ is similar.) Then, $\lama$ in the effective $U(1)$ theory over a generic region in the $u$-plane can be written as  
\be
\lama = (n - {\alpha \over 2}) \CH + (m + \half) \CB, \quad {\rm with} \, \, n, m, \in \IZ. 
\ee
By noting that $\lambda'_+ = \omega_+(\omega_+, \lama)$ and $(\lama)^2 = (\lambda'_+)^2 + (\lambda'_-)^2$, we find that the gauge kinetic action of $Z'_u$ is:
\be
\label{gauge kin}
\exp\biggl[- i \pi \bar \tau
(\lambda'_+)^2 - i \pi \tau (\lambda'_-)^2\biggr] =
\exp\biggl[  - \pi y \bigl[ (n - {\alpha \over 2})^2  + (m + \half)^2\bigr]
\cosh 2 \theta - i \pi x \bigl[ (n- {\alpha \over 2})^2 - (m + \half)^2 \bigr] \biggr].
\ee
Since $n,m$ are integral, and since $\cosh 2 \theta = (\epsilon^{-2} + \epsilon^2)/2$, (\ref{gauge kin}) always leads to an exponential suppression in the limit (\ref{limits}). 

The only other metric dependence in the integrand of $Z'_u$ comes from the terms:
\be
\label{metric dep}
\exp[{1 \over  8 \pi y}S_+^2/h^2] \exp[- i (S,\lambda_-)/h]
\biggl[ -i4\pi y h(\lama,\omega_+) +   (S,\omega_+) \biggr].
\ee
To obtain (\ref{metric dep}), we have used the fact that for $D \cap D > 0$, one must set $\eta =0$; we have also used the fact that $S \cap D =0$, and $h = da /du$. Now, 
\begin{align}
(S, \omega_+) & = \half {( \epsilon^{-1} + \epsilon )} \,  S\cdot F + \epsilon \, S\cdot B, \cr 
(S, \omega_-) & =-  \half {( \epsilon^{-1} - \epsilon )} \, S\cdot F + \epsilon \, S\cdot B. 
\end{align}
Consequently, we find that in the limit $\epsilon \to 0$, $(S, \lambda_-) \sim \epsilon^{-2}$, $(S, \omega_+) \sim \epsilon^{-1}$, and $(\lama, \omega_+) \sim \epsilon^{-1}$. Thus, in expanding the exponential terms in (\ref{metric dep}), we see that to any given order in $S$, this extra metric dependence contributes at most a power of $1/ \epsilon$. This is killed off by the exponential suppression in the first term on the RHS of (\ref{gauge kin}), which --  in the limit $\epsilon \to 0$ -- vanishes much faster than some power of $1/\epsilon$ grows.  Therefore, the contribution of the integration over any compact region vanishes when $\epsilon \to 0$.

To prove the vanishing of $Z'_u$ in the given limit, pointwise vanishing of the integrand is not quite enough.  This is because the integration region is non-compact and the convergence is not uniform throughout the $u$-plane. To complete the proof, we must therefore analyze what happens at the three cusps more carefully.   

Let us first consider the cusp at $\tau = \infty$. To study the behavior near the cusp, we can replace the integral over the fundamental domain $\CF$ by the integral over the strip $-\half\leq x \leq +\half, y\geq 1$. We shall focus on a term giving a fixed power $p^l S^r \tD^m$. The expression (\ref{gauge kin}) multiplies a modular form times a polynomial in $1/y$. After the  projection $\int dx$ the integral has an absolute value bounded above by:
\begin{align}\label{sum modular}
\sum_{n,m\in \IZ} \vert c(d(n,m, \alpha)) \vert
\sum_{N,M}
&
{ a_{N,M} (n,m,\alpha) \over
 \epsilon^{M}}
  \int^\infty_1 {d y \over  y^{1/2}}
  y^{-N}
\cr
\exp\bigl[  -2 \pi y (n - {\alpha \over 2})^2
& (\cosh \theta)^2 - 2 \pi y (m+\half)^2(\sinh
\theta)^2\bigr]
\end{align}
where the modular form is ${\cal C}(q) = \sum c(d) q^d$, $2d(n,m, \alpha) =(n - {\alpha \over 2})^2-(m+\half)^2$,   $N,M$ are nonnegative integers, and the number of terms in the sum $\sum_{N,M}$ is bounded by $r$ for the contribution to $p^l S^r \tD^m$. From  the estimate
\be
\int^\infty_1 {d y \over  y^{1/2}} y^{-N} \epsilon^{-M}
\exp\bigl[  -y {K \over \epsilon^2}   \bigr]
\sim { 1 \over  \epsilon^{M-2}} \exp\bigl[  - {K \over \epsilon^2}   \bigr]
\biggl( 1 + \CO(\epsilon^2) \biggr)
\ee
for some \emph{positive} number $K$, we can infer from (\ref{sum modular}) that $Z'_u$  at the cusp $\tau = \infty$ will vanish in the limit $\epsilon \to 0$. 

Let us now study the contribution of the monopole cusp. The modular transformation exchanges $w_2(E)$ with $- w_2(X)$. Moreover, recall that $\alpha_M = 0$ when $D \cap D > 0$. 
Then, working in the basis $\langle \CF, \CB \rangle$ for $H^2(X, \IZ)$, we have
\be
\label{mono lambda}
\lambda= n \CB + (m- \half ) \CF \qquad n,m \in \IZ
\ee
The gauge kinetic action is now
\be
\exp\biggl[
- i \pi \bar \tau \lambda_+^2 - i \pi \tau \lambda_-^2\biggr] = \exp\biggl[  - \pi y \bigl[
n^2  \cosh 2\theta + 2(m-\half)^2 e^{-2 \theta}
\bigr] - i \pi x \bigl[ n^2 - 2n (m - \half) \bigr] \biggr]
\ee
The metric dependence is as in (\ref{metric dep}) with $h \rightarrow h_M$, $\lama \to \lambda$ etc. For $n\not= 0 $, only a finite number of terms contribute to the integral and the argument is identical to that used for the cusp at $\tau = \infty$. However, for $n=0$, the entire
sum on $m$ survives the projection by $\int dx$. Before we proceed further, note that we have the estimate:
\be
\label{est}
\sum_{m\in\IZ} e^{-2 \pi y \epsilon^2 (m-\half)^2}
e^{ i \pi (m-\half)(1-\alpha)} (m - \half)^t
\sim \sum_k  a_k (\alpha)  \bigl( {1 \over  \epsilon^2 y} \bigr)^{k+ t +1/2} e^{-\pi/(8
\epsilon^2 y) },
\ee
where the $a_k(\alpha)$'s are constants that depend on $\alpha$.

Working at some fixed order in $S$ (and $\tD$), and noting that $h \sim 1/ 2i$ at the monopole cusp, the integral after the projection $\int dx$ and in the limit $\epsilon \to 0$, becomes a finite sum of terms proportional to:
\begin{align}
\label{com integral}
\int_1^\infty {dy \over  y^{3/2} } \bigl[ \tilde \mu(\tau_M) \bigr]_{q_M^0}
\sum_{m\in\IZ} e^{-2 \pi y \epsilon^2 (m-\half)^2}
e^{ i \pi (m-\half)(1 -\alpha)}
&
\biggl( { (S \cdot \CF)^2 \over \epsilon^{2}   y}\biggr)^{t_1}
 \cr
\biggl( {(S \cdot \CF)} (m-\half) \biggr)^{t_2} \cdot \biggl[ K_1
( S\cdot \CF)/ \epsilon
&
+ K_2 y (m-\half) \epsilon \biggr].
\end{align} 
Here $t_1, t_2$ are nonnegative integers, $\tilde \mu(\tau_M)$ is a certain modular form, and $K_1,K_2$ are constants.
But $\tilde \mu(\tau_M) \sim \vartheta_2^8/(\vartheta_3 \vartheta_4)^4 e^{2pu_M +S^2 T^M_S} \sim q_M + \cdots $ actually vanishes at the cusp, so in fact $[\tilde \mu(\tau_M)]_{{q_M^0}}$ vanishes. With this vanishing factor removed, and using the estimate (\ref{est}), we see that the rest of (\ref{com integral}) behaves, in the limit $\epsilon \to 0$, like the integral:
\be
\sum_k { A_k(\alpha) \over  \epsilon} \int_1^\infty {dy \over  y^{3/2} }
({1 \over  \epsilon^2 y})^{t_1} ({1 \over  \epsilon^2 y})^{k+ t_2+\half}
e^{-\pi/(8 \epsilon^2 y)} 
\ee
for some constant $A_k(\alpha)$. Making the change of variables $z=\epsilon^2 y$ shows that every term in the above sum is non-zero and finite as $\epsilon \rightarrow 0$. Hence, we conclude that $Z'_u$ will also vanish at the monopole cusp.  

At the dyon cusp, we have
\be
\label{dyon lambda}
\lama = n' \CB + (m' - \half) \CF,
\ee
where 
\be
n' = (n+1 - \alpha), \qquad m' = (m+1 - \alpha).
\ee
Comparing (\ref{dyon lambda}) with (\ref{mono lambda}), it is clear that the analysis for the dyon cusp follows that for the monopole cusp, albeit with $n$ and $m$ replaced by $n'$ and $m'$. However, since $n'$ (and $m'$), unlike $n$, can never be zero,  only a finite number of terms contribute to the integral such that the argument is identical to that used for the cusp at $\tau = \infty$. Thus, $Z'_u$ vanishes at the dyon cusp as well.  

In summary, we find,  after a careful analysis,  that $Z'_u$ will vanish throughout the $u$-plane when $\vol (\CF) \to 0$. Therefore, we conclude that $Z'_D$ must vanish in such a chamber. Notice that our above arguments for $X = \P^1 \times \P^1$ make no reference to the genus of the base $\CB$; as such, they are also applicable to other product rational ruled surfaces such as $X = \P^1 \times \Sigma_g$, where $\Sigma_g$ is a Riemann surface of genus $g$.  Moreover, the analysis holds for any value of $\alpha$ that is not identically zero; in particular, this means that when $\alpha \in 4 \IZ$, the corresponding generating function of the \emph{ordinary} Donaldson invariants is zero when $\vol(\CF) \to 0$. This is consistent with the fact that on any rational ruled surface whereby the fibers are very small compared to the base, the ordinary Donaldson invariants vanish since one cannot define stable bundles over it~\cite{ref 17}.

\newsection{The Blowup Formula}

Consider an arbitrary, simply-connected four-manifold $X$ with $b_1 =0, \, b^+_2 =1$ and positive scalar curvature such that $Z'_{SW}$ vanishes in certain chambers; if in addition, $X$ has $b^-_2 < 9$, there is no SW wall-crossing, and $Z'_{SW}$ will vanish for \emph{any} metric on $X$;  then $Z'_D = Z'_u$, and since  $\pi_1(X)=0$, one can use the homotopy invariance of $Z'_u$ to reduce the computation of $Z'_D$ to that of a rational ruled surface with a $\P^1$ base.  Any two such rational ruled surfaces, with any two given metrics, can be related to each other by a succession of blowups, blowdowns, and wall-crossings.  Hence, by starting with a four-manifold  $\IF_1$ or $\P^1\times \P^1$ in a chamber where $Z'_u=Z'_{SW}=Z'_D=0$, one can apply the wall-crossing  and blowup or blowdown formulas appropriately to compute the \emph{nonzero} ``ramified'' Donaldson invariants of $X$ in another chamber. An example of such an $X$ is $\P^2$, which can be obtained by ``blowing-down''  a single exceptional divisor in $\IF_1$. Another example of $X$ would be $\P^2$ ``blown-up'' at less than nine points. 

So far, we have obtained, for arbitrary four-manifolds with $b_1 =0, \, b^+_2 \geq 1$, the wall-crossing formulas for $Z'_D$ and established the vanishing of $Z'_D$ in certain chambers for  $\P^1\times \P^1$ and $\IF_1$. Therefore, let us, in this final section, ascertain the general blowup formula for $Z'_D$ of $X$, so that our results can be used to compute the ``ramified'' Donaldson invariants \emph{exactly} for some metric on $X$.

\bigskip\noindent{\it The Blowup Formula For $Z'_D$ Of $X$}

Let $\pi:\widehat X\to X$ be the blowdown map. Let $b$ be the exceptional curve contracted by $\pi$; it satisfies $b^2 = -1$, and the two-homology of $\widehat X$ is given by $H_2(X, \IZ) \oplus \langle b \rangle$. In addition, we have $b^+_2 (\widehat X) = b^+_2(X)$, $\chi(\widehat X) = \chi(X) + 1$, and $\sigma(\widehat X) = \sigma(X) -1$.  Let $I'_2(b)$ be the corresponding two-observable, and $t$ a real number. In the blowup formula, we seek to compute
\be
\langle \exp(2pu+I'_2(S)+tI'_2(b) )\rangle_{\widehat\xi,\widehat X}
\ee
in a limit in which the area of $b$ is small, in terms of
\be
\langle\exp(2pu+I'_2(S))\rangle_{\xi,X}.
\ee
As in the ordinary case, one can assume that  $\widehat\xi$ is a class that coincides with $\pi^*(\xi)$ away from $b$; this means that $\widehat\xi=\pi^*(\xi)+jb$ for $j=0$ or 1. Also, we are identifying the surfaces $S$ and $D$ in $X$ with their pullbacks to $\widehat X$; they satisfy $S \cap b = D \cap b =0$, since the class $\langle b \rangle$ is necessarily orthogonal to all classes inherited from $X$.

As in the ordinary case, let us start by explaining very generally, why a blowup formula should exist and what its general form should  be.
Firstly, let us introduce an ``impurity'' at a point in $X$.  Next, let us scale up the metric on $X$ by $\bar g\to t^2\bar g$, with very large $t$. The ``impurity'' will then be blown up with an area of $b$ (which is finite but very small) at that point. To a distant observer, it must be possible to simulate the effect of the impurity by some local, $\cq$-invariant observable.  But in the twisted $\CN=2$ theory, any local $\cq$-invariant observable (supported at a point as opposed to the $k$-form descendants) is a holomorphic function of $u$.  Together with the fact that $Z'_u$ is invariant under rescalings of the metric $\bar g$, we conclude that there must be holomorphic functions $F_j(u,t)$, for $j=0,1$, such that
\be
\label{blowup}
\langle\exp
\left(2pu+I'_2(S) +tI'_2(b)\right)\rangle_{\widehat\xi,\widehat
X}
=\langle\exp\left(2pu+I'_2(S)+F_j(u,t)\right)\rangle_{\xi,X}.
\ee
Thus, we see that a blowup formula ought to exist, whereby the blowup factor depends only on $u, t$ and not on the underlying four-manifold $X$.

To determine the explicit form of the blowup factor due to $F_j(u,t)$, one just needs to express the ``ramified'' u-plane integral corresponding to the LHS of (\ref{blowup}), in terms of the original ``ramified'' $u$-plane integral on $X$ corresponding to the correlation function $\langle\exp\left(2pu+I'_2(S)\right)\rangle_{\xi,X}$.   This entails replacing $S$ with $\widetilde S = S + tb$ and $w_2(E)$ with $w_2(\widetilde E) = w_2(E) + jb$ in the original ``ramified'' $u$-plane integral, while working in a chamber where $B_+  \to 0$ ($B$ being the cohomology class that is dual to $b$). For a \emph{nontrivially-embedded} surface operator, we have $\eta =0$; consequently, the only difference between the ``ramified'' lattice sum $\Psi'$ in (\ref{Psi'}) and the ordinary lattice sum $\Psi$ in eqn.~(3.18) of \cite{Moore-Witten}, is that $w_2(E)$ must be replaced by $(w_2(E) - \alpha \delta_D)$ in going from $\Psi$ to $\Psi'$. With this understood, we find that (cf. eqn.~(6.4) of~\cite{Moore-Witten})
\be
\label{Psi to Psi}
\Psi'_{\widehat X}  = \Psi'_{X}
\exp[ {   t^2 \over  8 \pi y   h^2}]
\sum_{n\in \IZ + \half w_2(\widetilde E)\cdot B }
\exp\bigl[  i \pi   \tau n^2 +   i n t  /  h ]
e^{- i \pi n},
\ee
where we have made use of the fact that $ \delta_D \cdot B = D \cap b =0$ in determining $n$. In fact, (\ref{Psi to Psi}) holds for a\emph{ trivially-embedded} surface operator as well: there is an extra factor of ${\rm exp}\{ -i \pi \eta (B \cdot \delta_D) \}$ on the RHS of (\ref{Psi to Psi}) when the surface operator is trivially-embedded, but since $ D \cap b =0$, it is simply equal to 1.  
 
Similarly, the measure factor for the blown-up manifold is related to that of the original manifold by:
\be
 \hat f_{\widehat X} = \hat f_X  \vartheta_4^{-1}
\exp \bigl[  -t^2 \hat T_S(u)  \bigr].
\ee
The $\vartheta_4^{-1}$ factor arises because the blowup changes $\chi$ and $\sigma$, and the other factor arises because $\widetilde S^2=S^2-t^2$. 

The blowup formula is thus given by 
\be
Z'_{D; \widehat \xi; \widehat X} = {\mathscr B}_{0, 1}(t \vert u) \, Z'_{D; \xi; X},
\ee
where in the case that $w_2(\widetilde E) \cdot B = 0 \,\, {\rm mod} \,\, 2$, 
\be
\label{blowup 1}
{\mathscr B}_0(t \vert u) = e^{-t^2 T_S(u)} {\vartheta_4\bigl({t\over  2 \pi h}  \vert \tau\bigr)
\over  \vartheta_4(0 \vert \tau) },
\ee
and in the case that $w_2(\widetilde E) \cdot B = 1 \,\, {\rm mod} \,\, 2$,
\be
\label{blowup 2}
{\mathscr B}_1(t \vert u) = e^{-t^2 T_S(u)} {\vartheta_1\bigl({t\over  2 \pi h}  \vert \tau\bigr)
\over  \vartheta_4(0 \vert \tau) }.
\ee
Note that the blowup factors (\ref{blowup 1}) and (\ref{blowup 2}) coincide with those for the generating function of the ordinary Donaldson invariants computed in~\cite{MW 10,stern}. In hindsight,  this is not at all surprising, since the blowup happens ``away'' from the surface $D$. Nevertheless, the corresponding ``ramified'' Donaldson invariants are  \emph{far from being the same} as the ordinary Donaldson invariants: the universal formula in (\ref{proof 1-final}) which tells us that the ``ramified'' Donaldson invariants are proportional to the ordinary Donaldson invariants, holds in the case that $b^+_2 > 1$, but here, $b^+_2 =1$.   

As the notation suggests, 
\be
{\mathscr B}_0(t \vert u)  = e^{-t^2 u /3} \sigma_3({4t \over \sqrt 2}), \qquad \, \,  {\mathscr B}_1(t \vert u)  = {\sqrt 2 \over 4}e^{-t^2 u /3} \sigma({4t \over \sqrt 2}),
\ee
where $\sigma_3(t)$ and $\sigma(t)$ are functions given by series expansions that truncate at finite powers of $t$ with coefficients being polynomials in $u$~\cite{Marcos}:
\be
\sigma_3(t) = 1 + {u \over 12} t + \CO(t^2), \qquad \, \sigma(t) = t - {1 \over 960} ({u^2 \over 3} - {1\over 4}) t^5 + \CO(t^7).
\ee
Thus, the blowup factors depend on $u,t$ only, as claimed.

\vspace{1.0cm}
\hspace{-1.0cm}{\large \bf Acknowledgements:}\\
\vspace{-0.5cm}

I would like to thank S.~Gukov, G.~Moore, N.~Seiberg, P.~Sulkowski, K.~Vyas and especially C.~LeBrun and  T.S.~Mrowka, for illuminating exchanges. This work is supported by the California Institute of Technology and the NUS-Overseas Postdoctoral Fellowship.

\appendix

\newsection{Elliptic Curves, Congruence Subgroups, And Modular Forms}

Here we collect some useful facts and notations related to various modular forms that appear in the paper.

The covariant Eisenstein function of weight two is
$\hat E_2$ where:
\begin{align}
E_2 & = 1 - 24 q + \cdots\cr
\hat E_2 & \equiv E_2 - { 3 \over  \pi y}.
\end{align}
Under $SL(2,\IZ)$ transformations, we have
\be
E_2(\tau) \to (c \tau + d)^2 \left(  E_2(\tau) + {12 c \over 2 \pi i(c \tau +d)}  \right).
\ee

Our conventions for the Jacobi theta functions are:
\begin{align}
\vartheta_1(\nu \vert \tau)& = i \sum_{n\in\IZ} (-1)^n q^{\half(n+\half)^2} e^{i \pi (2n +1) \nu} \cr
\vartheta_2(\nu \vert \tau)& =  \sum_{n\in\IZ}  q^{\half(n+\half)^2} e^{i \pi (2n +1) \nu} \cr
\vartheta_3(\nu \vert \tau)& =  \sum_{n\in\IZ}  q^{\half n^2} e^{i \pi 2n \nu}  \\
\vartheta_4(\nu \vert \tau)& =  \sum_{n\in\IZ} (-1)^n q^{\half n^2} e^{i \pi 2n \nu} \nonumber
 \end{align}
where $q = e^{2\pi i \tau}$. Also,
\begin{align}
\vartheta_2& = 2 q^{1/8}\prod (1-q^n)(1+q^n)^2\cr
& = \sum_{n\in\IZ} q^{\half(n+\half)^2} = 2 q^{1/8} + \cdots\cr
\vartheta_3& =  \prod (1-q^n)(1+q^{n-\half} )^2\cr
& = \sum_{n\in\IZ} q^{\half n^2} = 1 + 2 q^\half + \cdots\\
\vartheta_4& =  \prod (1-q^n)(1-q^{n-\half} )^2\cr
& = \sum_{n\in\IZ} q^{\half n^2}(-1)^n = 1 - 2 q^\half + \cdots \nonumber
\end{align}

The Seiberg-Witten curve is:
\begin{align}
\label{A5}
y^2 = x^3 -   u x^2 + {\Lambda^4\over  4} x
\end{align}
If we set $\Lambda=1$ (as we did in the main text), the  singularities will be at:
 $  u =1$ for the monopole cusp and $u=-1$ for the
dyon cusp. Then, the family of Seiberg-Witten curves defined by (\ref{A5}) will be parameterized by the modular curve of $\Gamma^0(4)$.

As explained in the main text, the $u$-plane can also be identified as the modular curve of $\Gamma(2)$,
which parametrizes a family of Seiberg-Witten curves defined by $y^2=(x^2-1)(x-u)$ instead.  The two families
of Seiberg-Witten curves differ by a two-isogeny.  The translation between the two descriptions is
given by:
\begin{align}
 u & = \tilde u \cr
 \tau & = 2 \tilde \tau \\
 a & = \tilde a/2 \cr
 a_d & = \tilde a_d \nonumber
\end{align}
where quantities in the $\Gamma(2)$ description are denoted with a tilde.

In terms of theta functions we have:
\begin{align}
 u & = \half { \vartheta_2^4 + \vartheta_3^4 \over  (\vartheta_2 \vartheta_3)^2
} \cr
u^2- 1 & = {1 \over  4} {\vartheta_4^8 \over  (\vartheta_2 \vartheta_3)^4 } = {
\vartheta_4^8 \over  64 h^4(\tau)} \cr
{ i \over  \pi} {du \over  d \tau} & = {1 \over  4}
 {\vartheta_4^8 \over  (\vartheta_2 \vartheta_3)^2 } \\
\Biggl( { ( {2i \over  \pi} {du \over  d \tau} )^2 \over  u^2-1 } \Biggr)^{1/8}
& =
\vartheta_4 \cr
h(\tau) & =
\dau = \half \vartheta_2 \vartheta_3 \nonumber
\end{align}

 The following $q$-expansions are also important:

\begin{align}
u = u_{(\infty,0)} & = { 1 \over  8 q^{1/4}} \biggl( 1 + 20 q^{1/2} - 62 q +
216 q^{3/2} + \cdots \biggr) \cr
& = {1\over {8\,{q^{{1\over 4}}}}} + {{5\,{q^{{1\over 4}}}}\over 2} -
  {{31\,{q^{{3\over 4}}}}\over 4} + 27\,{q^{{5\over 4}}} - {{641\,{q^{{7\over
4}}}}\over 8} +
  {{409\,{q^{{9\over 4}}}}\over 2} + \cdots 
\end{align}

\begin{align}
 u_{(\infty,1)} & = -{i \over  8 q^{1/4}}
   + {{5\,i}\over 2}\,{q^{{1\over 4}}} +
   {{31\,i}\over 4}\,{q^{{3\over 4}}} + 27\,i\,{q^{{5\over 4}}} +
   {{641\,i}\over 8}\,{q^{{7\over 4}}} + {{409\,i}\over 2}\,{q^{{9\over 4}}} +
\cdots 
\end{align}

\be
u_{M}(q_M) =
1 + 32\,q_M + 256\,{q_M^2} + 1408\,{q_M^3} + 6144\,{q_M^4} + 22976\,{q_M^5} +
76800\,{q_M^6}+\cdots
\ee

\begin{align}
 T(u) & = -{1 \over  24}  \biggl[ {  E_2\over  h(\tau)^2}  - 8 u\biggr]\cr
&
= q^{1/4} - 2\ q^{3/4} + 6\ q^{5/4} - 16\ q^{7/4} +
    37\ q^{9/4} - 78\ q^{11/4} \\
& + 158\ q^{13/4} - 312\ q^{15/4} +
    594\ q^{17/4}  + \cdots \nonumber
    \end{align}

\begin{align}    
T_M(q_M) & = \half + 8\ q_M + 48\ q_M^2 + 224\ q_M^3 + 864\ q_M^4 + 2928\ q_M^5
+ 9024\ q_M^6 + \cdots 
\end{align}

\begin{align}
h & = \half \vartheta_2 \vartheta_3 = {1\over 4}\vartheta_2^2(\tau/2)\cr
&=  q^{1/8} + 2\ q^{5/8} + q^{9/8} + 2\ q^{13/8} +
    2\ q^{17/8} + 3\ q^{25/8} + 2\ q^{29/8} + \cdots 
\end{align}

\begin{align}    
h_M& = {1 \over  2i} \vartheta_3 \vartheta_4 = {1\over  2i}\vartheta_4^2(2
\tau_M) \cr
&={1\over  2i} (1- 4 q_M + 4 q_M^2 + 4 q_M^4 -8 q_M^5 + \cdots ) 
\end{align}

Last but not least, near the monopole cusp we have:
\be
a_d(q_M) =
16i q_M( 1 + 6 q_M + 24 q_M^2 + 76 q_M^3 + \cdots)
\ee


\begin{thebibliography}{99}

\bibitem{structure}

P.B.~Kronheimer and T.S.~Mrowka,  ``Embedded Surfaces and the Structure of Donaldson's Polynomial Invariants'', J.~Differential~Geom. Vol.~41, No.~{{\bf 3}} (1995), 573-734.

\bibitem{structure 1}

P.B.~Kronheimer and T.S.~Mrowka, ``Recurrence relations and asymptotics for four-manifold invariants'', Bull. Amer. Math. Soc. (N.S.) {\bf 30} (1994) 215-221.


\bibitem{donaldson}

S.~Donaldson, ``Polynomial Invariants For Smooth Four-Manifolds,'' Topology {\bf 29} (1990) 257.


\bibitem{Witten}

E.~Witten, ``Topological Quantum Field Theory,'' Commun. Math. Phys. {\bf 117} (1988) 353.


\bibitem{SW}

N.~Seiberg and E.~Witten, ``Monopole Condensation, And Confinement In $N=2$ Supersymmetric Yang-Mills
Theory,'' [arXiv:hep-th/9407087]; Nucl. Phys. {\bf B426} (1994) 19; ``Monopoles, Duality and Chiral Symmetry Breaking in N=2 Supersymmetric QCD,'' [arXiv:hep-th/9408099]; Nucl. Phys. {\bf B431} (1994) 484.


\bibitem{monopoles}

E.~Witten, ``Monopoles and Four-Manifolds'', Math.Res.Lett. {\bf 1}:769-796,1994. [arXiv:hep-th/9411102].



\bibitem{Moore-Witten}

G.~Moore and E.~Witten, ``Integration Over the $u$-plane in Donaldson Theory'', Adv.Theor.Math.Phys.{\bf 1}:298-387,1998. [arXiv:hep-th/9709193].


\bibitem{sym}

E.~Witten, ``Supersymmetric Yang-Mills Theory On A Four-Manifold'', J.Math.Phys.{\bf 35}:5101-5135,1994. [arXiv:hep-th/9403195].


\bibitem{KM1}

P.B.~Kronheimer and T.S.~Mrowka, ``Gauge Theory for Embedded Surfaces: I'', Topology Vol. {\bf 32} (1993).


\bibitem{KM2}

P.B.~Kronheimer and T.S.~Mrowka, ``Gauge theory for embedded surfaces: II'', Topology {{\bf 34}} (1995) 37-97.




\bibitem{Preskill}

J.~Preskill and L.M.~Krauss, ``Local Discrete Symmetry and Quantum Mechanical Hair,.'' Nucl. Phys. {\bf{B341}} (1990) 50-100.

\bibitem{Alford}

M.G.~Alford, K.-M.~Lee, J.~March-Russell and J.~Preskill, ``Quantum Field Theory of Non-Abelian Strings and Vortices,'' Nucl. Phys. {\bf{B384}} (1992) 251-317, [arXiv:hep-th/9112038].

\bibitem{Bucher}

M.~Bucher, K.-M.~Lee, and J.~Preskill, ``On Detecting Discrete Chesire Charge,'' Nucl. Phys. {\bf{B386}} (1992) 27-42, [arXiv:hep-th/9112040].


\bibitem{Braverman}

A.~Braverman, ``Instanton counting via affine Lie algebras I: Equivariant J-functions of (affine) flag manifolds and Whittaker vectors'', [arXiv:math/0401409];
A.~Braverman, P.~Etingof, ``Instanton counting via affine Lie algebras II: from Whittaker vectors to the Seiberg-Witten prepotential'', [arXiv:math/0409441].
M.~Henningson, ``Commutation relations for surface operators in six-dimensional (2, 0) theory'', JHEP 0103 (2001) 011; [arXiv:hep-th/0012070].



\bibitem{n=4} 

S.~Gukov, ``Surface Operators and Knot Homologies'', [arXiv:0706.2369]; S.~Gukov, E.~Witten, ``Rigid Surface Operators'', [arXiv:0804.1561]; 
A. Di Giacomo, V.I. Zakharov, ``Surface operators and magnetic degrees of freedom in Yang-Mills theories'', [arXiv:0806.2938]; N.~Wyllard, ``Rigid surface operators and S-duality: some proposals'', [ arXiv:0901.1833]. 

\bibitem{Gukov-Witten}

S.~Gukov and E.~Witten, ``Gauge Theory, Ramification, And The Geometric Langlands Program'',  Current Developments in Mathematics Volume 2006 (2008), 35-180. [arXiv:hep-th/0612073]. 

\bibitem{langlands}

A.~Beilinson and V.~Drinfeld, ``Quantization Of Hitchin's
Integrable System And Hecke Eigensheaves,'' preprint (ca. 1995),
http://www.math.uchicago.edu/~arinkin/langlands/.


\bibitem{2}

E.~Koh and S.~Yamaguchi, ``Holography of BPS surface operators'', JHEP 0902:012,2009, [arXiv:hep-th/0812.1420].

\bibitem{3}

N.~Drukker, J.~Gomis and S.~Matsuura, `` Probing N=4 SYM With Surface Operators'', JHEP0810:048,2008, [arXiv:hep-th/0805.4199].


\bibitem{4}

E.~Buchbinder, J.~Gomis and F.~Passerini, ``Holographic Gauge Theories in Background Fields and Surface Operators'', JHEP0712:101,2007, [arXiv:hep-th/0710.5170].

\bibitem{5}

E.~Koh and S.~Yamaguchi, ``Surface operators in the Klebanov-Witten theory'',  JHEP 0906:070,2009; [arXiv:hep-th/0904.1460].



\bibitem{8}

J.~Gomis and S.~Matsuura, `` Bubbling Surface Operators And S-Duality'', JHEP0706:025,2007, [arXiv:hep-th/0704.1657].


\bibitem{mine}

M.C.~Tan, ``Surface Operators in N = 2 Abelian Gauge Theory'', 	JHEP 0909:047,2009. [arXiv:0906.2413].

\bibitem{loop}

L.F.~Alday et al., ``Loop And Surface Operators In N=2 Gauge Theory And Liouville Modular Geometry''. [arXiv:0909.0945].

\bibitem{davide} 

D.~Gaiotto, ``Surface Operators in N=2 4d Gauge Theories'', [arXiv:0911.1316]. 


\bibitem{Appendix 1}

M.C.~Tan, ``Notes On The ``Ramified'' Seiberg-Witten Equations And Invariants''. [arXiv:hep-th/0912.1891].


\bibitem{Marcos}

J.~Labastida and M.~Marino, ``Topological Quantum Field Theory And Four-Manifolds'', Mathematical Physics Studies, vol.~{\bf 25}, Springer.



\bibitem{QFT2}

``Quantum Fields and Strings, A Course for Mathematicians. Vol. 2'', AMS IAS. 


\bibitem{Seiberg}

N.~Seiberg,``Supersymmetry and Non-Perturbative Beta Functions'' Phys. Lett. {\bf 206}{\bf B} (1988) 75.

\bibitem{divecchia}

D'Adda, A., DiVecchia, P., ``Supersymmetry and Instantons''. Phys. Lett. {\bf 73B}, 162 ( 1978).


\bibitem{coleman}

S.~Coleman,  ``Aspects Of Symmetry: Selected Erice Lectures'', Cambridge University Press.  

\bibitem{Lerche}

W.~Lerche, ``Introduction to Seiberg-Witten Theory and its Stringy Origin''. [arXiv:hep-th/9611190] 



\bibitem{hori}

A.~Hannany, K.~Hori, ``Branes and N =2 Theories in Two Dimensions'', 	Nucl.Phys. B{\bf 513} (1998) 119-174; [arXiv:hep-th/9707192]. 


\bibitem{MS}

K.~Hori et al., ``Mirror Symmetry'', Clay Mathematics Monographs, V. 1.  


\bibitem{Witten-sol}

E.~Witten, ``Solutions Of Four-Dimensional Field Theories Via M Theory'', Nucl.Phys.B {\bf 500} :3-42,1997. [arXiv:hep-th/9703166].

\bibitem{vafa}

M.~Bershadsky, V.~Sadov, C.~Vafa,``D-Branes and Topological Field Theories'', Nucl.Phys. B{\bf 463} (1996) 420-434. [arXiv:hep-th/9511222].

\bibitem{mine first}

M.C. Tan, ``Surface Operators in Abelian Gauge Theory'', JHEP 0905:104.2009. [arXiv:0904.1744]. 

\bibitem{Borcherds}

R.~Borcherds, ``Automorphic forms with singularities on Grassmannians''; Invent. Math. {\bf 132}, 491-562 (1998). [arXiv:alg-geom/9609022].

\bibitem{MW 15}   

L. G\"ottsche, ``Modular forms and Donaldson
invariants for 4-manifolds with $b_+=1$,'' [arXiv:alg-geom/9506018].



\bibitem{MW 10}

L. G\"ottsche and D. Zagier,
``Jacobi forms and the structure of Donaldson
invariants for 4-manifolds with $b_+=1$,''
[arXiv:alg-geom/9612020].




\bibitem{ref 17}

H. J. Hoppe and H. Spindler, ``Modulraume Stabiler 2-Bundel
Auf Regeflachen,'' Math. Ann. {\bf 249} (1980) 127.

\bibitem{stern}

R. Fintushel and R.J. Stern,
``The blowup formula for Donaldson invariants,''
[arXiv:alg-geom/9405002]; Annals of Math. {\bf 143} (1996) 529.










\end{thebibliography}
\end{document}